\documentclass[fleqn,11pt,oneside]{scrbook}
\usepackage{fullpage}
\usepackage{multirow}
\usepackage{mathptmx}
\usepackage{helvet}
\usepackage{courier}
\usepackage{type1cm}         

\usepackage{makeidx}         
\usepackage{graphicx}        
\usepackage{multicol}        
\usepackage[bottom]{footmisc}

\usepackage{amsmath,amsfonts,amssymb}
\usepackage[affil-it]{authblk}
\usepackage{enumerate}
\usepackage{subfigure}
\usepackage{longtable}
\usepackage{array}
\usepackage{tabularx}
\usepackage{bm}
\usepackage{hyperref}

\begin{document}

\title{\textbf{Bursty Human Dynamics}}






\begin{titlepage}
\fontfamily{phv}\selectfont
    \begin{center}
        \vspace*{2cm}

        {\huge\bfseries Bursty Human Dynamics\par}
   
        \vspace{1.5cm}
        
	\large{M\'arton Karsai$^{1,*}$, Hang-Hyun Jo$^{2,3,4}$, Kimmo Kaski$^{4}$} \\
	
        \vspace{.5cm}
        
        \footnotesize{$^{1}$Laboratoire de l'Informatique du Parall\'elisme, ENS de Lyon, France \\
	$^{2}$Asia Pacific Center for Theoretical Physics, Pohang, Republic of Korea\\
	$^{3}$Department of Physics, Pohang University of Science and Technology, Pohang, Republic of Korea\\
	$^{4}$Department of Computer Science, School of Science, Aalto University, Espoo, Finland\\
  	$^{*}$Corresponding author: marton.karsai@ens-lyon.fr}
        
        \vspace{0.8cm}
\parbox{0.8\linewidth}{\textbf{Abstract:} Bursty dynamics is a common temporal property of various complex systems in Nature but it also characterises the dynamics of human actions and interactions. At the phenomenological level it is a feature of all systems that evolve  heterogeneously over time by alternating between periods of low and high event frequencies. In such systems, bursts are identified as periods in which the events occur with a rapid pace within a short time-interval while these periods are separated by long periods of time with low frequency of events. As such dynamical patterns occur in a wide range of natural phenomena, their observation, characterisation, and modelling have been a long standing challenge in several fields of research. However, due to some recent developments in communication and data collection techniques it has become possible to follow digital traces of actions and interactions of humans from the individual up to the societal level. This led to several new observations of bursty phenomena in the new but largely unexplored area of human dynamics, which called for the renaissance to study these systems using research concepts and methodologies, including data analytics and modelling. As a result, a large amount of new insight and knowledge as well as innovations have been accumulated in the field, which provided us a timely opportunity to write this brief monograph to make an up-to-date review and summary of the observations, appropriate measures, modelling, and applications of heterogeneous bursty patterns occurring in the dynamics of human behaviour.}
 
\vspace{5cm}

\normalsize{2018}

\vspace{1cm}
   
\scriptsize{The final publication is available at Springer via http://dx.doi.org/10.1007/978-3-319-68540-3.}
\end{center}
\end{titlepage}








\frontmatter
\tableofcontents
\include{acronym}



\mainmatter

\chapter{Introduction}
\label{chapter:intro}

To begin with, one defines bursty behaviour or burstiness of a system as intermittent increases and decreases in the activity or frequency of events. Such a dynamical system showing large temporal fluctuations cannot be characterised by a Poisson process with a single temporal scale. Rather it can be considered as a result of non-Poissonian dynamics with strong temporal heterogeneities on various temporal scales\footnote{Non-Poissonian bursty dynamics is in general characterised by the heterogeneous distribution of inter-event times passing between the consecutive occurrences of a given type of event. In contrast a system with Poissonian dynamics, inter-event times are distributed exponentially. However, many empirical inter-event time distributions are broad and follow a log-normal, Weibull, or power-law form, implying that the underlying mechanisms behind them maybe different than a Poisson process. See more about this question in Chapter~\ref{chapter:meas}.}.

\begin{figure}[!ht]
\centering
  \includegraphics[width=.8\textwidth]{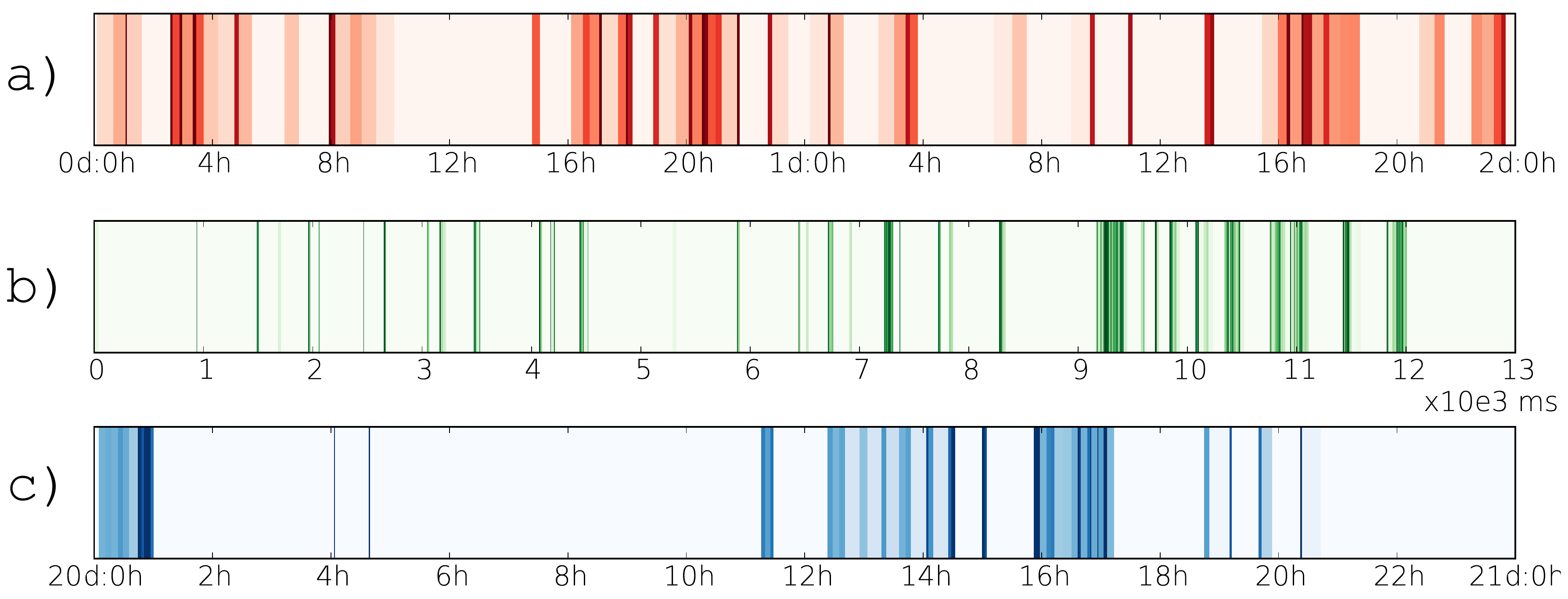}
  \caption{(a) Sequence of earthquakes with magnitude larger than two at a single location (south of Chishima Island, 8th--9th October 1994). (b) Firing sequence of a single neuron from a rat's hippocampal. (c) Outgoing mobile phone call sequence of an individual. The shorter the time between the consecutive events are, the darker color is coded. (\emph{Source:} Adapted from Ref.~\cite{Karsai2012Universal} under a Creative Commons Attribution-NonCommercial-ShareAlike 3.0 Unported License.)}
\label{fig:BurstySignals}
\end{figure}

There are a number of systems in Nature that evolve following non-Poissonian dynamics. One of the commonly known examples is the emergent dynamics of earthquakes~\cite{Corral2004LongTerm, Davidsen2013Earthquake,Bak2002Unified, deArcangelis2006Universality,Smalley1987A}, in which the times of shocks occurring at a given location show bursty temporal patterns, as illustrated  in Fig.~\ref{fig:BurstySignals}(a). The occurrence of such events is governed by the modified Omori's Law~\cite{Powell1975Statistical}, which states that the frequency of aftershocks decreases as a power law and can lead to a broad inter-event time distribution of shocks, when observed over a longer period of time. Another example of a natural phenomenon exhibiting bursty temporal patterns is solar flares induced by huge and rapid releases of energy~\cite{McAteer2007The,Wheatland1998The}. It has been shown that the stochastic processes underlying these apparently different phenomena show such universal properties that lead to the same distributions of event sizes, inter-event times, and temporal clustering~\cite{deArcangelis2006Universality}. These kinds of heterogeneities in the behaviour of systems emerging from different origins have been explained in the frame of self-organised criticality (SOC)~\cite{Bak1996How}, which provides a commonly accepted example of a theory for describing the burstiness of a system.

Also in case of neuronal firing its sequences are featured as having bursty temporal patterns ~\cite{Tsubo2012Powerlaw,Kepecs2003Information,Grace1984The,Kemuriyama2010A}, as depicted in Fig.~\ref{fig:BurstySignals}(b) illustrating a firing sequence of a single neuron observed in-vitro in a rat's hippocampus. Consecutive firings of a single neuron but also of groups of neurons evolve in spike trains, in which the short high-activity periods are separated by periods without any activity. Moreover, it has been suggested that neuronal firing patterns might be the result of integrate-and-fire mechanism~\cite{Hesse2014Self}, commonly assumed to occur in self-organised critical systems. This theory accounts for bursty patterns evolving at the single neuron level, but at the same time could explain collective firing patterns in a connected network of neurons.

Further examples of burstiness have been observed in the context of biological evolution passing through bursty patterns~\cite{Uyeda2011The, Pagel1999Inferring}  both on short and long time-scales with consistent patterns in the times of divergence across taxonomic groups.  Here it has been argued~\cite{Uyeda2011The} that changes for short temporal scales (of the order of $\sim1$ million years) are constrained fluctuations and do not accumulate over time, while for long temporal scales ($\sim1$--$360$ million years) the evolution yields bursty patterns of increasing divergence due to radial phenotypic changes.

Burstiness is also seen in the contexts of ecology and animal dynamics where heterogeneous temporal patterns characterise the dynamics of single animal movements or even the evolution of larger ecological systems~\cite{Blonder2012Temporal,Bohorquez2009Common}. Several examples have shown~\cite{Proekt2012Scale,Sorribes2011Origin,Wearmouth2014Scaling} that the dynamics of animals, e.g., in initiating conflicts \cite{Proekt2012Scale}, communication, foraging~\cite{Sorribes2011Origin}, predators waiting in ambush~\cite{Wearmouth2014Scaling}, or the displacement of monkeys or mice~\cite{Boyer2012Nonrandom,Nakamura2008Of} form complex self-similar temporal patterns reproduced on multiple time scales very similarly to examples observed in human behaviour. In addition bursty temporal patterns of switching between contrasting activities have been found in case of humans~\cite{Barabasi2010Bursts} and animals as well as in mammalian wake-sleep patterns and in the stop-start motion of fruit flies~\cite{Reynolds2011On}. Based on these observations, it has been proposed~\cite{Sorribes2011Origin,Reynolds2011On} that such dynamics can be commonly investigated with priority queuing models, which were developed primarily to understand human behaviour, but could also be used  more generally to make an association between the activity dynamics of animals and humans.

Apart from the above examples, scale-invariant bursty temporal patterns have also been found in several man-made systems. One example is written text, in which successive occurrences of the same word display deviations due to burstiness from the Poisson behaviour and are well characterised by a stretched exponential or Weibull distribution function~\cite{Altmann2009Beyond}. In case of engineering systems perhaps some of the best examples of bursty behaviour are in the context of package-based traffic and wireless communication signals, which were found to evolve through non-Poissonian dynamics~\cite{Chlebus1995Is,Janevski2003Traffic,Lee2013Mobile,Paxson1995WideArea}. Due to their importance in package overload and resource management, thorough methodology has been developed to detect and predict bursty package arrival patterns, while several communication protocols were proposed to avoid such situations~\cite{Kouvatsos2009Traffic}. As a final example of man-made system behaving burstily we mention financial markets, in which non-Poissonian dynamics characterises time series of returns of financial assets, stock sales, order books, and other transactions. The characterisation of such phenomena falls within the scope of econophysics~\cite{Mantegna2007Introduction}, which has successfully applied methods borrowed from statistical physics and signal processing to understand the dynamics of financial systems.

Although the above mentioned examples represent vastly different systems they are all similar in showing bursty dynamical patterns at the phenomenological level. Due to these apparent similarities the expectation is that these systems can be studied with similar methodologies in terms of measuring and analysing their properties as well as developing analytical theories and modelling to describe their behaviour. Perhaps some of the best examples of these commonly applied developments are the concepts of self-organised criticality for $1/f$ noise~\cite{Bak1996How, Jensen1998Self,deArcangelis2006Universality,Ramos2011Self,Brunk2001Self,Beggs2003Neuronal}, priority queuing processes~\cite{Cobham1954Priority,Barabasi2005Origin}, and self-exciting point processes~\cite{Hawkes1971Spectra, Mehrdad2015Hawkes}. These have been successfully used to model several of the above-mentioned systems, suggesting common mechanisms, like integrate-fire, prioritising, or reinforcement processes, to be acting in the background. Moreover, the burstiness can appear at different scales of the system or levels of its organisation. In some cases it characterises the dynamics of single units, like the firing of a neuron, movement of an animal, earthquakes at a given location, or bursty overload of a single communication router. However, in other examples burstiness appears as the mesoscale or system-level phenomena, like the collective firing of networked neurons, collective migration of animals, emerging earthquakes in a larger area, or correlated bursty traffic in communication networks. All in all, these examples provide evidence of some form of universality and multi-scale feature of burstiness, which commonly appears in Nature, man-made systems, and also in human dynamics, as we will next explain in more detail.

\section{Bursty human dynamics}

Let us now shift our focus from bursty behaviour in physical, biological, ecological and man-made technological systems, on burstiness  appearing in human behaviour, social relations, and various other endeavours of human sociality. In the observation of these systems the technological development plays an ever-increasing role, by having facilitated novel means for people to connect, communicate and interact with each other, at the same time leaving behind digital footprints of these events. All these have already affected and continue to mold our social actions and behaviour including the functions and services of our societies to the level that we can speak about the techno-social behaviour of people.  In addition, the vast amounts of digital footprint data, which people generate using information communication technology (ICT), reflect their social interactions as part of their life course and as members of society.

In studying social systems the researchers have earlier confronted a quite insurmountable obstacle of the lack of data on human behaviour at multiple scales and channels of communication. The availability of large-scale data and recent advances in complex systems research, computational and data science, computer science, network science, and social science would facilitate the quantitative analysis and description of individual and social behaviour in a rather unprecedented way and detail. Advances in these areas were limited by the difficulties of getting access or collecting large amounts of detailed data (Big Data), which is  necessary for validating theories and developing quantitative approach. However, we are more and more in the position to follow the dynamics of multiple and simultaneous actions and interactions of individuals, the interaction dynamics of groups and communities, and even the evolution of large-scale social systems. All this is possible with access to large amount of anonymised data (for privacy preservation) collected from communication logs or personal electronic devices. This in turn allows us to observe directly the dynamics of millions of individuals or even to detect the emergence of collective behaviour ``in vivo'' with minimal observational bias or intervention. As we will discuss below these advances have already led  to various observations of bursty temporal patterns in several aspects of human dynamics.

Bursty behaviour has been found at different levels of human dynamics. At the behavioural level the timings of actions by individuals were shown to present heterogeneous temporal patterns, while similar dynamics have been observed in dyadic social interactions, or even in collective social phenomena. Among the first observations was the study of Eckmann~\emph{et al.}~\cite{Eckmann2004Entropy}, who reported a broad inter-event time distribution with a power-law tail by analysing the dataset for email correspondence recorded at a university domain. A few months after that Barab\'asi published his paper entitled by ``The origin of bursts and heavy tails in human dynamics''~\cite{Barabasi2005Origin}, where he proposed a priority queuing model to explain the broad inter-event time distributions. This seminal paper initiated an avalanche of studies to observe, characterise, and model bursty phenomena detected in a number of human activities. Various examples of burstiness were found, like emails~\cite{Eckmann2004Entropy,Barabasi2005Origin}, letter correspondence~\cite{Oliveira2005Human}, mobile phone calls and short messages~\cite{Karsai2012Universal}, web browsing~\cite{Dezso2006Dynamics}, printing~\cite{Harder2006Correlated}, library loans~\cite{Vazquez2006Modeling}, job submission to computers~\cite{Kleban2003Hierarchical}, and file transfer in computer network~\cite{Paxson1995WideArea}, or even in arm movements of human subjects~\cite{Coley2008Arm}, just to mention a few. To demonstrate a typical signal of bursty activity we show the outgoing mobile phone call activity of a single person in Fig.~\ref{fig:BurstySignals}(c). In addition, further examples were identified at the group or societal level, such as the emergence of causal temporal motifs~\cite{Kovanen2013Temporal}, the evolution of mass demonstrations, revolutions, global information cascades, and wars~\cite{Bouchaud2013Crises,Tang2010Stretched}. For further information about these phenomena we refer the reader to a popular science book by Barab\'asi~\cite{Barabasi2010Bursts}, which gives an entertaining summary about several of these observations.

In his original modelling study Barab\'asi suggested that bursty activity patterns could be the consequence of prioritising tasks~\cite{Barabasi2005Origin,Oliveira2005Human,Vazquez2006Modeling}. In other words people do not execute their ``to-dos'' in a random fashion but assign importance to each task at hand. This induces intrinsic correlations between different tasks and results in bursty patterns of completed activities. Since then alternative and fundamentally different approaches have been proposed. One of the main alternative concepts was suggested by Malmgren~\emph{et al.}~\cite{Malmgren2008Poissonian,Malmgren2009Universality}, who argued that ``human behaviour is primarily driven by external factors such as circadian and weekly cycles, which introduces a set of distinct characteristic time scales, thereby giving rise to heavy tails''. This approach assumes no intrinsic correlations in human activities but models the dynamics as alternating homogeneous and non-homogeneous Poisson processes. The third main modelling concept assumes strong correlations between consecutive events and employs memory functions~\cite{Vazquez2006Impact, Han2008Modeling}, self-exciting point processes~\cite{Masuda2013SelfExciting,Jo2015Correlated}, or reinforcement mechanisms~\cite{Karsai2012Correlated,Wang2014Modeling} in simulating bursty activity patterns. Finally, several other modelling ideas were suggested assuming self-organised criticality~\cite{Tang2010Stretched}, local structural correlations~\cite{Myers2014Bursty}, some dynamical process like random walk~\cite{Goetz2009Modeling}, contact process~\cite{Odor2014Slow}, or voter model~\cite{FernandezGracia2013Timing} to introduce heterogeneous temporal patterns at the individual or system levels.

All these efforts lead to the situation in which bursty human dynamics became a well-recognised research area, with wide-ranging studies, a rich set of methodologies, and several modelling concepts. Based on these advancements more far-reaching scientific questions have been addressed about the effects of non-Poissonian patterns of individuals on collective dynamical processes, whether they are ongoing or co-evolving with bursty action and interaction patterns of individuals. A typical example is the diffusion of information in a temporal social network where individuals interact in a bursty fashion but are connected together in a network where information can diffuse globally. The main question here is whether bursty dyadic interactions enhance or slow down the speed and/or control the emergence of globally spreading processes, like information diffusion, epidemics, or random walk~\cite{Holme2012Temporal}. Beyond the conventional modelling and simulation techniques of such processes, data-driven models and random reference systems~\cite{Karsai2011Small,Miritello2011Dynamical} were recently shown to be very successful in addressing such questions.

\section{About this monograph}

As we briefly summarised above the fascinating phenomenon of bursty dynamics of various human activities has been investigated widely over the last decade. All these studies contributed to this field that emerged with a broad set of observations, methodologies, modelling, and applications. Although there are still several open questions, this field became specialised enough to benefit from a structured review of already established results. This has been the main reason to motivate us to write this monograph. Over the last ten years categorically different interpretations were proposed to explain bursty patterns in human dynamics. Thus to inform the reader about all the concepts and ideas, beyond a categorical summary of earlier results, our secondary aim has been to introduce various explanations objectively and report the related scientific discussions.

After this brief introduction we have organised our work in five chapters. First in Chapter~\ref{chapter:meas} we summarise the relevant methodologies developed to observe, characterise, and measure the non-Poissonian dynamics of human activities. After the reader is familiarised with these techniques, in Chapter~\ref{chapter:emp} we turn to collect a number of related empirical observations of human bursty phenomena in various systems and at different organisational levels. In Chapter~\ref{chapter:model} we give a systematic summary of modelling concepts and principles, and finally in Chapter~\ref{chapter:processes} we discuss several studies addressing the effects of bursty behaviour on different dynamical processes. We close the monograph with a Chapter to summarise, discuss, and conclude as well as to propose some directions for future research.

To the reader of this monograph we want to emphasise that we focus exclusively on heterogeneous temporal patterns in human dynamics. Thus the observations and methodologies herein for studying other systems are out of our scope. More precisely we focus on systems where the observed phenomena directly reflect the dynamics of human actions or interactions. Hence we do not discuss the dynamics of systems that are only indirectly related to human actions, like in the case of financial or transportation systems. We also remark that although our aim has been to complete as comprehensive review as possible of the field of human bursty behaviour, we might have unintentionally missed some related articles, which we apologise for. Also note that a review paper has been written recently about related topics~\cite{Zhou2013Statistical}, however using a language which is not common in the international scientific community. Thus we hope that our work gives a valuable contribution to the field and helps students and experts who are interested to learn about bursty human dynamics.


\chapter{Measures and characterisations}
\label{chapter:meas}


In order to investigate the dynamics of human social behaviour quantitatively, we first introduce it as a time series and we show how it is characterised by means of various techniques of time series analysis. According to Box~\emph{et al.}~\cite{Box2008Time}, a time series is a set of observations that are made sequentially in time. The timing of an observation denoted by $t$ can be either continuous or discrete. Since most datasets of human dynamics have recently been recorded digitally, we will here focus on the case of discrete timings. In this sense, the time series can be called an event sequence, where each event indicates an observation. In this series the $i$th event takes place at time $t_i$ with the result of the observation $z_i$ that can denote a number, a symbol, or even a set of numbers, depending on what has been measured. The sequence of $\{(t_i,z_i)\}$ can be simply denoted by $z_t$. Some events could occur in a time interval or with duration. For example, a phone call between two individuals may last from few minutes to hours~\cite{Holme2012Temporal}. In many cases as the time scale for event duration is much smaller than that of our interest, the event duration will be ignored in our monograph unless stated otherwise.

In most cases a time series refers to observations made at regular timings. For a fixed time interval $t_{\textrm{int}}$, the timings are set as $t_i=t_0+ t_{\textrm{int}}i$ for $i=0, 1, 2,\cdots$. In many cases, $t_0$ and $t_{\textrm{int}}$ are fixed at the outset thus they can be ignored for time series analysis. An example of a time series with regular observations is the daily price of a stock in the stock market, constituting a financial time series~\cite{Mantegna2007Introduction}. Such time series are often analysed by using traditional techniques like the autocorrelation function with the aim to reveal the dependencies between observed values, which often show inhomogeneities and large fluctuations in them.

One also finds many cases in which the timings of observations are inhomogeneous, like in case of emails sent by a user~\cite{Barabasi2005Origin}. The fact that the occurrence of events is not regular in time leads to temporally inhomogeneous time series, potentially together with the variation of observed value $z_t$. In these cases we can talk about two kinds of inhomogeneities in observed time series. On the one hand, fluctuations are associated both with temporal inhomogeneities and with the variation of observations. On the other hand, inhomogeneities can be associated only with the timings of events, not with observation values. This is the case of several recent datasets, e.g., those related to communication or individual transactions. In such datasets events are typically not assigned with content due to privacy reasons, thus only their timings are observable. In the following Sections we will mainly focus on the latter type of time series.

We remark that the time series with regular timings but with irregular observed values could be translated into time series with irregular timings. This can be done, e.g., by considering only the observations with $z_t \geq z_{\textrm{th}}$, where $z_{\textrm{th}}$ denotes some threshold value. Then the time series can be generated, which contains only observations with extreme values, like crashes in the financial markets. In the opposite direction, the time series with irregular timings can be also translated into that with regular timings, e.g., by binning the observations over a sufficiently large time window $t_w$. More precisely, one can obtain the time series with regular timings as follows: 
\begin{equation}
    \tilde z_k\equiv \sum_{kt_w\leq t<(k+1)t_w} z_t
\end{equation}
for all possible integers $k$. This constitutes a coarse-graining process for the time series.

\section{Point processes as time series with irregular timings}
\label{subsect:point}

A time series with irregular timings can be interpreted as the realisation of a point process on the time axis. To introduce these interpretations, let us first disregard the information contained in the observation results $z_t$, as it is not generally accessible, and consider only the timings of events. On the one hand, the event sequence with $n$ events can be represented by an ordered list of event timings, i.e., $ev(t_i)=\{t_0,t_1,\cdots,t_{n-1}\}$, where $t_i$ denotes the timing of the $i$th event. On the other hand, the event sequence can be depicted as a binary signal $x(t)$ that takes a value of $1$ at time $t=t_i$, or $0$ otherwise. For discrete timings, one can write the signal as
\begin{equation}
x(t)=\sum_{i=0}^{n-1}\delta_{t,t_i},
\label{eq:xt}
\end{equation}
where $\delta$ denotes the Kronecker delta.

\subsection{The Poisson process}

The temporal Poisson process is a stochastic process, which is commonly used to model random processes such as the arrival of customers at a store, or packages at a router. It evolves via completely independent events, thus it can be interpreted as a type of continuous-time Markov process. In a Poisson process, the probability that $n$ events occur within a bounded interval follows a Poisson distribution
\begin{equation}
P(n)=\frac{\lambda^{n}e^{-\lambda}}{n!},
\label{eq:PoissEvRate}
\end{equation}
where $\lambda$ denotes the average number of events per interval, which is equal to the variance of the distribution in this case. Since these stochastic processes consist of completely independent events, they have served as reference models when studying bursty systems. As we will see later, bursty temporal sequences emerge with fundamentally different dynamics with strong temporal heterogeneities and temporal correlations. Any deviation in their dynamics from the corresponding Poisson model can help us to indicate patterns induced by correlations or other factors like memory effects.

Throughout the monograph we are going to refer to two types of Poisson processes. One type, called the \emph{homogeneous Poisson process}, is characterised by a constant event rate $\lambda$, while the other type, called the \emph{non-homogeneous Poisson process}, is defined such that the event rate varies over time, denoted by $\lambda(t)$. For more precise definitions and discussion on the characters of Poisson processes we suggest the reader to study the extended literature addressing this process, e.g., Ref.~\cite{Grimmett2009Probability}. We remark that the Poisson processes and their variants have been studied in terms of shot noise in electric conductors and related systems~\cite{Blanter2000Shot, Lowen1990Powerlaw, Bulashenko2000Suppression}.

\subsection{Characterisation of temporal heterogeneities}

\begin{figure}[!t]
    \center
    \includegraphics[width=\columnwidth]{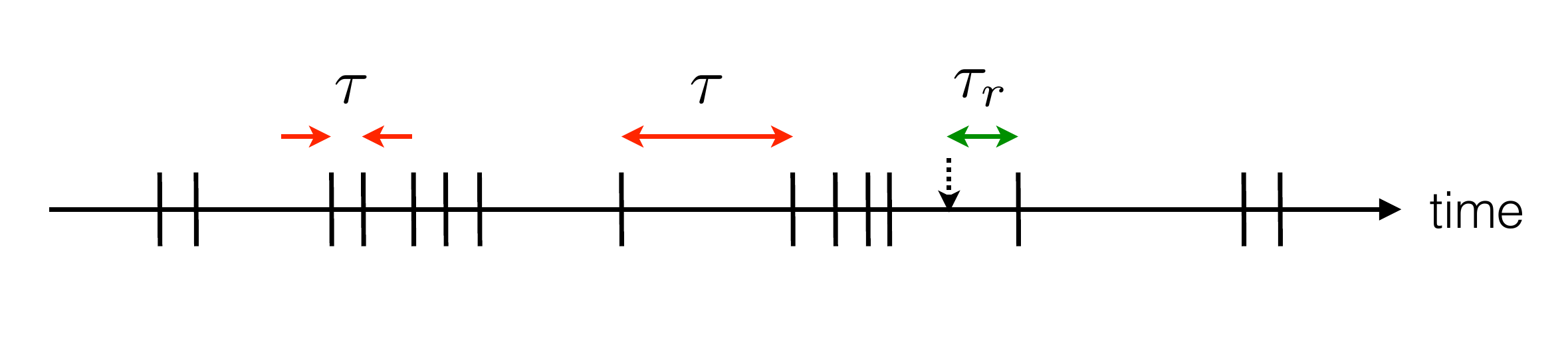}
    \caption{Schematic diagramme of an event sequence, where each vertical line indicates the timing of the event. The inter-event time $\tau$ is the time interval between two consecutive events. The residual time $\tau_r$ is the time interval from a random moment (e.g., the timing annotated by the vertical arrow) to the next event. In most empirical datasets, the distributions of $\tau$ are heavy-tailed.}
    \label{fig:scheme1}
\end{figure}

The temporal irregularities of an event sequence can be characterised in terms of various quantities. For this, a schematic diagramme and a realistic example of such event sequences are respectively depicted in Fig.~\ref{fig:scheme1} and Fig.~\ref{fig:example_model}(a), where the example has been generated using a model for bursty dynamics~\cite{Jo2015Correlated}. 

\subsubsection{The inter-event time distribution}

In order to formally introduce these measures let us first consider an event sequence $ev(t_i)$ and define the inter-event time as
\begin{equation}
    \tau_i\equiv t_i-t_{i-1},
\end{equation}
which is the time interval between two consecutive events at times $t_{i-1}$ and $t_i$ for $i=1,\cdots,n-1$. Then we obtain a sequence of inter-event times, i.e., $iet(\tau_i)=\{\tau_1,\cdots,\tau_{n-1}\}$, where $n\geq 2$ is assumed. By ignoring the order of $\tau_i$s, we can compute the probability density function of inter-event times, i.e., the inter-event time distribution $P(\tau)$. For completely regular time series, all inter-event times are the same and equal to the mean inter-event time, denoted by $\langle\tau\rangle$, thus the inter-event time distribution reads as follows:
\begin{equation}
P(\tau)=\delta(\tau-\langle\tau\rangle),
\end{equation}
where $\delta(\cdot)$ denotes the Dirac delta function. Here the standard deviation of inter-event times, denoted by $\sigma$, is zero. 

For the completely random and homogeneous Poisson process, it is easy to derive~\cite{Grimmett2009Probability} that the inter-event times are exponentially distributed as follows:
\begin{equation}
    P(\tau)=\frac{1}{\langle\tau\rangle} e^{-\tau/\langle\tau\rangle},
\end{equation}
where $\sigma=\langle\tau\rangle$. Note that the event rate introduced in Eq.~(\ref{eq:PoissEvRate}) is $\lambda=1/\langle \tau \rangle$.

\begin{figure}[!t]
   \includegraphics[width=\columnwidth]{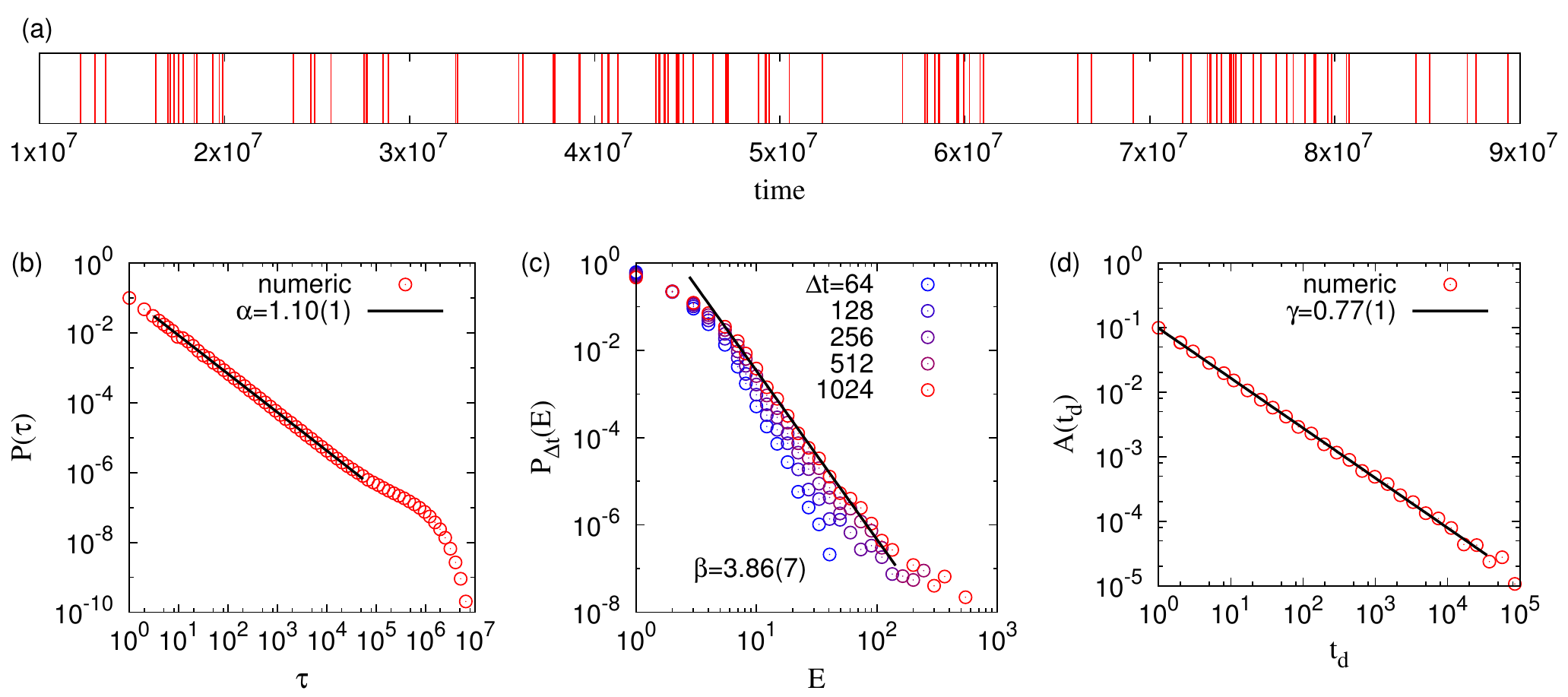}
    \caption{(a) An example of the realistic event sequence generated by a model with preferential memory loss mechanism for correlated bursts~\cite{Jo2015Correlated} using the parameter values of $\mu=0.1$, $\nu=2$, and $\epsilon=\epsilon_L=10^{-6}$. The bursty behaviour of the event sequence can be characterised by (b) inter-event time distribution $P(\tau)$, (c) bursty train size distribution $P_{\Delta t}(E)$ for time window $\Delta t$, and (d) autocorrelation function $A(t_d)$ with time delay $t_d$. In addition, the burstiness parameter and memory coefficient of the event sequence were estimated as $B\approx 0.483$ and $M\approx 0.038$, respectively.}
    \label{fig:example_model}
\end{figure}

Finally, in many empirical processes in nature and society, inter-event time distributions have commonly been observed to be broad with heavy tails ranging over several magnitudes. In such bursty time series the fluctuations characterised by $\sigma$ are much larger than $\langle\tau\rangle$, indicating that $P(\tau)$ is rather different from an exponential distribution, as it would derive from Poisson dynamics. Bursty systems evolve through events that are heterogeneously distributed in time. It leads to a broad $P(\tau)$, which can be fitted with either power law, log-normal, or stretched exponential distributions, just to name a few candidates. Most commonly, many empirical analyses show that $P(\tau)$ could be described in the power-law form with an exponential cutoff as
\begin{equation}
    P(\tau)\simeq C\tau^{-\alpha}e^{-\tau/\tau_c},
\end{equation}
where $C$ denotes a normalisation constant, $\alpha$ is the power-law exponent, and $\tau_c$ sets the position of the exponential cutoff. Refer to an example of the power-law $P(\tau)$ in Fig.~\ref{fig:example_model}(b). The power-law scaling of $P(\tau)$ indicates the lack of any characteristic time scale, but the presence of strong temporal fluctuations, characterised by the power-law exponent $\alpha$. Power-law distributions are also associated to the concepts of scale-invariance and self-similarity as demonstrated in Ref.~\cite{Newman2005Power}. In this sense, the value of $\alpha$ is deemed to have an important meaning, especially in terms of universality classes in statistical physics~\cite{Plischke2006Equilibrium}. Interestingly, as will be discussed in Chapter~\ref{chapter:emp}, a number of recent empirical researches have reported power-law inter-event time distributions with various exponent values.

Nevertheless, we note that although recent studies disclosed several bursty systems with broad inter-event time distributions, it is not trivial to identify the best functional form of distribution fitting the data points and to estimate its parameters like the value of power-law exponent. For the related statistical and technical issues, one can see Ref.~\cite{Clauset2009Powerlaw} and references therein. In addition, the effect of finite size of the observation period on the evaluation of inter-event time distributions has recently been discussed in Ref.~\cite{Kivela2015Estimating}.

\subsubsection{The burstiness parameter}

The heterogeneity of the inter-event times can be quantified by a single measure introduced by Goh and Barab\'asi~\cite{Goh2008Burstiness}. The burstiness parameter $B$ is defined as the function of the coefficient of variation (CV) of inter-event times $r\equiv \sigma/\langle \tau\rangle$ to measures temporal heterogeneity as follows:
\begin{equation}
    B\equiv \frac{r-1}{r+1}=\frac{\sigma-\langle\tau\rangle}{\sigma+\langle\tau\rangle}.
    \label{eq:burstiness_param}
\end{equation}
Here $B$ takes the value of $-1$ for regular time series with $\sigma=0$, and it is equal to $0$ for random, Poissonian time series where $\sigma=\langle \tau\rangle$. In case when the time series appears with more heterogeneous inter-event times than a Poisson process, the burstiness parameter is positive ($B>0$), while taking the value of $1$ only for extremely bursty cases with $\sigma \rightarrow \infty$. This measure has found a wide range of applications because of its simplicity, e.g., in analysing earthquake records, heartbeats of human subjects, and communication patterns of individuals in social networks, as well as for testing models of bursty dynamics~\cite{Goh2008Burstiness, Jo2012Circadian, Yasseri2012Dynamics, Kim2014Index, Wang2015Temporal, Zhao2015Empirical, Jo2015Correlated, Gandica2016Origin, Li2016Collective}.

However, it was recently shown that the range of $B$ is strongly affected by the number of events $n$ especially for bursty temporal patterns~\cite{Kim2016Measuring}. For the regular time series, the CV of inter-event times, $r$, has the value of $0$ irrespective of $n$ as all the inter-event times are the same. For the random time series, one gets $r=\sqrt{(n-1)/(n+1)}$ by imposing the periodic boundary condition to the time series. This case basically corresponds to the Poisson process. Finally, for the extremely bursty time series, one has $r=\sqrt{n-1}$, corresponding to the case when all events occur asymptotically at the same time. This implies the strong finite-size effect on the burstiness parameter for time series with moderate number of events. We also remark that $B=1$ is realised only when $n\to \infty$. Let us assume that one compares the degrees of burstiness of two event sequences but with different numbers of events in them. If the measured values of $B$ are the same for both event sequences, does it really mean that those event sequences are equally bursty? This is not a trivial issue. Thus, in order to fix these strong finite-size effects, an alternative measure has been introduced for the burstiness parameter in Ref.~\cite{Kim2016Measuring}:
\begin{equation}
    B_n\equiv \frac{\sqrt{n+1} r-\sqrt{n-1}}{(\sqrt{n+1}-2)r +\sqrt{n-1}},
\end{equation}
which was devised to have the value of $1$ for $r=\sqrt{n-1}$, $0$ for $r=\sqrt{(n-1)/(n+1)}$, and $-1$ for $r=0$, respectively. The authors claimed that using this measure, one can distinguish the finite-size effect from the intrinsic burstiness characterising the time series.

\subsubsection{The memory coefficient}

So far, we have ignored any possible correlation between inter-event times for the sake of simple description. As a first approximation to quantify dependencies between consecutive inter-event times, a joint distribution $P(\tau_i,\tau_{i+1},\cdots,\tau_{i+k})$ of arbitrary number of consecutive inter-event times can be directly studied in a non-trivial fashion as introduced in Ref.~\cite{Jo2015Correlated}. For a simpler description of such dependencies, Goh and Barab\'asi~\cite{Goh2008Burstiness} introduced the memory coefficient $M$ to measure two-point correlations between consecutive inter-event times as follows:
\begin{equation}
    M\equiv \frac{1}{n-2}\sum_{i=1}^{n-2}\frac{(\tau_i-\langle \tau\rangle_1)(\tau_{i+1}-\langle\tau\rangle_2)}{\sigma_1\sigma_2},
    \label{eq:memory_coeff}
\end{equation}
with $\langle\tau\rangle_1$ (respectively $\langle\tau\rangle_2$) and $\sigma_1$ (respectively $\sigma_2$) being the average and the standard deviation of inter-event times $\{\tau_i | i=1,\cdots, n-2\}$ (respectively $\{\tau_{i+1} | i=1,\cdots, n-2\}$). Beyond only considering consecutive inter-event times, this measure can be extended to capture correlations between inter-event times separated by exactly $m-1$ intermediate inter-event times ($m\geq 1$). As a general form, the memory coefficient can be written as follows:
\begin{equation}
  M_m\equiv \frac{1}{n-m-1}\sum_{i=1}^{n-m-1}\frac{(\tau_i-\langle \tau\rangle_1)(\tau_{i+m}-\langle\tau\rangle_2)}{\sigma_1\sigma_2}
\label{eq:Mgen}
\end{equation}
with the corresponding definitions of $\langle\tau\rangle_1$, $\langle\tau\rangle_2$, $\sigma_1$, and $\sigma_2$. Then, the set of $M_m$ for all possible $m$ may fully characterise the memory effects between inter-event times.

Note that an alternative measure, called the local variation, was introduced originally in neuroscience~\cite{Shinomoto2003Differences}. The local variation is defined as
\begin{equation}
    {\textrm{LV}} \equiv \frac{1}{n-2}\sum_{i=1}^{n-2}\frac{3(\tau_i-\tau_{i+1})^2}{(\tau_i+\tau_{i+1})^2},
\end{equation}
which takes the values of $0$, $1$, and $3$, respectively, for the regular, random, and extremely bursty time series. This measure has also been used to analyse datasets describing human bursty patterns~\cite{Aoki2016Inputoutput}.

We also introduce an entropy-based measure for the correlations between consecutive inter-event times that applies only to the power-law inter-event time distribution~\cite{Baek2008Testing}. If the inter-event time distribution is a power law as $P(\tau)\propto \tau^{-\alpha}$ for $\tau\geq \tau_{\textrm{min}}$, to each inter-event time $\tau_i$ one can assign a number $r_i$ as follows:
\begin{equation}
r_i=1-\left(\frac{\tau_i}{\tau_{\textrm{min}}}\right)^{1-\alpha},
\end{equation}
which will be uniformly distributed between $[0,1)$. Then the correlation between consecutive inter-event times is measured in terms of the mutual information using the joint probability density function $P(r_i, r_{i+1})$:
\begin{equation}
I(r_i;r_{i+1})\equiv \sum_{r_i}\sum_{r_{i+1}}P(r_i, r_{i+1}) \log\left[\frac{P(r_i, r_{i+1})}{P(r_i)P(r_{i+1})}\right].
\end{equation}
If $\tau_i$ and $\tau_{i+1}$ are fully uncorrelated, so are $r_i$ and $r_{i+1}$, leading to the zero value of the mutual information defined above.

\subsubsection{The autocorrelation function}

The conventional way for detecting correlations in time series is to measure the autocorrelation function. For this, we use the representation of event sequences as binary signals $x(t)$ as defined in Eq.~(\ref{eq:xt}). In addition, for a proper introduction we need to define the delay time $t_d$, which sets a time lag between two observations of the signal $x(t)$. Then the autocorrelation function with delay time $t_d$ is defined as follows:
\begin{equation}
  A(t_d)\equiv \frac{ \langle x(t)x(t+t_d)\rangle_t- \langle x(t)\rangle^2_t}{ \langle x(t)^2\rangle_t- \langle x(t)\rangle^2_t},
\end{equation}
where $\langle \cdot \rangle_t$ denotes the time average over the observation period. For more on the autocorrelation function, see Ref.~\cite{Box2008Time}. In the time series with temporal correlations, $A(t_d)$ typically decays as a power law:
\begin{equation}
A(t_d)\sim t_d^{-\gamma}
\end{equation}
with decaying exponent $\gamma$. One can see an example of the power-law decaying $A(t_d)$ in Fig.~\ref{fig:example_model}(d). In addition, note that one can relate $A(t_d)$ to the power spectrum or spectral density of the signal $x(t)$ as follows:
\begin{equation}
    \label{eq:powerSpectrum}
    P(\omega)=\left|\int x(t)e^{i\omega t}dt\right|^2 \propto \int A(t_d)e^{-i\omega t_d}dt_d,
\end{equation}
which appears as the Fourier transform of autocorrelation function. We are mostly interested in the power-law decaying power spectrum as
\begin{equation}
P(\omega)\sim\omega^{-\alpha_\omega}
\end{equation}
with $0.5<\alpha_\omega<1.5$, then the time series is called $1/f$ noise. $1/f$ noise has been ubiquitously observed in various complex systems~\cite{Bak1987Selforganized}, hence extensively studied for the last few decades.

The scaling relation between $\alpha$ and $\gamma$ has been studied both analytically and numerically. Let us first mention the relation between $\alpha_\omega$ and $\gamma$. If $A(t_d) \sim t_d^{-\gamma}$ for $0<\gamma<1$, then from Eq.~(\ref{eq:powerSpectrum}) one finds the scaling relation:
\begin{equation}
    \label{eq:alpha_omega_gamma}
    \alpha_\omega=1-\gamma. 
\end{equation}
When the inter-event times are i.i.d. random variables with $P(\tau)\sim \tau^{-\alpha}$, implying no interdependency between inter-event times, the power-law exponent $\alpha_\omega$ is obtained as a function of $\alpha$ as follows~\cite{Lowen1993Fractal, Allegrini2009Spontaneous}:
\begin{equation}
    \label{eq:alpha_omega_alpha}
    \alpha_\omega=\left\{\begin{tabular}{ll}
            $\alpha-1$ & for $1<\alpha\leq 2$,\\
            $3-\alpha$ & for $2<\alpha\leq 3$,\\
            $0$ & for $\alpha>3$.
        \end{tabular}\right.
\end{equation}
For this result, the following inter-event time distribution was used:
\begin{equation}
    P(\tau)=\left\{\begin{tabular}{ll}
            $\frac{\alpha-1}{a^{1-\alpha}-b^{1-\alpha}}\tau^{-\alpha}$ & for $0<a<\tau <b$,\\
            $0$ & otherwise.
        \end{tabular}\right.
\end{equation}
Combining Eqs.~(\ref{eq:alpha_omega_gamma}) and~(\ref{eq:alpha_omega_alpha}), we have
\begin{eqnarray}
    \label{eq:alpha_gamma}
    \begin{tabular}{ll}
        $\alpha+\gamma=2$ & for $1<\alpha\leq 2$,\\
        $\alpha-\gamma=2$ & for $2<\alpha\leq 3$,
    \end{tabular}
\end{eqnarray}
which have also been derived in Ref.~\cite{Vajna2013Modelling}. The above power-law exponents can be related via the Hurst exponent $H$, i.e., $\gamma=2-2H$~\cite{Kantelhardt2001Detecting} or $\alpha_\omega=2H-1$~\cite{Allegrini2009Spontaneous, Rybski2012Communication}. This indicates that the power-law decaying autocorrelation function could be explained solely by the inhomogeneous inter-event times, not by the interdependency between inter-event times. In fact, the observed autocorrelation functions measure not only the inhomogeneities in inter-event times themselves but also correlations between consecutive inter-event times of arbitrary length. Thus, it is required to distinguish these effects from each other, if possible, for better understanding of bursty behaviour. For this, another measurement has recently been introduced, called bursty train size distribution, to be discussed below.

\subsubsection{The bursty train size distribution}
\label{sec:PE}

\begin{figure}[!t]
    \center
    \includegraphics[width=\columnwidth]{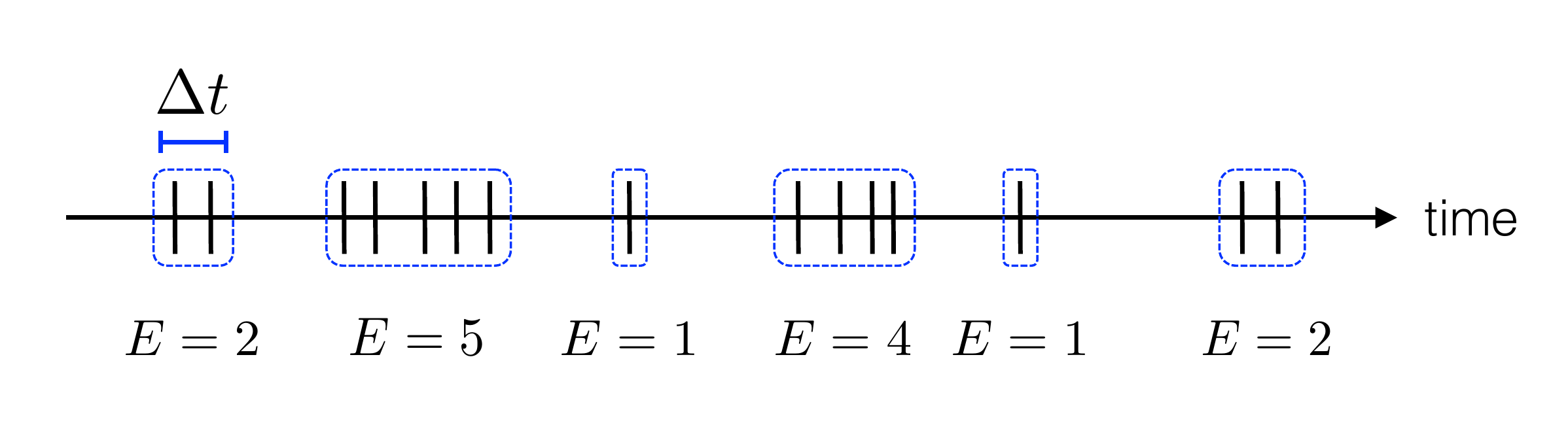}
    \caption{Schematic diagramme of an event sequence, where each vertical line indicates the timing of the event. For a given time window $\Delta t$, a bursty train is determined by a set of events separated by $\tau\leq \Delta t$, while events in different trains are separated by $\tau>\Delta t$. The number of events in each bursty train, i.e., bursty train size, is denoted by $E$. In most empirical datasets, the distributions of $E$ are heavy-tailed.}
    \label{fig:scheme2}
\end{figure}

\begin{figure}[!t]
    \centering
    \includegraphics[width=\columnwidth]{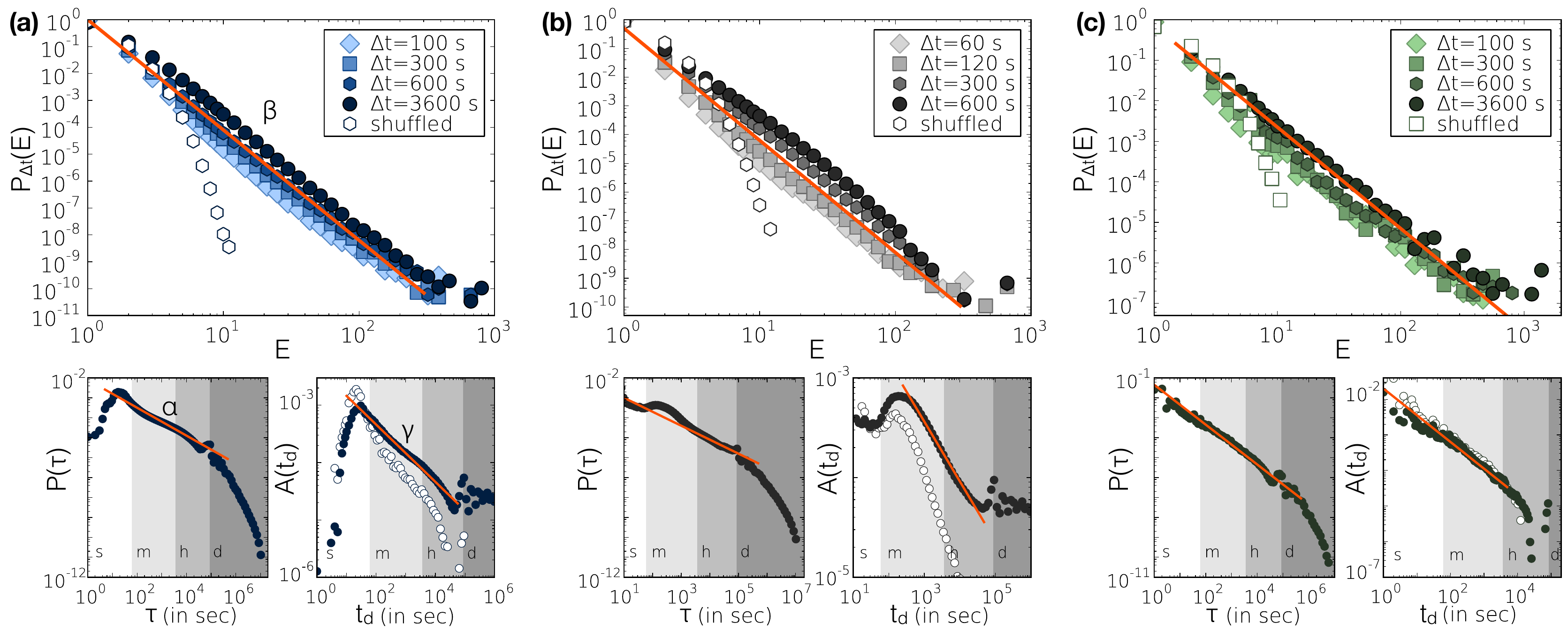}
    \caption{The characteristic functions of human communication event sequences. The bursty train size distribution $P_{\Delta t}(E)$ with various time windows $\Delta t$ (main panels), the inter-event time distribution $P(\tau)$ (left bottom panels), and autocorrelation functions $A(t_d)$ (right bottom panels) are calculated for different communication datasets. (a) Mobile phone call dataset: The scale-invariant behaviour was characterised by power-law functions with exponent values $\alpha\simeq 0.7$, $\beta\simeq 4.1$, and  $\gamma\simeq 0.5$ (b) Almost the same exponents were estimated for short message sequences taking values $\alpha\simeq 0.7$, $\beta\simeq 3.9$ and  $\gamma\simeq 0.6$. (c) Email event sequence with estimated exponents $\alpha\simeq 1.0$, $\beta\simeq 2.5$ and $\gamma=0.75$. A gap in the tail of $A(t_d)$ on figure (c) appears due to logarithmic binning and slightly negative correlation values. Empty symbols assign the corresponding calculation results on independent sequences. Lanes labeled with s, m, h and d are denoting seconds, minutes, hours and days, respectively. (\emph{Source:} Adapted from Ref.~\cite{Karsai2012Universal} under a Creative Commons Attribution-NonCommercial-ShareAlike 3.0 Unported License.)}
    \label{fig:PE}
\end{figure}

The above mentioned ambiguity of the autocorrelation function called for another way to indicate correlations between consecutive inter-event times. A method has been proposed by detecting correlated bursty trains as introduced in Ref.~\cite{Karsai2012Universal}. A bursty train is a sequence of events, where each event follows the previous one within a time window $\Delta t$. $\Delta t$ actually defines the maximum time between consecutive events, which are assumed to be causally correlated. In this way, an event sequence can be decoupled into a set of causal event trains in which each pair of consecutive events in a given train is closer than $\Delta t$, while trains are separated from each other by an inter-event time $\tau>\Delta t$. To obtain the size of each bursty train, denoted by $E$, we can count the number of events they contain, as depicted in Fig.~\ref{fig:scheme2}. Note that this notion assigns a bursty train size $E=1$ to standalone events, which occurs independently from any of the previous or following events, according to this definition. The relevant measure for temporal correlation is the bursty train size distribution $P_{\Delta t}(E)$ for a fixed $\Delta t$. If events are independent, $P_{\Delta t}(E)$ must appear as follows:
\begin{eqnarray}
    P_{\Delta t}(E) &=& \left[ \int_0^{\Delta t}P(\tau)d\tau \right]^{E-1}\left[1- \int_0^{\Delta t} P(\tau)d\tau\right]\\
    & \approx & \frac{1}{E_c(\Delta t)}e^{-E/E_c(\Delta t)},
    \label{eq:PEindep}
\end{eqnarray}
where $E_c(\Delta t)\equiv -1/\ln F(\Delta t)$ with the cumulative distribution of inter-event times $F(\Delta t)\equiv \int_0^{\Delta t}P(\tau)d\tau$. Since $F(\Delta t)$ is not a function of $E$ in this case, the functional form of $P(\tau)$ is irrelevant to the functional form of $P_{\Delta t}(E)$, which appears with an exponential distribution for any independent event sequences. Thus any correlation between inter-event times may lead to different forms of $P_{\Delta t}(E)$, implying that any deviation from an exponential form of $P_{\Delta t}(E)$ indicates correlations between inter-event times. Interestingly, several empirical cases have been found to  show the power-law distributed train sizes as
\begin{equation}
P_{\Delta t}(E)\sim E^{-\beta},
\end{equation}
with the power-law exponent $\beta$ for a wide range of $\Delta t$~\cite{Karsai2012Universal, Karsai2012Correlated, Jiang2013Calling, Kikas2013Bursty}. For the demonstration of such observations, see Fig.~\ref{fig:PE}(a--c) adopted from Ref.~\cite{Karsai2012Universal}. This phenomenon, called \emph{correlated bursts}, has been shown to characterise several systems in nature and human dynamics~\cite{Karsai2012Universal}.

Finally, we mention the possible effects of interdependency between inter-event times on the scaling relations between power-law exponents of inter-event times and autocorrelation function as presented in Eq.~(\ref{eq:alpha_gamma}). For example, one can compare the autocorrelation function calculated for an empirical event sequence with that for the shuffled event sequence, where correlations between inter-event times are destroyed, as shown in the lower right panels in each of Fig.~\ref{fig:PE}(a--c). By doing so, the effects of interdependency between inter-event times can be tested. Such effects of correlation between inter-event times on the scaling relation should be studied more rigorously in the future as they are far from being fully understood. So far only a few studies have tackled this issue, e.g., see Refs.~\cite{Rybski2012Communication, Karsai2012Universal, Vajna2013Modelling, Jo2015Correlated}.

\subsubsection{Memory kernels}

We also introduce the memory kernel as one of the measurements for the bursty temporal patterns~\cite{Crane2008Robust, Masuda2013SelfExciting, Aoki2016Inputoutput}. The memory kernel $\phi(t)$ relates the past events, either being endogenous or being exogenous, to the future events. This measure, which represents the effect by the past events, has been empirically found to have the power-law form as 
\begin{equation}
    \phi(t)\sim t^{-(1+\theta)},
\end{equation}
where $t$ is the elapsed time from the past event and $\theta$ denotes the power-law exponent characterising the degree of memory effects. However, in general, memory kernels are also assumed to follow different functional forms, e.g., hyperbolic, exponential~\cite{Masuda2013SelfExciting}, or power-law~\cite{Jo2015Correlated}. They are commonly applied in modelling bursty systems using self-exciting point processes~\cite{Mehrdad2015Hawkes}: For a given set of past events occurred before the time $t$, the event rate at time $t$ reads as follows:
\begin{equation}
    \lambda(t)=V(t)+\sum_{i,t_i\leq t} \phi(t-t_i),
\end{equation}
where $V(t)$ is the exogenous source, and $t_i$ denotes the timing of the $i$th event. We are going to discuss more in details in Section~\ref{sec:SelfExcPP}.

\subsubsection{Other characteristic measures}

In addition to the conventional measures of bursty behaviour that we already introduced, we here mention some less recognised ones. There indeed exist a number of traditional measures and techniques in nonlinear time series analysis~\cite{Kantz2004Nonlinear, Mantegna2007Introduction}. Among them we here introduce the detrended fluctuation analysis (DFA), originally devised for analysing DNA sequences~\cite{Peng1994Mosaic}. For a given time series $x(t)$ for $0\leq t<T$, with its average value $\langle x\rangle$, the cumulative time series is constructed by 
\begin{equation}
y(t)\equiv \int_0^t (x(t')-\langle x\rangle)dt'.
\end{equation}
The total time period $T$ is divided into segments of size $w$. For each segment, the cumulative time series is fit to a polynomial $y_w (t)$. Using the fit polynomials for all segments, the mean-squared residual for the entire range of time series is calculated as follows:
\begin{equation}
F(w)\equiv \sqrt{\int_0^T [y(t)-y_w (t)]^2 dt},
\end{equation}
which typically scales with the segment size $w$ as $w^H$. Here the scaling exponent $H$ is called the Hurst exponent~\cite{Bryce2012Revisiting}.

\section{Inter-event time, residual time, and waiting time}
\label{sec:iet_rt_wt}

As for the terminology for burstiness, there is a common confusion between the definitions of inter-event time, waiting time, and residual time. Here we would like to clarify their definitions and relations to each other.

For a given event sequence, the \emph{inter-event time} $\tau$ is defined as the time between two consecutive events. However, the observations of an event sequence always cover a finite period of time, which has to be considered in the terminology. So let us assume an observer who begins to observe the time series of events at a random moment of time, and waits for the next, firstly observed event to take place. The time interval between the beginning time of the observation period and the next event has been called the \emph{residual time} $\tau_r$, also often called the \emph{residual waiting time} or \emph{relay time}~\cite{Kampen2007Stochastic}. A similar definition of the residual time is found in queuing theory in a situation when a customer arrives at a random time and waits for the server to become available~\cite{Cox1972Theory, Cooper1998Some}. The residual time then is the time interval between the time of arrival and the time of being served, thus it corresponds to the remaining or residual time to the next event after a random arrival. The residual time distribution can be derived from the inter-event time distribution as
\begin{equation}
    P(\tau_r)=\frac{1}{\langle \tau\rangle}\int_{\tau_r}^\infty P(\tau)d\tau,
\label{eq:rst_iet}
\end{equation}
and the average residual time can be calculated as
\begin{equation}
    \langle \tau_r \rangle = \int_0^\infty \tau_r P(\tau_r)d\tau_r =  \frac{\langle \tau^2\rangle}{2\langle \tau\rangle}.
\label{eq:taurderiv}
\end{equation}
This result explains a phenomenon called the \emph{waiting-time paradox}, which has important consequences on dynamical processes evolving on bursty temporal systems that we will discuss in details later in Section \ref{sec:wtp}. As we mentioned earlier, a common reference dynamics to quantify the heterogeneity of a bursty sequence is provided by a Poisson process. Thus we may consider a normalised average residual time after dividing $\langle \tau_r \rangle$ by the corresponding residual time of a Poisson process $\langle \tau_{r}^{P} \rangle$, which is simply $\langle \tau\rangle$. This can then be written as
\begin{equation}
\frac{\langle \tau_r \rangle}{\langle \tau_{r}^{P} \rangle}=\frac{\langle \tau^2 \rangle}{2\langle \tau \rangle^2}=\frac{1}{2}\left[ \left( \frac{\sigma}{\langle \tau \rangle}\right)^2+1\right]=\frac{B^2+1}{(B-1)^2},
\label{r_definition}
\end{equation}
where $\sigma$ is the standard deviation of $P(\tau)$ and $B$ is the burstiness parameter as defined in Eq.~(\ref{eq:burstiness_param}). Consequently this ratio can equally well be seen as a measure of burstiness.

Contrary to the above definitions, \emph{waiting times} are not necessarily derived from series of consecutive events, but they can rather characterise the lifespan of single tasks. The tasks wait to be executed for a period depending on their priorities as well as on the newly-coming other tasks. In this way the \emph{waiting time} $\tau_w$, also often called \emph{response time} or \emph{processing time}, is defined as the time interval a newly arrived task needs to wait before it is executed. For example, in an editorial process, each submitted manuscript gives rise to one waiting time until the decision is made~\cite{Mryglod2012Editorial, Jo2012Timevarying,Hartonen2013How} and the waiting time distribution is obtained from a number of submitted manuscripts. However, the heavy tail of the waiting time distribution, $P(\tau_w)$, implies the heterogeneity of the editorial system, but not necessarily the bursty dynamics of the process itself. On the other hand, the waiting time can be deduced from an event sequence, e.g., of directed interactions, like the time between receiving and responding to an email or letter. In these cases, a close relation between $P(\tau)$ and $P(\tau_w)$ seems to appear. Actually, it has been argued that in case of a process with heterogeneous waiting time distribution, the inter-event time distribution is also heterogeneous and vice versa, and can be characterised by the same exponent~\cite{Barabasi2005Origin, Vazquez2006Modeling, Li2008Empirical, Formentin2015New}. Waiting times will be duly addressed later in Section~\ref{sec:PriQueMod},
where they appear as the central quantity in the definition of priority queuing models~\cite{Abate1997Asymptotics, Barabasi2005Origin}.

\section{Collective bursty phenomena}\label{subsect:system}

So far we have been discussing measures to characterise bursty behaviour at the level of single individuals. However, individuals form egocentric networks and connected to a larger social system, which could show bursty dynamics and be characterised at the system level. Since individual dynamics is observed to be bursty, it may effect the system-level dynamics and the emergence of any collective phenomena, while also the contrary is true: If the collective dynamics is bursty, it must affect the temporal patterns of each individual. The structure of social systems has been commonly interpreted as social networks~\cite{Borgatti2009Network, Wasserman1994Social}, where nodes are identified as individuals and links assign their interactions. Thanks to the recent access to a huge amount of digital datasets related to human dynamics and social interaction, a number of empirical findings have been cumulated to study the structure and dynamics of social networks. Researchers have analysed various social networks of face-to-face interactions~\cite{Eagle2006Reality, Zhao2011Social, Fournet2014Contact}, emails~\cite{Eckmann2004Entropy, Klimt2004Enron}, mobile phone communication~\cite{Onnela2007Structure, Blondel2015Survey}, online forums~\cite{Hric2014Community, Eom2015Tailscope}, Social Networking Services (SNS) like Facebook~\cite{Ugander2011Anatomy} and Twitter~\cite{Kwak2010What}, as well as even massive multiplayer online games~\cite{Szell2010Measuring, Szell2010Multirelational}. These studies of social networks show that there are commonly observed features or \emph{stylised facts} characterising their structures~\cite{Jackson2010Social, Murase2015Modeling, Kertesz2016Multiplex}, see also the summary in Table~I in Ref.~\cite{Jo2017Stylized}. For example, one finds broadly distributed network quantities like node degree and link weight~\cite{Albert2002Statistical, Onnela2007Analysis}, homophily~\cite{McPherson2001Birds, Newman2002Assortative}, community or modular structure~\cite{Granovetter1973Strength, Fortunato2010Community}, multilayer nature~\cite{Kivela2014Multilayer, Boccaletti2014Structure}, and geographical and demographic correlations~\cite{Onnela2011Geographic, Palchykov2012Sex, Jo2014Spatial} to mention a few. All these characters play important roles in the dynamics of social interactions.

At the same time, such datasets lead to the observation of mechanisms and correlations driving the interaction dynamics of people. This is the subject of the recent field of \emph{temporal networks}~\cite{Holme2012Temporal, 2013Temporal, Holme2015Modern, Masuda2016Guide}, which identifies social networks as temporal objects, where interactions are time-varying, and code the static structure after aggregation over an extensive period. Temporal networks are commonly interpreted as a sequence of events, which are defined as triplets $(i,j,t)$, indicating that a node $i$ interacts with a node $j$ at time $t$. The analysis of event sequences of large number of individuals can disclose the mesoscopic structure of bursty interaction patterns, and enable us to characterise burstiness at the system level as well.

\subsection{Bursty patterns in egocentric networks}
\label{sec:egoburst}

The interaction dynamics of a focused individual or an ego can be exploited from the global temporal network by extracting all event sequence where the ego $i$ participates as:
\begin{equation}
    x_i(t)\equiv \sum_{j\in\Lambda_i} x_{ij}(t),
\end{equation}
where $\Lambda_i$ denotes the neighbour set of the ego $i$. In other works, the event sequence $x_i(t)$ builds up from interaction sequences on single links, $x_{ij}(t)$, which together define the dynamics of the egocentric network. Our first question is how the bursty interactions of an ego are distributed among the different neighbours.

We have already discussed that by observing an individual, her bursty activities may evolve in trains where several events follow each other within a short time window $\Delta t$. This is especially true for communication dynamics, where interactions like mobile calls, SMSs or emails sent or received by an ego, exhibit such patterns. However, the question remains whether such bursty communication trains are the consequences of some collective interaction patterns in the larger egocentric network, e.g., to organise an event or to process information, or on the contrary, they evolve on single links induced by discussions between only two people. One can easily figure this out by decoupling the entangled egocentric dynamics to single links and see how the bursty train size distribution $P(E)$ changes before and after this process. If the first hypothesis is true, as long trains of an ego are distributed among many links, after decoupling the trains should fall apart and their size distribution should change radically. On the other hand, if the second hypothesis is true, their size distribution should not change considerably. Using mobile phone call and SMS sequences, it has been shown in Ref.~\cite{Karsai2012Correlated} that after decoupling, $P(E)$ measured on single links are almost identical to ones observed in individual activity sequences. In support of this observation it has been found that $\sim 80\%$ of trains evolve on single links, almost independently from the train size. Consequently, this suggests that long correlated bursty trains are more like the property of links rather than nodes and are commonly induced by dyadic interactions. This study further discusses the difference between call and SMS sequences and finds that call (respectively SMS) trains are more imbalanced (respectively balanced) than one would expect from the overall communication balance of the social tie.

\begin{figure}[!t]
    \center
    \includegraphics[width=\columnwidth]{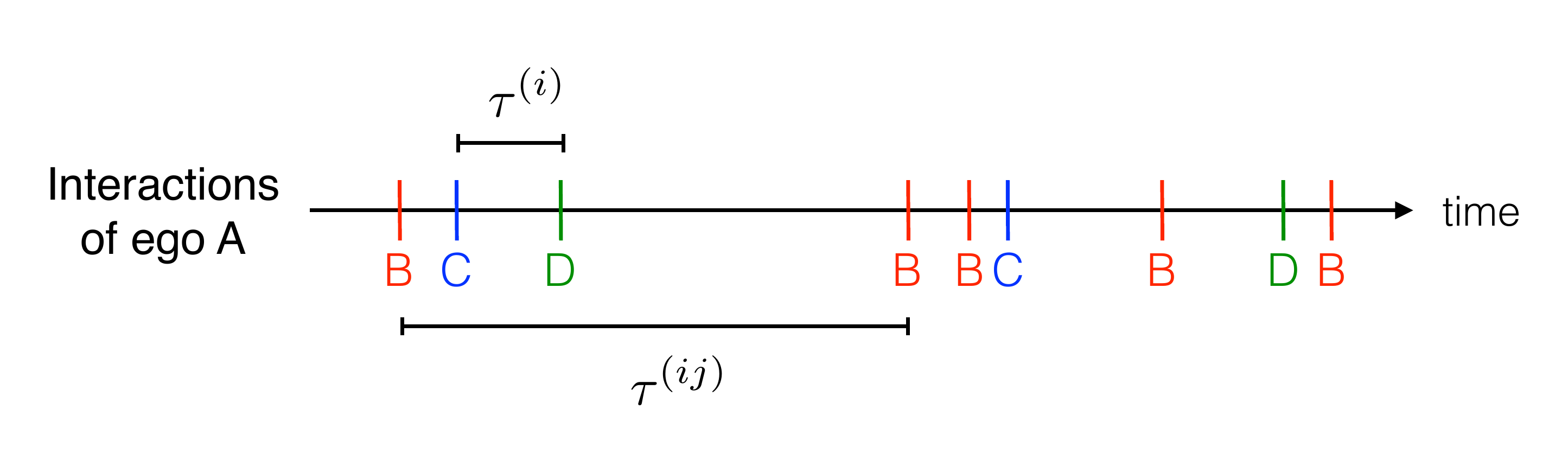}
    \caption{Schematic example of the event sequence of an individual $A$ with her various contexts (neighbours) $B$, $C$, and $D$. The collective inter-event time $\tau^{(i)}$ is defined as the time interval between consecutive events of any contexts of the ego $i=A$. The contextual inter-event time $\tau^{(ij)}$ is defined between events of the same context, e.g., $j=B$.}
    \label{fig:contextualBursts}
\end{figure}

One can adopt the same picture to understand the contribution of bursty patterns on links to the overall inter-event time distribution of an ego. This question was addressed by Jo \emph{et al.}, who proposed an alternative explanation for bursty links related to contextual dependence of behaviour. In their interpretation, the context of an event~\cite{Jo2012Spatiotemporal, Jo2013Contextual} is the circumstance in which the event occurs and can be a person, a social situation with some convention, or a place. In case of social interactions, for an ego $i$ the context of social interactions can be associated to a neighbour $j$ in the egocentric network. Then the question is how much \emph{contextual bursts}, which evolve in the interaction sequences of single links $x_{ij}(t)$, determine \emph{collective bursts} observable in the overall interaction sequence $x_i(t)$ of the ego $i$. This question can be addressed on the level of inter-event times. As depicted in Fig.~\ref{fig:contextualBursts}, let us denote collective inter-event times in $x_i(t)$ as $\tau^{(i)}$, while contextual inter-event times in $x_{ij}(t)$ as $\tau^{(ij)}$. It is straightforward to see that a contextual inter-event time comprises typically of multiple collective inter-event times as follows:
\begin{equation}
    \label{eq:tauij_taui}
    \tau^{(ij)} =\sum_{k=1}^n \tau^{(i)}_k,
\end{equation}
where $n-1$ is the number of events with contexts other than $j$ between two consecutive events with $j$. For example, one finds $n=3$ in Fig.~\ref{fig:contextualBursts} between the first and second observed interactions with context $B$. The relation between $P(\tau^{(ij)})$ and $P(\tau^{(i)})$ for uncorrelated inter-event times has been studied analytically and numerically in Ref.~\cite{Jo2013Contextual}, where both $P(\tau^{(ij)})$ and $P(\tau^{(i)})$ are assumed to have power-law forms with exponents $\alpha'$ and $\alpha$, respectively. For deriving the scaling relation between $\alpha'$ and $\alpha$, another power-law distribution is assumed for $n$ in Eq.~(\ref{eq:tauij_taui}), i.e., the number of collective inter-event times for one contextual inter-event time, as $P(n)\sim n^{-\eta}$. The distribution of $n$ is related to how the ego distributes her limited resource like time to her neighbours. Then one can write the relation between distribution functions as follows:
\begin{eqnarray}
    P(\tau^{(ij)})&=&\sum_{n=1}^\infty P(n) F_n(\tau^{(ij)}),\\
    F_n(\tau^{(ij)})&\equiv &\prod_{k=1}^n \int_{\tau_0}^\infty d\tau_k^{(i)} P(\tau_k^{(i)}) \delta\left(\tau^{(ij)}-\sum_{k=1}^n \tau_k^{(i)}\right),
\end{eqnarray}
where $F_n$ denotes the probability of making one $\tau^{(ij)}$ as a sum of $n$ $\tau^{(i)}$s, and $\tau_0$ is the lower bound of inter-event times $\tau^{(i)}$. By solving this equation, the scaling relation between $\alpha'$, $\alpha$, and $\eta$ is obtained~\cite{Jo2013Contextual}:
\begin{equation}
    \alpha'=\min\{(\alpha-1)(\eta-1)+1,\alpha,\eta\}.
\end{equation}
This result provides a condition under which the statistical properties of the ego's own temporal pattern could be described similarly to those of the ego's relationships.

Note that this terminology can be generalised for event sequences not only on links but for an arbitrary set of neighbours associated to the same context $\Lambda$. In this case the contextual event sequence can be written as
\begin{equation}
    x_\Lambda (t)\equiv \sum_{i,j\in \Lambda} x_{ij}(t),
\end{equation}
where the summation considers individuals $i$ and $j$ who both belong to the same context or group of $\Lambda$. Then one can study the relation between statistical properties at different levels of contextual grouping. For example, empirical analysis using online forum dataset was recently performed to relate individual bursty patterns to the forum-level bursty patterns in Ref.~\cite{Panzarasa2015Emergence}. 

In another work Song \emph{et al.}~\cite{Song2013Connections} proposed scaling relations between power-law exponents characterising structural and temporal properties of temporal social networks. In terms of structure they concentrate on the distribution of node degrees and link weights observed over a finite time period. Here the node degree indicates the number of neighbours of a node, while the link weight is defined as the number of interaction events between two neighbouring nodes. Both of these distributions can be approximated as power-laws with exponents $\epsilon_k$ and $\epsilon_w$. To characterise the dynamics of the network they consider individual activity $a_i$, defined as the total number of interactions of an ego $i$ within a given period, and inter-event time distributions, but not in real time but event times and not of egos but of social ties. In this case inter-event time is defined as the number of events between two consecutive interaction of the ego and one specific neighbour (similar to $n$ in Eq.~(\ref{eq:tauij_taui})). Distributions of these dynamical quantities can be also approximated by power-laws with exponents assigned as $1+\alpha_{a}$ for activity and $1+\alpha_{\tau}$ for inter-event times. They first show that the degree of a node $i$, denoted by $k_i$, observed for a period $[ t_1, t_2 ]$ is increasing as
\begin{equation}
k_i(t_1,t_2)\sim a_i(t_1,t_2)^{\kappa_i}.
\end{equation}
They argue that the power-law exponent $\kappa_i$ measured for an ego $i$, what they call the sociability, satisfies the condition
\begin{equation}
\kappa_i+\alpha_{\tau,i}=1,
\end{equation}
where $\alpha_{\tau,i}$ denotes the inter-event time exponent observed in the interaction sequence of the ego $i$. They further argue that the degree and weight distribution exponents can be determined by the dynamical parameters as
\begin{equation}
    \epsilon_k=1+\min \left\{ 
    \frac{\alpha_a}{1-\overline{\alpha}_{\tau}}, \frac{u}{\overline{\alpha}_{\tau} \ln \overline{a}} \right\},
\hspace{.3in} \epsilon_w=2-\overline{\alpha}_{\tau},
\end{equation}
where $\overline{\alpha}_{\tau}$ and $\overline{a}$ denote average values, while $u$ is a parameter capturing the variability of sociability $\kappa$. The authors support these scaling relations by introducing the scaling functions to scale the corresponding distributions obtained from various human interaction datasets.

\subsection{Bursty temporal motifs}

Taking off from the egocentric point of view, bursty temporal interaction patterns can appear not only centered around a single ego but also between a larger number of people. Such patterns are formed by causally correlated sequence of interactions, which appear within a short time window between two or more people. These \emph{temporal motifs} are arguably induced by group conversations, information processing, or organisation like a common event, etc., and can be associated to burstiness at the mesoscopic level of networks. The emergence of such group-level bursty events is rather rare and it strongly depends on the observed communication channel and the type of induced events. However, it has been shown that some of them appear with a significantly larger frequency as compared to random reference models.

\begin{figure}[!t]
    \center
    \includegraphics[width=\columnwidth]{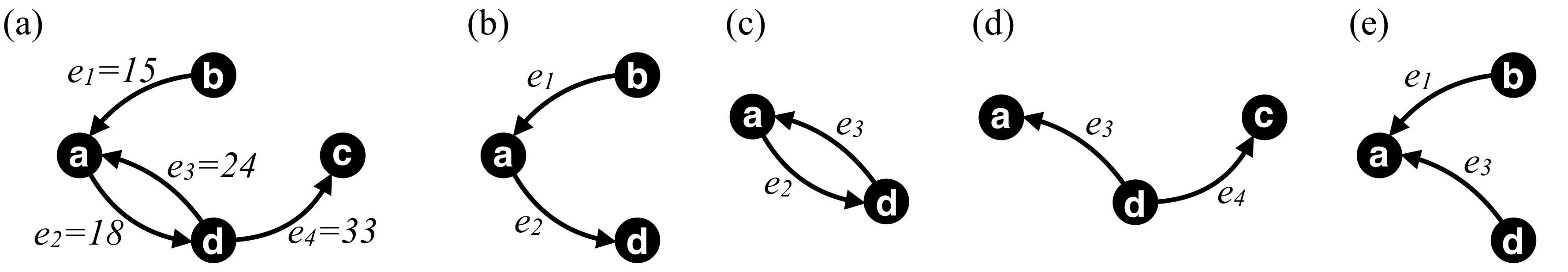}
    \caption{(a) A directed temporal network between four nodes $a$, $b$, $c$, and $d$ with four events, $e_1$, $e_2$, $e_3$, $e_4$, respectively at $t=15$, $18$, $24$, and $33$. By assuming that $\Delta t = 10$, $e_2$ end $e_4$ are adjacent but not $\Delta t$-adjacent. (b--d) All 2-event valid temporal subgraphs. (e) An invalid subgraph because it skips the event $e_2$ that for node $a$ takes place between $e_1$ and $e_3$.}
    \label{fig:tempmotill}
\end{figure}

Temporal motifs are defined in temporal networks. For a schematic example, see Fig.~\ref{fig:tempmotill}(a). Here interactions between nodes occur in different timings and they are interpreted as events assigned with time stamps. For a more detailed definition and characterisation of temporal networks we refer the reader to Refs.~\cite{Holme2012Temporal, Masuda2016Guide}. Temporal motifs consist of \emph{$\Delta t$-adjacent events} in the temporal network, which share at least one common node and happens within a time window $\Delta t$. Two events that are not directly $\Delta t$-adjacent might be \emph{$\Delta t$-connected} if there is a sequence of events connecting the two events, which are successive in time and $\Delta t$-adjacent. A connected temporal subgraph is then a set of events where each pair of events are $\Delta t$-connected, as depicted in Fig~.\ref{fig:tempmotill}(b--e). To define temporal motifs we further restrict our definition on \emph{valid temporal subgraph} where for each node in the subgraph the events involving the node must be consecutive, e.g., as in Fig.~\ref{fig:tempmotill}(b--d). Note that for the final definition of temporal motifs we consider only \emph{maximal valid temporal subgraphs}, which contain all events that are $\Delta t$-connected to the included events. For a more precise definition, see Refs.~\cite{Kovanen2011Temporal, Kovanen2013Temporal}. Also note that an alternative definition of temporal motifs has been proposed recently, where motifs are defined by events which all appear within a fixed time window~\cite{Paranjape2017Motifs}.

One way to detect temporal motifs is by interpreting them as static directed colored graphs and find all isomorphic structures with equivalent ordering in a temporal network~\cite{Kovanen2011Temporal}. The significance of the detected motifs can be inferred by comparing the observed frequencies to those calculated in some reference models, where temporal and causal correlations were removed. Such analysis has shown~\cite{Kovanen2011Temporal} that the most frequent motifs in mobile phone communication sequences correspond to dyadic bursty interaction trains on single links. On the other hand the least frequent motifs are formed by non-causal events, suggesting strong dependence between causal correlations and bursty phenomena.

\subsection{System level characterisation}

Finally, we discuss methods to characterise bursty phenomena at the level of the whole social network. Temporal inhomogeneity at the system level can be measured in terms of \emph{temporal network sparsity}~\cite{Perotti2014Temporal}. This measure counts the number of microscopic configurations associated with the macroscopic state of a temporal networks. This concept of multiplicity has been known in statistical physics. More specifically, in a temporal network for a given time window, events can be distributed over the links of the corresponding static structure. Here we denote a link between nodes $i$ and $j$ as $ij$, and the set of all links as $L$. Thus, for a time window one can measure the fraction of events on a given link $ij$, denoted by $p_{ij}$, and compute the Shannon entropy considering each link $ij\in L$ as:
\begin{equation}
    H_{L} = -\sum_{ij\in L} p_{ij}\ln p_{ij},
\end{equation}
which quantifies how heterogeneously events are distributed among different links. After computing an average entropy $\langle H_L \rangle$ over several time windows, one can estimate the \emph{effective number of links} as
\begin{equation}
    L^{\textrm{eff}} \equiv \exp(\langle H_L \rangle),
\end{equation}
which gives the number of links in a given time window assuming that the event rate per a link is constant. Simultaneously measuring the effective number of links in the empirical temporal network and in a random reference model where events are uniformly distributed in time, one can introduce the notion of \emph{temporal network sparsity}:
\begin{equation}
    \zeta_{\textrm{temp}} \equiv \frac{L^{\textrm{eff}}}{L^{\textrm{eff}}_{\textrm{ref}}}.
\end{equation}
This measure indicates the overall distribution of events within a given time window as compared to the case with homogeneously distributed events. The smaller value $\zeta_{\textrm{temp}}$ has, the more severe heterogeneities characterise the event sequence and the more ``temporally sparse'' the network is. This measure turns out to have some explanatory power for spreading dynamics on various temporal networks~\cite{Perotti2014Temporal}.

\begin{figure}[!t]
    \center
    \includegraphics[width=\columnwidth]{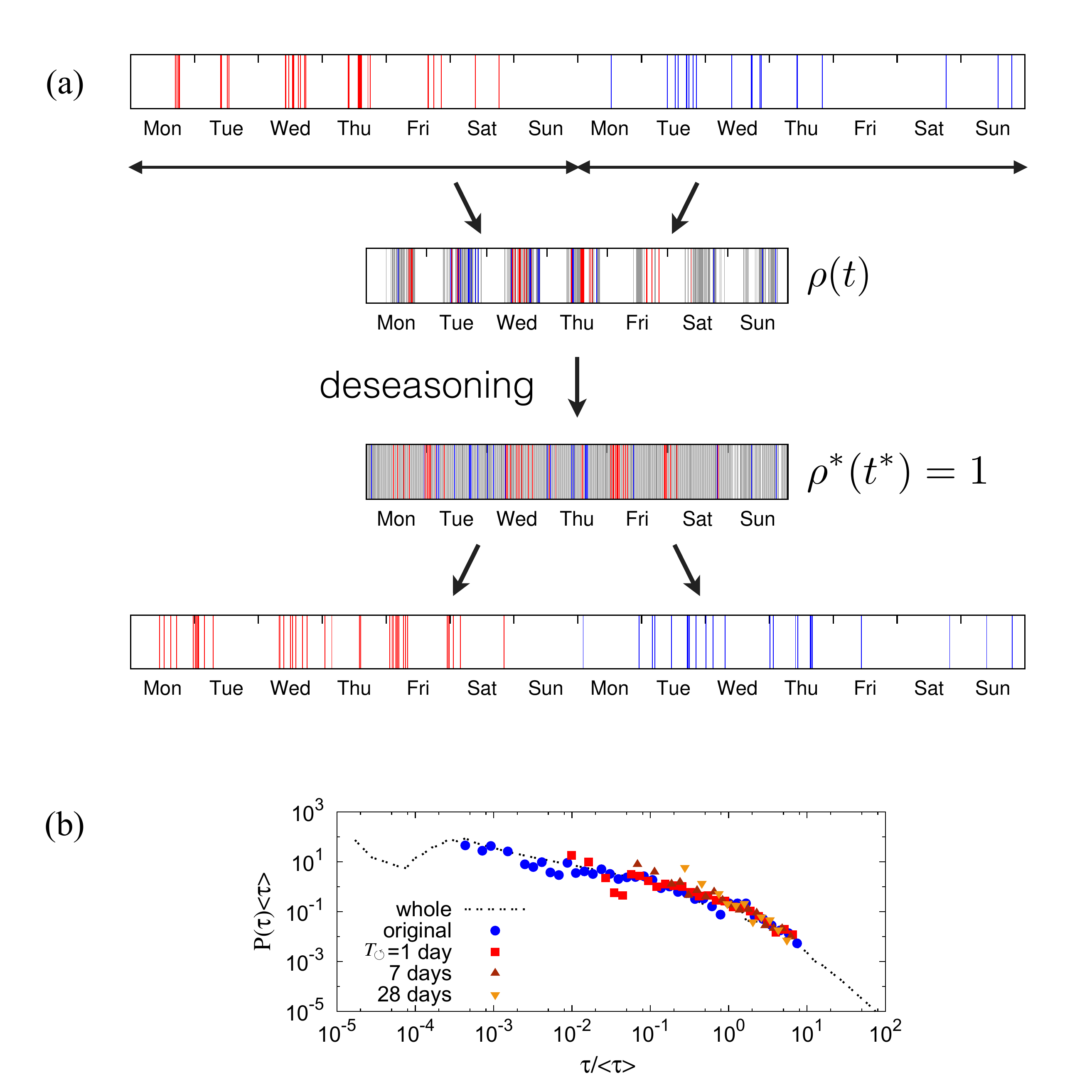}
    \caption{(a) An example of the deseasoning method applied to a mobile call series of a user, with $T_{\circlearrowleft}=1$ week. The top shows the first two weeks of the call series colored in red (the first week) and blue (the second week). Events for all weeks are collected in one week period to obtain the event rate $\rho(t)$ for $0\leq t< T_{\circlearrowleft}$. After deseasoning, the events in each week are put back to their original slot. (b) The original inter-event time distribution for individuals with $200$ calls is compared to the distributions with deseasoned inter-event times for various values of $T_{\circlearrowleft}$. (\emph{Source:} Adapted from Ref.~\cite{Jo2012Circadian} under $\textcopyright$ IOP Publishing \& Deutsche Physikalische Gesellschaft (CC BY-NC-SA).)}
    \label{fig:deseason}
\end{figure}

\section{Cyclic patterns in human dynamics}\label{subsect:cyclic}

It is evident that humans follow intrinsic periodic patterns of circadian, weekly, and even longer cycles~\cite{Malmgren2008Poissonian, Jo2011Circadian, Jo2012Circadian, Aledavood2015Daily}. Such cycles clearly contribute to the inhomogeneities of temporal patterns, and they often result in an exponential cutoff to the inter-event time distributions. Identifying and filtering out such cyclic patterns from a time series can reveal bursty behaviour of different origins than those cycles. In order to characterise such cyclic patterns, let us consider a time series, i.e., the number of events at time $t$, denoted by $x(t)$, for the entire period of $0\leq t< T$. One may be interested in a specific cycle, like daily or weekly ones, with period denoted by $T_{\circlearrowleft}$. Then, for a given period of $T_{\circlearrowleft}$, the event rate with $0\leq t <T_{\circlearrowleft}$ can be defined as
\begin{equation}
    \rho(t)\equiv \frac{T_{\circlearrowleft}}{X}\sum_{k=0}^{T/T_{\circlearrowleft}} x(t+kT_{\circlearrowleft}),\vspace{.2in} X\equiv \int_0^{T} x(t)dt.
\end{equation}
Such cycles turn out to be also apparent in the inter-event time distributions $P(\tau)$. For example, one finds peaks of $P(\tau)$ corresponding to multiples of one day in mobile phone calls and blog posts~\cite{Jo2011Circadian, Kim2013Microscopic}. Note that such periodicities could be characterised by means of a power spectrum analysis in Eq.~(\ref{eq:powerSpectrum}), however here we take another way.

Once such cycles are identified in terms of the event rate $\rho(t)$, we can filter them by deseasoning the time series~\cite{Jo2012Circadian}. First, we extend indefinitely the domain of $\rho(t)$ by $\rho(t+kT_{\circlearrowleft})=\rho(t)$ with an arbitrary integer $k$. Then using the identity of $\rho(t)dt=\rho^*(t^*)dt^*$ with the deseasoned event rate of $\rho^*(t^*)=1$, we can get the deseasoned time $t^*(t)$ as
\begin{equation}
    t^*(t)\equiv \int_0^t \rho(t')dt'.
\end{equation}
For the schematic example of the deseasoning method, see Fig.~\ref{fig:deseason}(a). In plain words, the time is dilated (respectively contracted) at the moment of the high (respectively low) event rate. Then the deseasoned event sequence of $\{t^*(t_i)\}$ is compared to the original event sequence of $\{t_i\}$ to see how strong signature of burstiness or memory effects remained in the deseasoned sequence. This reveals whether the empirically observed temporal heterogeneities can (or cannot) be explained by the intrinsic cyclic patterns, characterised in terms of the event rate. For example, if one obtains the deseasoned inter-event time $\tau_i^*$ corresponding to the original inter-event time $\tau_i=t_i-t_{i-1}$ as
\begin{equation}
    \tau_i^* \equiv t^*(t_i)-t^*(t_{i-1})=\int_{t_{i-1}}^{t_i} \rho(t')dt',
\end{equation}
then the deseasoned inter-event time distribution $P(\tau^*)$ can be compared to the original inter-event time distribution $P(\tau)$. This method was applied to the mobile phone call series~\cite{Jo2012Circadian}, as partly depicted in Fig.~\ref{fig:deseason}(b), where the inter-event time distributions for the original and deseasoned event sequences show almost the same shape for various values of $T_{\circlearrowleft}$. This indicates that there could be other origins for the human bursty dynamics than the circadian and weekly cycles of humans. In order to quantitatively study the effects of deseasoning, the burstiness parameter $B$ has been measured for both original and deseasoned mobile phone call series to find the overall decreased yet positive values of $B$, implying that the bursts remain after deseasoning. In addition, the memory coefficients $M_m$, bursty train size distributions $P_{\Delta t}(E)$, and autocorrelation function $A(t_d)$ can be also measured by using the deseasoned event sequence of $\{t^*(t_i)\}$ for the comparison to the original ones.

It is straightforward to extend this method for the aggregated time series at different levels of activity groups, including the whole population. For a set of individuals $\Lambda$, the number of events in time $t$ is denoted by 
\begin{equation}
    x_{\Lambda}(t)\equiv \sum_{i\in \Lambda} x_i(t),
\end{equation}
where $x_i(t)$ is the number of events of an individual $i$ at time $t$. Then, for a given period of $T_{\circlearrowleft}$, the event rate with $0\leq t <T_{\circlearrowleft}$ is defined as
\begin{equation}
    \rho_{\Lambda}(t)\equiv \frac{T_{\circlearrowleft}}{X_{\Lambda}}\sum_{k=0}^{T/T_{\circlearrowleft}} x_{\Lambda}(t+kT_{\circlearrowleft}),\ X_{\Lambda}\equiv \int_0^{T} x_{\Lambda}(t)dt.
\end{equation}
Using this event rate for the actual set of individuals $\Lambda$, one can get the deseasoned time $t^*_{\Lambda}(t)$ as follows:
\begin{equation}
    t^*_{\Lambda}(t)\equiv \int_0^t \rho_{\Lambda}(t')dt'.
\end{equation}

We remark that the fully deseasoned time series, i.e., for $T_{\circlearrowleft}=T$, corresponds to the time series represented in the ordinal time-frame, where real timings of events are replaced by the orders of events. Now if $T_{\circlearrowleft}=T$, we have the event rate for a node $i$ as $\rho_i(t) =\frac{T}{X_i}x_i(t)$ with $X_i$ denoting the total number of events of the node $i$. We assign the timing of the $k$th event between $i$ and $j$ by $t^{(ij)}_k$ and get the deseasoned inter-event time corresponding to $\tau^{(ij)}_k=t^{(ij)}_k-t^{(ij)}_{k-1}$ as 
\begin{equation}
    {\tau^*}^{(ij)}_k \equiv \frac{T}{X_i} \int_{t^{(ij)}_{k-1}}^{t^{(ij)}_k} x_i(t')dt' = \frac{T}{X_i}n^{(ij)}_k.
\end{equation}
Here $n^{(ij)}_k$ is the contextual ordinal inter-event time, i.e., the number of events of contexts other than $j$ between two consecutive events with the context $j$. Thus, the fully deseasoned real time-frame is simply translated into the ordinal time-frame. The characterisation of bursts in terms of the ordinal time-frame has also been studied in other contexts, e.g., in terms of activity clock~\cite{Gauvin2013Activity}, relative clock~\cite{Zhou2012Relative}, and ``proper time''~\cite{Formentin2014Hidden, Formentin2015New}. In these works, the elapsed time is counted in terms of the number of events instead of the real time.

\subsection{Remark on non-stationarity}

So far, the time series has been assumed to be stationary, either explicitly or implicitly. As the stationarity by definition indicates the symmetry under the time translation, all non-Poissonian processes could be considered non-stationary, hence various time series analysis methods mentioned cannot be applied to the bursty temporal patterns. However, the definition of the stationarity can be relaxed by allowing a non-stationary behaviour only for some specific time scale: For example, human individuals can show a daily cycle in their temporal patterns, while they might keep their daily routines for several months or longer. Then, their temporal patterns can be considered stationary only at time scales that are longer than one day and shorter than several months. This relaxed definition of stationarity could be yet misleading, given the fact that most bursty phenomena show scale-free, hierarchical nature in terms of time scales, while we can apply various time series analysis methods as long as the time series looks stationary at least at some specific time scales. In this sense, the deseasoning method or detrended fluctuation analysis and its variants can be useful for removing the non-stationary temporal patterns from the original time series, hence for allowing us to investigate the bursty nature of time series without being concerned with non-stationarity issue. This is an important issue but has been largely ignored in many works, except for some recent studies mostly in relation to the dynamic processes on networks~\cite{Horvath2014Spreading, Holme2016Temporal}.


\chapter{Empirical findings in human bursty dynamics}\label{chapter:emp}


There are a number of natural phenomena that show complex structural and dynamical patterns as results of self-organisation and adaptive response to the environment. Such fundamental characteristics are also found in social systems, in which the behaviour of large number of interacting individuals induces complex and heterogeneous patterns at different organisational scales. Therefore, we find a number of empirical evidences showing temporal inhomogeneities or bursty behaviour in human dynamics, mostly due to the recent development of information-communication technology (ICT) and a number of accessible large-scale digital datasets. In this Chapter, we provide a systematic introduction of empirical findings from diverse sources of data. We will conduct the discussion at two main levels of organisation, (i) at the level of individual activities and (ii) at the level of interaction-driven collective activities. The first category includes individual activities that do not necessarily concern with direct interactions between individuals. This category also includes activities by individuals but collected at a population or system level, in which individuals do not explicitly coordinate or cooperate with others for their own actions. However, as there is no clear-cut distinction between these two types of activities due to the intrinsic sociality of humans, these categories must not be considered exclusive. We will show the empirical findings in interaction-driven collective activities mainly according to the interaction or communication channels. Finally, we will discuss other bursty patterns that are not covered by the above two categories, i.e., the bursts observed for financial activities, in human mobility patterns, and in the behavioural patterns of animals like monkeys or fruit flies.

\begin{table}[!t]
    \caption{Empirical findings of individual activities. First column collects the paper where observations were reported, second column summarises the analysed dataset, and the last column provides information about the analysis results, mostly for the statistics of inter-event times and waiting times. In case of power-law distributions, $\alpha$ ($\alpha_w$) denotes the corresponding exponent of inter-event times (or waiting times), with errors in parentheses whenever available. The exponent of bursty train size distribution is denoted by $\beta$, while the decaying exponent of autocorrelation function is denoted by $\gamma$.}
    \label{table:individual}
    \begin{tabular}{m{3.5cm} | m{6.1cm} | m{4.7cm}}
	\hline
        Reference & Dataset & Finding \\
        \hline
        Paxson \emph{et al.}~\cite{Paxson1995WideArea} & TCP connection packets from Bellcore, U.C.B. & $\alpha\approx 2$ \\ \hline
        Kleban \emph{et al.}~\cite{Kleban2003Hierarchical} & Job submissions to supercomputers, Blue Mountain and Blue Pacific & stretched exponential $P(\tau)$ \\ \hline
        Harder \emph{et al.}~\cite{Harder2006Correlated} & Print requests to the printer at Imperial College London & $\alpha=1.76$ for different thresholds of file size, $\alpha=1.3$ for individuals \\ \hline
        Vazquez \emph{et al.}~\cite{Vazquez2006Modeling} & Library loans by the faculty at University of Notre Dame & $\alpha$ distributed around $1$ \\ \hline
        Nakamura \emph{et al.}~\cite{Nakamura2007Universal} & Locomotor activity, e.g., resting periods, from 14 patients and 11 healthy control subjects & $\alpha=1.92(6)$ for controls, $1.72(11)$ for patients \\ \hline
        Alfi \emph{et al.}~\cite{Alfi2007Conference, Alfi2009How} & Statphys23 registration statistics & logarithmic singularity up to the deadline \\ \hline
        Coley \emph{et al.}~\cite{Coley2008Arm} & Inter-movement intervals of arms of human subjects & power-law $P(\tau)$ \\ \hline
        Goh \emph{et al.}~\cite{Goh2008Burstiness} & Various datasets in human dynamics, texts, and cardiac rhythms & high $B$ and negligible $M$ for human activities \\ \hline
        Baek \emph{et al.}~\cite{Baek2008Testing} & Linux command histories of six users & $\alpha\in [1.47,1.74]$ \\ \hline
        Bohorquez \emph{et al.}~\cite{Bohorquez2009Common} & Conflicts from media, government and non-governmental organization, and academic studies & heterogeneous numbers of conflicts per day \\ \hline
        Bogachev \emph{et al.}~\cite{Bogachev2009Occurrence} & Outgoing traffic of 3 HTTP servers: two Canadian universities and NASA Kennedy Space Center & stretched exponential $P(\tau)$ \\ \hline
        Jo \emph{et al.}~\cite{Jo2012Timevarying} & Paper updating intervals in \url{arXiv.org} & $\alpha_w\in [0.76,1.16]$ depending on the number of authors \\ \hline
        Mryglod \emph{et al.}~\cite{Mryglod2012Editorial} & Paper processing times in Physica A and others & log-normal $P(\tau_w)$ with power-law tails of $\alpha_w=1$ \\ \hline
        Hartonen \emph{et al.}~\cite{Hartonen2013How} & Paper processing times in JSTAT and JHEP & no power-law in $P(\tau_w)$ \\ \hline
        Lee \emph{et al.}~\cite{Lee2013Mobile} & WiFi connectivity of iPhone users in urban areas & $\alpha=1.63$ \\ \hline
        Hasan \emph{et al.}~\cite{Hasan2013Spatiotemporal} & Stay times from smart card transaction dataset in London, UK & heavy-tailed $P(\tau)$ \\ \hline
        Wang \emph{et al.}~\cite{Wang2015Temporal} & Emergency calls in a Chinese city & $\alpha\in [0.86,1.19]$, $\beta=2.21$ \\ \hline
    \end{tabular}
\end{table}

\begin{figure}[!t]
    \center
    \includegraphics[width=\columnwidth]{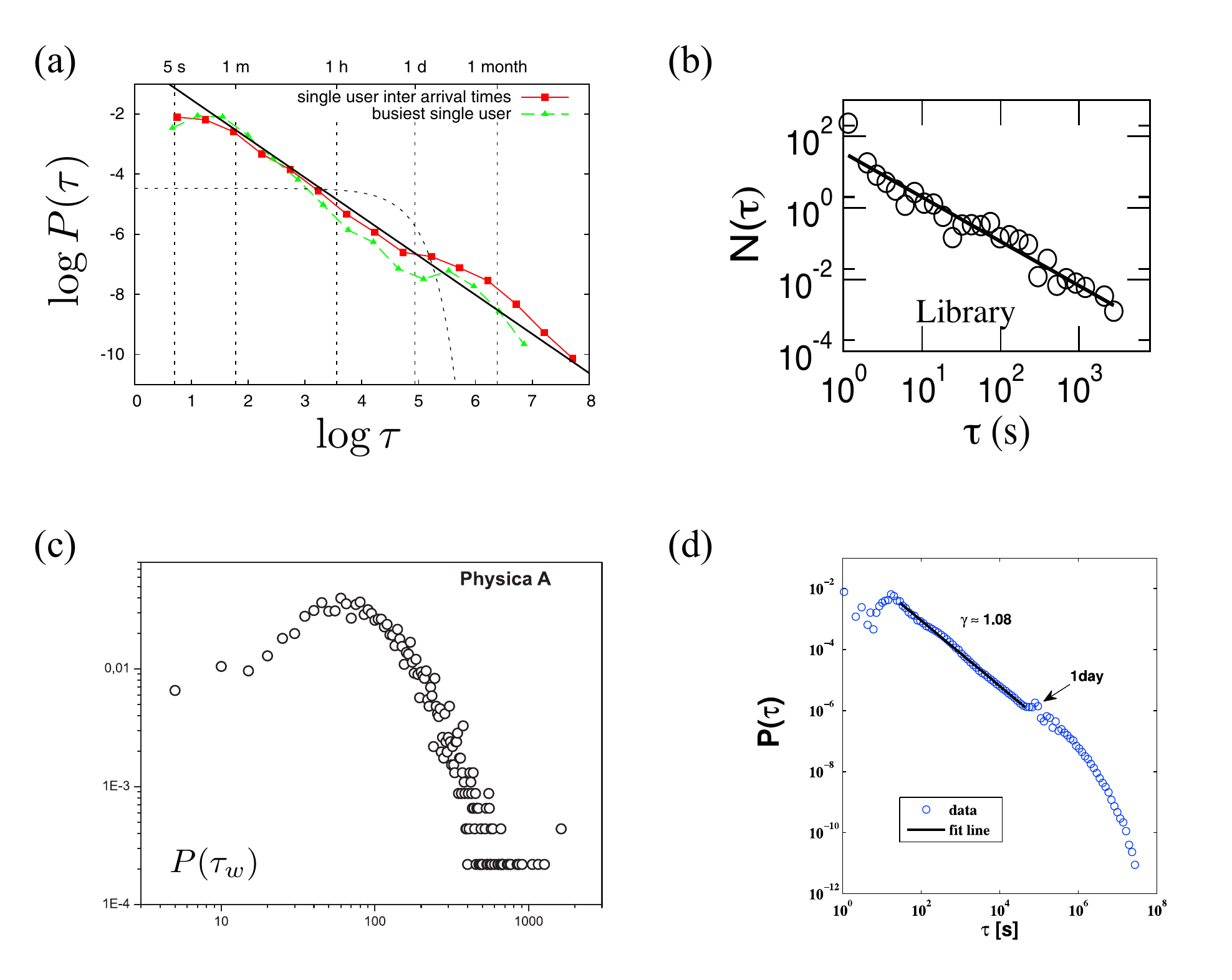}
    \caption{Individual activities: Examples of inter-event time distribution $P(\tau)$ from the datasets for (a) job submissions to a printer (\emph{Source:} Adapted from Ref.~\cite{Harder2006Correlated} under Copyright $\textcopyright$ 2005 Elsevier B.V. All rights reserved.), (b) library loans made by a single individual (\emph{Source:} Adapted from Ref.~\cite{Vazquez2006Modeling} under Copyright (2006) by the American Physical Society.), (c) manuscripts submitted to Physica A (\emph{Source:} Adapted from Ref.~\cite{Mryglod2012Editorial}.), and (d) emergency calls in a Chinese city (\emph{Source:} Adapted from Ref.~\cite{Wang2015Temporal} under Copyright $\textcopyright$ 2015 Elsevier B.V. All rights reserved.). In all cases, heavy tail behaviour in $P(\tau)$ is observed, most of which have been fitted with the power-law form in the mentioned references.}
    \label{fig:empirical_individual}
\end{figure}

\section{Individual activities}\label{subsect:individual}

We first overview the empirical findings of inhomogeneous temporal patterns or bursts in individual activities not necessarily involving direct interactions between individuals. We also include observations of bursts in individual activities at the population or system level. Such findings and observations range from everyday life to professional activities, e.g., including job submissions to supercomputers~\cite{Kleban2003Hierarchical}, print requests to the printer~\cite{Harder2006Correlated}, and library loans by the faculty at a university~\cite{Vazquez2006Modeling}. Events like paper processing or updating times~\cite{Jo2012Timevarying, Mryglod2012Editorial, Hartonen2013How} and human arm movements~\cite{Nakamura2007Universal, Coley2008Arm} have also been analysed to show the existence of bursty behaviour. In Table~\ref{table:individual} we have summarized a number of empirical results of bursty behaviour although it should not be considered as an exhaustive list.

In most cases mentioned above, the distributions $P(\tau)$ and $P(\tau_w)$ have been reported to have a heavy or power-law tail. To demonstrate such cases a few examples of inter-event time distribution are presented in Fig.~\ref{fig:empirical_individual}. Whenever the distribution is described in terms of a power law, the power-law exponent is provided and its value turns out to be quite diverse, i.e., ranging from $0.7$ to $2$. It should be noted, however, that even in case of the same dataset one can find that the value of the power-law exponent can vary from one individual to another, in other words describing the behaviour in terms of a distribution of exponent values. This observation indicates that the power-law behaviour in human dynamics is rather sensitive to the details of the phenomenon in question, and hence it seems not supporting the perspective of universality classes in statistical physics. It has been argued that the large variance of individual characters may induce a heterogeneous inter-event time distribution at the population level~\cite{Jiang2016Twostate}. Researchers have also found other functional forms of $P(\tau)$ that match with the empirical datasets, like stretched exponential~\cite{Kleban2003Hierarchical, Bogachev2009Occurrence} and log-normal ones~\cite{Mryglod2012Editorial}, which implies that different bursty phenomena may have different origins or follow different mechanisms. 

Heavy-tailed $P(\tau)$ for individual activities may give some hints about the origin of bursty dynamics of human individuals. For this reason one can ask a question: Can the bursty dynamics observed in the individual activities be understood in terms of the ``purely'' intrinsic property of those individuals, or in terms of the interaction-driven extrinsic property? In other words, one can ask if the bursty dynamics of individuals is the consequence of node burstiness or link burstiness. As human beings are social, it is hard to say how much the behavioural datasets reflect purely individual actions, when compared to the interaction-driven activities. This is an important yet unresolved issue for understanding the origin of bursts in human dynamics.

\section{Interaction-driven collective activities}\label{subsect:interaction}

\begin{figure}[!t]
    \includegraphics[width=\columnwidth]{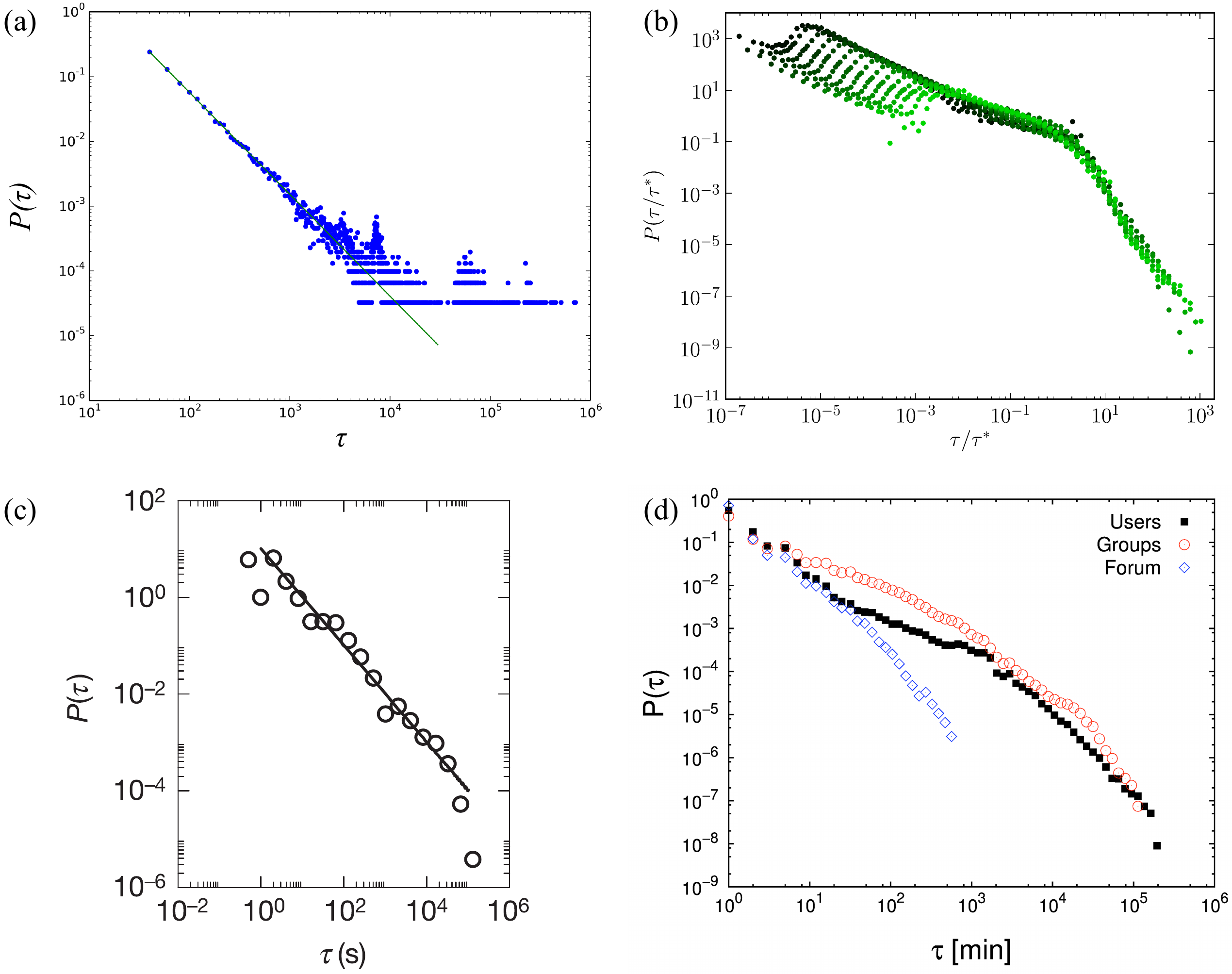}
    \caption{Interaction-driven collective activities: Examples of inter-event time distribution $P(\tau)$ from the datasets for (a) face-to-face interaction among high school students (\emph{Source:} Adapted from Ref.~\cite{Fournet2014Contact}.), (b) mobile phone calls in a European country (\emph{Source:} Adapted from Ref.~\cite{Karsai2011Small} under Copyright (2011) by the American Physical Society.), (c) email communications in a university (\emph{Source:} Adapted from Ref.~\cite{Barabasi2005Origin} by permission from Macmillan Publishers Ltd: Nature, 435:207--211, copyright (2005).), and (d) messages on an online forum (\emph{Source:} Adapted from Ref.~\cite{Panzarasa2015Emergence} under Copyright (2015) by the American Physical Society.). In all cases, heavy tails of $P(\tau)$ are observed, most of which have been fitted with power-law form in the mentioned references.}
    \label{fig:empirical_interaction}
\end{figure}

Next we present some empirical findings of interaction-driven collective activities of several subcategories mostly according to the communication channel used in the interaction. Among these cases face-to-face interaction is considered to be the most direct and natural way of communication or interaction between human individuals as in this case people must be spatially close to each other and the communication takes place in real time. As for the other means of communication nearest to face-to-face interactions are those based on real time video and voice links, like Skype and Hangout, as they provide the feeling of closeness or even intimacy between the individuals, although they are not spatially close to each others. Then comes the communication by using phones, especially mobile ones in recent years, as the interaction still takes place in real time over the voice link providing some feeling of closeness or intimacy though the communication is location-independent. After mobile phone communication come services like traditional posted letters, emails, and text messages or SMSs, which are not means of real time interactions. The recently introduced web-based messaging services, e.g., proposed by Social Networking Services (SNSs), have become very popular lately especially for younger generations. It should be noted, however, that people nowadays interact with each other using many of these communication channels in parallel, simultaneously or intermittently. In Fig.~\ref{fig:empirical_interaction} we have collected a few examples of inter-event time distributions.

\subsection{Face-to-face interactions}\label{subsubsect:face2face}

To collect data from face-to-face interaction at large scale is challenging. However, today's communication technology and smart devices provide a solution as these devices are able  to communicate with each other, thus enabling to collect data from face-to-face interactions, containing information of the proximity between individuals. Examples of this approach are the SocioPatterns (\url{www.sociopatterns.org}) and other similar projects~\cite{Obadia2015Detailed}, in which wearable sensors with Radio-Frequency Identification (RFID) are used to collect datasets of individual contacts in real environments, such as schools, museums, hospitals, and academic conferences~\cite{Cattuto2010Dynamics, Stehle2011Simulation, Stehle2011HighResolution, Isella2011Whats, Vanhems2013Estimating}. Other studies have used Bluetooth devices~\cite{Hui2005Pocket}, infrared modules~\cite{Takaguchi2011Predictability}, or motes~\cite{Hashemian2010Flunet} that can also communicate with each other. Despite the advantages of directly measuring the proximity between individuals, the number of nodes in those datasets is relatively small, i.e., of the order of hundreds. Thus, large-scale conclusions may not be deduced from this approach.

A number of empirical findings concerning face-to-face interaction are summarised in Table~\ref{table:face2face}. In some analyses, it was found that the inter-event times or inter-contact times are power-law distributed with exponent between $1.4$ and $1.6$~\cite{Hui2005Pocket, Takaguchi2011Predictability, Fournet2014Contact}. The distributions of contact times or durations have been observed to show heavy tails in their distributions~\cite{Hui2005Pocket, Cattuto2010Dynamics, Stehle2011Simulation, Stehle2011HighResolution, Isella2011Whats, Fournet2014Contact}. Since the ``typical'' timescale of contact durations is much shorter than that of inter-contact times, the contact durations can be ignored in the analyses of bursty patterns. Note that the contact durations could be affected by the inter-contact times just before or after the contacts, of which the latter resembles the recovery time of neurons after firing.

\begin{table}[!t]
    \caption{Empirical findings of interaction-driven activities based on face-to-face interaction. The notations are the same as in Table~\ref{table:individual}.}
    \label{table:face2face}
    \begin{tabular}{m{3.5cm} | m{6.1cm} | m{4.7cm}}
        \hline
        Reference & Dataset & Finding \\
        \hline
        Hui \emph{et al.}~\cite{Hui2005Pocket} & Face-to-face interaction logs in an IEEE conference & $\alpha=1.4$ \\ \hline
        Takaguchi \emph{et al.}~\cite{Takaguchi2011Predictability} & Face-to-face interaction logs in offices in Japan & $\alpha=1.52$ \\ \hline
        Starnini \emph{et al.}~\cite{Starnini2012Random, Starnini2013Modeling}, Zhao \emph{et al.}~\cite{Zhao2011Social} & Face-to-face proximity datasets using RFID in the frame of SocioPatterns project & heavy-tailed $P(\tau)$\\ \hline
        Sun \emph{et al.}~\cite{Sun2013Understanding} & Users' encounters in public transit transaction in Singapore & daily peaks in $P(\tau)$ \\ \hline
        Fournet \emph{et al.}~\cite{Fournet2014Contact} & Face-to-face encounters between high school students in France & $\alpha=1.57$ \\ \hline
    \end{tabular}
\end{table}

\subsection{Mobile phone-based interactions}\label{subsubsect:mobilephone}

Recently, mobile phones or handsets are utilised to accurately measure or sense human behaviour. These personal devices, being equipped with a variety of sensors like GPS and WiFi, are carried around by the users everyday and all day through, thus they are capable to collect precise information about the communications, whereabouts, and online activities of their owners. Moreover, since the number of users or phone numbers in some datasets is up to several millions or even larger~\cite{Onnela2007Structure, Miritello2011Dynamical, Aoki2016Inputoutput}, they provide ways to overcome the issues due to the small sampling sizes.

The reliability of datasets collected from mobile phones was tested in the series of studies conducted within the frame of Reality Mining project~\cite{Eagle2006Reality, Pentland2009Reality, Eagle2009Inferring, Aharony2011Social}. It was shown that the behavioural data are at least comparable to self-report survey data in terms of friendship network and even capturing information that were missing from self-reports~\cite{Eagle2009Inferring}. Similar approaches were taken in the OtaSizzle project at Aalto University~\cite{Mantyla2008OtaSizzle, Karikoski2010Measuring, Jo2012Spatiotemporal} and Copenhagen Networks Study~\cite{Stopczynski2014Measuring, Sekara2016Fundamental, Sapiezynski2016Inferring}, where multiple kinds of individual activities and interactions were recorded simultaneously but from a relatively small group of volunteers. For other studies using mobile phone datasets, see Ref.~\cite{Blondel2015Survey} and references therein.

\begin{table}[!t]
    \caption{Empirical findings for interaction-driven activities using mobile phones, i.e., mobile phone calls and Short Message Services (SMSs). The notations are the same as in Table~\ref{table:individual}.}
    \label{table:mobilePhone}
    \begin{tabular}{m{3.5cm} | m{6.1cm} | m{4.7cm}}
        \hline
        Reference & Dataset & Finding \\
        \hline
        Candia \emph{et al.}~\cite{Candia2008Uncovering} & Mobile phone calls (source not mentioned) & $\alpha=0.9(1)$ \\ \hline
        Hong \emph{et al.}~\cite{Hong2009HeavyTailed} & SMS records of volunteers in the university & $\alpha\in [1.52,2.09]$ \\ \hline
        Wu \emph{et al.}~\cite{Wu2010Evidence} & SMSs of individual users from a mobile phone operator & bimodal $P(\tau)$ with power-law regimes, where $\alpha$ ($\alpha_w$) centred at $1.5$ ($2$) \\ \hline
        Miritello \emph{et al.}~\cite{Miritello2011Dynamical} & Mobile phone calls from a European operator in a single country & heavy-tailed $P(\tau_w)$ \\ \hline
        Zhao \emph{et al.}~\cite{Zhao2011Empirical} & SMSs in China & $\alpha\in [1.1,1.3]$ depending on activity level \\ \hline
        Karsai \emph{et al.}~\cite{Karsai2011Small, Karsai2012Correlated, Karsai2012Universal} & Mobile phone calls and SMSs from a European operator & $\alpha=0.7$, $\beta=4.1$, $\gamma=0.5$ for calls and $\alpha=1.0$, $\beta=3.9$, $\gamma=0.6$ for SMSs \\ \hline
        Kivela \emph{et al.}~\cite{Kivela2012Multiscale} & Mobile phone calls from a European operator & heavy-tailed $P(\tau)$ and $P(\tau_w)$ \\ \hline
        Jo \emph{et al.}~\cite{Jo2012Circadian} & Mobile phone calls and SMSs from a European operator & heavy-tailed $P(\tau)$ with daily and weekly peaks \\ \hline
        Jiang \emph{et al.}~\cite{Jiang2013Calling, Jiang2016Twostate} & Mobile phone call dataset from a Chinese cell phone operator & $\alpha=0.873$ for all users, stretched exponential or $\alpha \in [1.5,2.6]$ for individuals, exponential $P_{\Delta t}(E)$ for a majority of individuals \\ \hline
        Schneider \emph{et al.}~\cite{Schneider2013Unravelling} & Surveys and mobile phone data from Paris and Chicago & heavy-tailed $P(\tau)$ for home/work, $\alpha=0.49$ for other locations \\ \hline
        Aoki \emph{et al.}~\cite{Aoki2016Inputoutput} & Mobile phone calls and SMSs from a European cellphone service provider & $\alpha=1.176$ (calls) and $1.388$ (SMSs) \\ \hline
    \end{tabular}
\end{table}

Bursty dynamics has been observed in both mobile phone calls and Short Message Services (SMSs). In a number of empirical results, one finds the heavy-tailed distribution $P(\tau)$, in particular with power-law scaling regime. The values of power-law exponent $\alpha$ turned out to be dependent on communication channels, whether they are for calls or for SMSs. We found that for calls the exponent value of $\alpha\approx 0.7$~\cite{Karsai2012Universal, Karsai2012Correlated} ($\approx 1.2$ in Ref.~\cite{Aoki2016Inputoutput}) tends to be smaller than that observed in SMS sequences with $\alpha=1.0$~\cite{Karsai2012Universal, Karsai2012Correlated} ($\approx 1.4$ in Ref.~\cite{Aoki2016Inputoutput}). It was also found in Fig.~\ref{fig:empirical_interaction}(b) that $P(\tau)$s for different activity groups collapse onto one curve when being normalised by the average inter-event time for each $P(\tau)$, which implies a strong similarity in human behaviour across the different activity levels.

From another mobile phone dataset, Jiang~\emph{et al.} found that although the aggregate $P(\tau)$ follows a power law, a majority of individual users, i.e., more than 73\%, show Weibull distributions for inter-event times~\cite{Jiang2013Calling}. For other users in the ``power-law'' group, the values of $\alpha$ varied from $1.5$ to $2.6$. In addition, bimodal distributions of inter-event times have been observed in SMS datasets~\cite{Wu2010Evidence}: The distributions are power-law for $\tau<\tau_c$, and exponential for $\tau>\tau_c$. This functional form is different from the power-law distribution with exponential cutoff, hence implying different mechanisms to act in the background. For the power-law regime, the values of $\alpha$ obtained at the individual level are distributed around $1.5$. As shown in Refs.~\cite{Karsai2012Universal, Karsai2012Correlated}, long-range memory effects have also been observed in terms of heavy-tailed burst size distributions and power-law decaying autocorrelation functions both in calls and SMSs. We note that such long-range memory effects have been investigated only recently and phenomenologically. For more fundamental understanding, one might need to obtain more information about the mobile phone users as typically done in sociology, e.g., in Ref.~\cite{Rettie2009Mobile}.

\begin{table}[!t] 
    \caption{Empirical findings of interaction-driven activities by letters and emails. The notations are the same as in Table~\ref{table:individual}.}
    \label{table:letteremail}
    \begin{tabular}{m{3.5cm} | m{6.1cm} | m{4.7cm}}
        \hline
        Reference & Dataset & Finding \\
    \hline
    Oliveira \emph{et al.}~\cite{Oliveira2005Human}, Vazquez \emph{et al.}~\cite{Vazquez2006Modeling} & Letter correspondence of Darwin, Einstein, and Freud & $\alpha_w \approx 1.5$ \\ \hline
        Li \emph{et al.}~\cite{Li2008Empirical} & Letter correspondence of a Chinese scientist & $\alpha=\alpha_w=2.1(1)$ \\ \hline
        Malmgren \emph{et al.}~\cite{Malmgren2009Universality} & Letter correspondence of 16 writers, performers, politicians, and scientists & heavy-tailed $P(\tau)$ \\ \hline
        Formentin \emph{et al.}~\cite{Formentin2014Hidden, Formentin2015New} & Letters, emails, SMSs from diverse sources & $\alpha_w \approx 1.5$ \\ \hline
        Barab\'asi~\cite{Barabasi2005Origin}, Eckmann \emph{et al.}~\cite{Eckmann2004Entropy}, Johansen~\cite{Johansen2004Probing} & Emails in a university (Universite de Geneve or Weizmann Institute of Science) & $\alpha=1$, $\alpha_w=1$ \\ \hline
        Malmgren \emph{et al.}~\cite{Malmgren2008Poissonian} & Email dataset as in~\cite{Barabasi2005Origin} & heavy-tailed $P(\tau)$  \\ \hline
        Gao \emph{et al.}~\cite{Gao2011Network} & Individuals in Enron email dataset & $\alpha\in [0.8,1.8]$ \\ \hline
        Iribarren \emph{et al.}~\cite{Iribarren2009Impact, Iribarren2011Branching} & Campaign propagation dataset by emails & heavy-tailed $P(\tau_w)$ \\ \hline
    \end{tabular}
\end{table}

\subsection{Communication by posted letters and emails}\label{subsubsect:letter}

In contrast to the face-to-face interaction and mobile phone call communication, communication by posted letters, electronic mails or emails, and text messages or SMSs does not take place synchronously in real time and may depend on the location and distance between the senders and receivers. Hence their interaction patterns could be very different from those in face-to-face and mobile phone call communications. As we already discussed SMSs in the previous Subsection, here we consider posted letters and emails.

Traditionally posted letters were one of the most important communication channels between people outside their daily proximity before various ICT-based communication channels like emails and mobile phones emerged. There are a few examples of such datasets of letters being exchanged by historic figures, such as Darwin, Einstein, and Freud, that have been analysed and found to have the heavy-tailed distribution of waiting times between letters being sent~\cite{Oliveira2005Human}. In these cases the waiting time distributions have often been fitted using power-law forms~\cite{Oliveira2005Human, Li2008Empirical}, while alternative mechanisms excluding power-law forms have been studied in terms of cascading non-homogeneous Poisson processes~\cite{Malmgren2009Universality}. See the summary of empirical findings in Table~\ref{table:letteremail}.

Recently, however, the usage of posted letters has dramatically dropped due to people using emails for various purposes and contexts, e.g., for communications between colleagues or friends, etc. This development has provided an unprecedented amount of email data and plethora of email datasets rich of detailed and useful information of social interactions and temporal patterns for the researchers to investigate. Many of these datasets have been analysed to investigate the origin of bursts in human dynamics, which has led to some debates on the issue. In the beginning, Barab\'asi claimed that the inter-event time and waiting time distributions for email users show power-law tails as $P(\tau)\sim\tau^{-\alpha}$ with $\alpha=1$~\cite{Barabasi2005Origin}. Similar analyses using the same email dataset were previously performed in Refs.~\cite{Eckmann2004Entropy,Johansen2004Probing}. Since then, there are debates between different research groups about the origin of bursts~\cite{Stouffer2005Comment, Barabasi2005Reply, Stouffer2006Lognormal}. Malmgren~\emph{et al.} suggested Poissonian explanation for heavy tails in the email communication patterns~\cite{Malmgren2008Poissonian} to argue that the bursts are the consequence of daily and weekly cycles of humans with cascading behaviour whenever the email session is initiated. Later they also argued about the universality in human activity~\cite{Malmgren2009Universality}. Despite these debates, many issues were left unresolved, such as how cyclic patterns intrinsic in human behaviour interplays with other human factors like task executions. For resolving this, a deseasoning method was applied to the mobile phone calls and SMSs from a European operator, leading to the conclusion that the burstiness is robust with respect to the deseasoning of circadian and weekly cycles~\cite{Jo2012Circadian}. Here we should remark that the bursty dynamics observed in one communication channel, e.g., emails, could be driven by different mechanisms as ones observed other communication channels, e.g., mobile phone calls and SMSs. Thus one needs to be careful whenever translating the conclusions from one dataset into those for another dataset. 

\begin{table}[!t]
    \caption{Empirical findings of interaction-driven and some individual activities using web services (part I). The notations are the same as in Table~\ref{table:individual}.}
    \label{table:web-based1}
    \begin{tabular}{m{3.5cm} | m{6.1cm} | m{4.7cm}}
        \hline
        Reference & Dataset & Finding \\
    \hline
        Henderson \emph{et al.}~\cite{Henderson2001Modelling} & Logging of users in networked games, Quake and Half-Life & $\alpha=2.15$  \\ \hline
        Dewes \emph{et al.}~\cite{Dewes2003Analysis} & Web-chat messages in the University of Saarland & power-law $P(\tau)$ \\ \hline 
        Vazquez \emph{et al.}~\cite{Vazquez2006Modeling}, Dezs\H{o} \emph{et al.}~\cite{Dezso2006Dynamics} & Web browsing in Hungarian news and entertainment portal, \url{www.origo.hu} & $\alpha$ distributed around $\approx 1.1$ \\ \hline
        Kujawski \emph{et al.}~\cite{Kujawski2007Growing} & Online forums and news groups in Poland & $\alpha=1.25$ for a forum, $1.33$ for a news group \\ \hline
        Goncalves \emph{et al.}~\cite{Goncalves2008Human} & Logs of individuals to the web server at Emory University & $\alpha=1$ for one URL, $\alpha=1.25$ for all pages \\ \hline
        Zhou \emph{et al.}~\cite{Zhou2008Role} & Rating by users in Netflix & $\alpha=2.08$ for the whole, $\in [1.5,2.5]$ depending on activity level \\ \hline
        Hu \emph{et al.}~\cite{Hu2008Empirical} & Online music service in a Chinese university & heavy-tailed $P(\tau)$ \\ \hline
        Crane \emph{et al.}~\cite{Crane2008Robust} & YouTube video views & time series into 4 classes: exogenous/endogenous critical/subcritical \\ \hline
        Altmann \emph{et al.}~\cite{Altmann2009Beyond} & Frequent words in USENET discussion groups & stretched exponential $P(\tau)$ \\ \hline
        Rybski \emph{et al.}~\cite{Rybski2009Scaling} & Messages in online community of men having sex with men, messages between teenagers & power-law decaying $A(t_d)$ and Hurst exponent $>0.5$ \\ \hline
        Radicchi~\cite{Radicchi2009Human} & Feedback messages in Ebay and queries in America On Line & $\alpha=1.9$ for both datasets  \\ \hline
        Radicchi~\cite{Radicchi2009Human} & Logging to English Wikipedia & $\alpha=1.2$ for the whole, $\in [1.1,2.3]$ depending on activity level \\ \hline
        Rocha \emph{et al.}~\cite{Rocha2010Information} & Online posts by buyers and sellers in the prostitution network & $\alpha=1.49(4)$ for sellers, $1.5(1)$ for buyers \\ \hline
        Ratkiewicz \emph{et al.}~\cite{Ratkiewicz2010Characterizing} & Two traffic datasets of Wikipedia & $\alpha=0.8$ (events if $\frac{\Delta k}{k}>1$) \\  \hline
\end{tabular}
\end{table}

\begin{table}[!t] 
    \caption{Empirical findings of interaction-driven and some individual activities using web services (part II). The notations are the same as in Table~\ref{table:individual}.}
    \label{table:web-based2}
    \begin{tabular}{m{3.5cm} | m{6.1cm} | m{4.7cm}}
        \hline
        Reference & Dataset & Finding \\
    \hline
        Wang \emph{et al.}~\cite{Wang2011Heterogenous} & Edits of articles in Chinese Wikipedia and blog posting in website of Nanjing University & $\alpha$ depending on activity level \\ \hline
        Guo \emph{et al.}~\cite{Guo2011Weblog} & Logging of bloggers at \url{sciencenet.cn} & $\alpha\in [0.2,0.5]$\\ \hline
        Szell \emph{et al.}~\cite{Szell2012Understanding} & Intervals between jumps in massive multiplayer online game, Pardus & $\alpha=2.2$ \\ \hline
        Rybski \emph{et al.}~\cite{Rybski2012Communication} & Messages in an online community POK & $\alpha=1.5$ in days, $\alpha=2.25$ in seconds \\ \hline
        Jo \emph{et al.}~\cite{Jo2012Spatiotemporal} & Web domain visits by all users in OtaSizzle project & heavy-tailed $P(\tau)$ \\ \hline
        Yan \emph{et al.}~\cite{Yan2012Human, Yan2013Social} & Messages in Chinese microblog & $\alpha=1.231$ for one user, $1.323$ for others \\ \hline
        Yasseri \emph{et al.}~\cite{Yasseri2012Dynamics, Yasseri2012Circadian} & Edits on $20$ highly disputed articles of Wikipedia & daily patterns, $\alpha=0.97$, $\gamma=0.56$ \\ \hline
        Garas \emph{et al.}~\cite{Garas2012Emotional} & Posts in Internet Relay Chat (IRC) & $\alpha=1.53$ \\ \hline
        Zhou \emph{et al.}~\cite{Zhou2012Relative} & Datasets from AOL, Delicious, SMS, and Twitter & $\alpha=1.31$, $1.12$, $1.33$, $1.05$ for each dataset \\ \hline
        Zhao \emph{et al.}~\cite{Zhao2012Empirical} & Netflix, MovieLens, Delicious, Ebay, FriendFeed, and Twitter & $\alpha\in [1.17,2.15]$ \\ \hline
        Gaito \emph{et al.}~\cite{gaito2012bursty} & Link creation in Renren online social networks & $\alpha$ broadly distributed around $1$ \\ \hline
        Mathiesen \emph{et al.}~\cite{Mathiesen2013Excitable} & Tweets mentioning brand names & $1/f$ noise \\ \hline
        Kikas \emph{et al.}~\cite{Kikas2013Bursty} & Social link creation and removal in Skype & $\alpha\approx 0.85$ \\ \hline
        Zhao \emph{et al.}~\cite{Zhao2013Emergence} & E-commerce (Douban and Taobao), and MPR & $\alpha\in[1.41,2.04]$ for individuals \\ \hline
        Karimi \emph{et al.}~\cite{Karimi2014Structural} & Posts and messages in Sweden's online movie community & Broad $P(\tau)$ \\ \hline
        Panzarasa \emph{et al.}~\cite{Panzarasa2015Emergence} & Messages on an online forum at the University of California, Irvine & $\alpha=1.53(11)$, $0.71(4)$, $1.87(5)$ for different scaling regimes at the user level\\ \hline
        Kwon \emph{et al.}~\cite{Kwon2016Double} & Edits on $418$ featured articles in English Wikipedia & double power-law with $\alpha=1$ for small $\tau$ and $2$ for large $\tau$ \\ \hline
        Zhang \emph{et al.}~\cite{Zhang2016Multiscale} & Chatting messages at Tencent QQ in China & $\alpha\in [1.3,1.5]$ \\ \hline
\end{tabular}
\end{table}

\subsection{Web-based activities and social interactions}\label{subsubsect:web} 

Since the World Wide Web (WWW) was invented by Tim Berners-Lee in 1989~\cite{BernersLee2000Weaving}, it has grown enormously for the past few decades to find over one billion websites today~\cite{InternetLiveStatsTotal}. Nowadays it is not just a network made of hyperlinks between web pages, but it functions as the platform for e-commerce, online forums~\cite{Ugander2011Anatomy, Eom2015Tailscope}, and SNSs like Twitter~\cite{Kwak2010What} and Facebook~\cite{Ugander2011Anatomy}, etc. More recently the websites are accessible not only from desktop computers but also from various mobile devices including mobile phones and tablets. In this sense, the web-based datasets can be considered to reflect well human behaviour of people's information acquisition, entertainment, and maintaining relationships, etc., despite the fact that the available datasets are typically reflecting only some aspects of the reality. As all the interactions on the web can be in principle recorded, such datasets, along with those from mobile phones, opened a new avenue for social sciences. The analysis of the collected large data corpus called for methodologies borrowed from other disciplines like computational and even physical sciences. This is even more true for the modelling the possible underlying mechanisms in producing the structure of the system and modelling its dynamics~\cite{Lazer2009Computational}.  

In Tables~\ref{table:web-based1} and~\ref{table:web-based2}, we present a number of empirical findings for bursts in various web-based activities including several individual activities. Here the interaction between individual users could be message exchanges~\cite{Dewes2003Analysis, Rybski2009Scaling, Radicchi2009Human, Rybski2012Communication, Yan2012Human, Garas2012Emotional, Karimi2014Structural, Zhang2016Multiscale} and discussions in forums, news groups, and Internet Relay Chat (IRC) channels~\cite{Kujawski2007Growing, Rocha2010Information, Garas2012Emotional, Panzarasa2015Emergence, Zhang2016Multiscale}. Individual activities include logging actions to online games~\cite{Henderson2001Modelling}, web servers at universities~\cite{Goncalves2008Human}, Wikipedia~\cite{Radicchi2009Human}, and blogs~\cite{Vazquez2006Modeling, Dezso2006Dynamics}, as well as online queries~\cite{Radicchi2009Human}, edits of articles on the web~\cite{Wang2011Heterogenous, Yasseri2012Dynamics, Kwon2016Double}, and jumps in the online game~\cite{Szell2012Understanding} to name a few.

On the basis of these studies we observe that the values of power-law exponent of inter-event time distributions are very diverse, i.e., ranging from $0.2$ to $2.5$. It turns out that the power-law behaviour depends on the activity level of users or on the observation scale. For example, Zhou~\emph{et al.} showed by analysing the rating patterns in Netflix that the group of more active users shows a larger value of $\alpha$, i.e., less bursty temporal patterns~\cite{Zhou2008Role}. More active users may have smaller averages of inter-event times, not necessarily leading to the larger values of $\alpha$. Thus, this tendency or dependence of power-law behaviour on the activity level of users must be investigated rigorously. For this, the interaction or network structure of individual users can be relevant in understanding the complex bursty dynamics. The effect of the observation scale on the power-law behaviour, from individuals to the groups they form, and to the whole population they belong, was observed by Panzarasa and Bonaventura~\cite{Panzarasa2015Emergence}. By analysing messages posted in an online forum at a university, they found that the inter-event time distribution at the individual level shows three scaling regimes, i.e., for short, intermediate, and long inter-event times, as depicted in Fig.~\ref{fig:empirical_interaction}(d). The scaling regime for the short inter-event times coincides with that of the inter-event time distribution at the population level. As for the inter-event time distribution at the group level, they argued that the group dynamics is governed by a nontrivial reciprocal mechanism between users. In addition to the inter-event time distributions, the long-range memory effects have also been measured in terms of autocorrelation function, power spectrum, or Hurst exponent, e.g., in Refs.~\cite{Rybski2009Scaling, Rybski2012Communication, Yasseri2012Dynamics, Jo2012Circadian, Panzarasa2015Emergence}. Finally, as for the interplay between bursty dynamics and network evolution, Gaito \emph{et al.}~\cite{gaito2012bursty} identified bursty link creation patterns in online social networks and characterised the individual link creation dynamics in four different phases (as acceleration, deceleration, cruising, and inactive). In another work Kikas~\emph{et al.}~\cite{Kikas2013Bursty} analysed the correlation between service adoption and bursty link creation of Skype users, while Myers and Leskovec~\cite{Myers2014Bursty} analysed Twitter datasets to find that information diffusion also creates sudden bursts of new links that in turn affect the users' local network structure.

\begin{table}[!t]
    \caption{Empirical findings of financial activities. The notations are the same as in Table~\ref{table:individual}.}
    \label{table:economic}
    \begin{tabular}{m{3.5cm} | m{6.1cm} | m{4.7cm}}
        \hline
        Reference & Dataset & Finding \\
        \hline
        Mainardi \emph{et al.}~\cite{Mainardi2000Fractional} & Financial market datasets for BUND future & stretched exponential $P(\tau)$ for small $\tau$, power-law $P(\tau)$ for large $\tau$ \\ \hline
        Raberto \emph{et al.}~\cite{Raberto2002Waitingtimes} & GE stock prices & stretched exponential $P(\tau)$ \\ \hline
        Masoliver \emph{et al.}~\cite{Masoliver2003Continuoustime} & US dollar-Deutsche mark future exchange & $\alpha=3.47$ \\ \hline
        Vazquez \emph{et al.}~\cite{Vazquez2006Modeling} & Trade transactions by a stock broker at Central European bank & $\alpha=1.3$ \\ \hline
        Wang \emph{et al.}~\cite{Wang2010Human} & Time intervals between contracts and payments of a logistics company in Shanghai & $\alpha_w=2.5$ \\ \hline
    \end{tabular}
\end{table}

\section{Other bursty patterns}\label{sect:others}

As we remarked in the beginning of this Chapter, we primarily concentrate on direct observations of bursty phenomena in human dynamics. Yet we would like to briefly mention some related sets of observations, which may not be directly taken on human activities or temporal behaviour, but certainly have relevance in the scope of bursty human dynamics.

\subsection{Financial activities}\label{subsect:econo}

We present the bursty patterns in financial interactions in Table~\ref{table:economic}. Examples include financial trades in markets for future, stocks, and foreign exchange. In these cases, the events are mostly defined by transactions, implying that the inter-event time measures the time interval between two consecutive transactions. It has been found that the inter-event time distributions are heavy-tailed. In case with power-law inter-event time distributions, the values of the power-law exponent was found to range from $1.3$ to $3.47$~\cite{Vazquez2006Modeling, Masoliver2003Continuoustime}. 

The inhomogeneous temporal patterns in economic and financial systems have been extensively studied but mostly from the macroscopic perspective~\cite{Mantegna2007Introduction}. For the microscopic approach, one can refer to the economic perspective that time is considered as tradable goods or cost~\cite{Nichols1971Discrimination, Barzel1974Theory}. Based on this concept, one can discuss about the optimal waiting time of agents when they must wait for the service or goods, for example, as modelled in Ref.~\cite{Jo2012Optimized}. Hence such agents might be driven by objectives like maximising the profit or utility, which may lead to different bursty patterns than those for other human activities like communications. Or the economic constraints like the limited time resource can also account for the bursts in communications. These issues can be investigated for better understanding the origin of bursts in human dynamics.

\begin{figure}[!t]
    \includegraphics[width=\columnwidth]{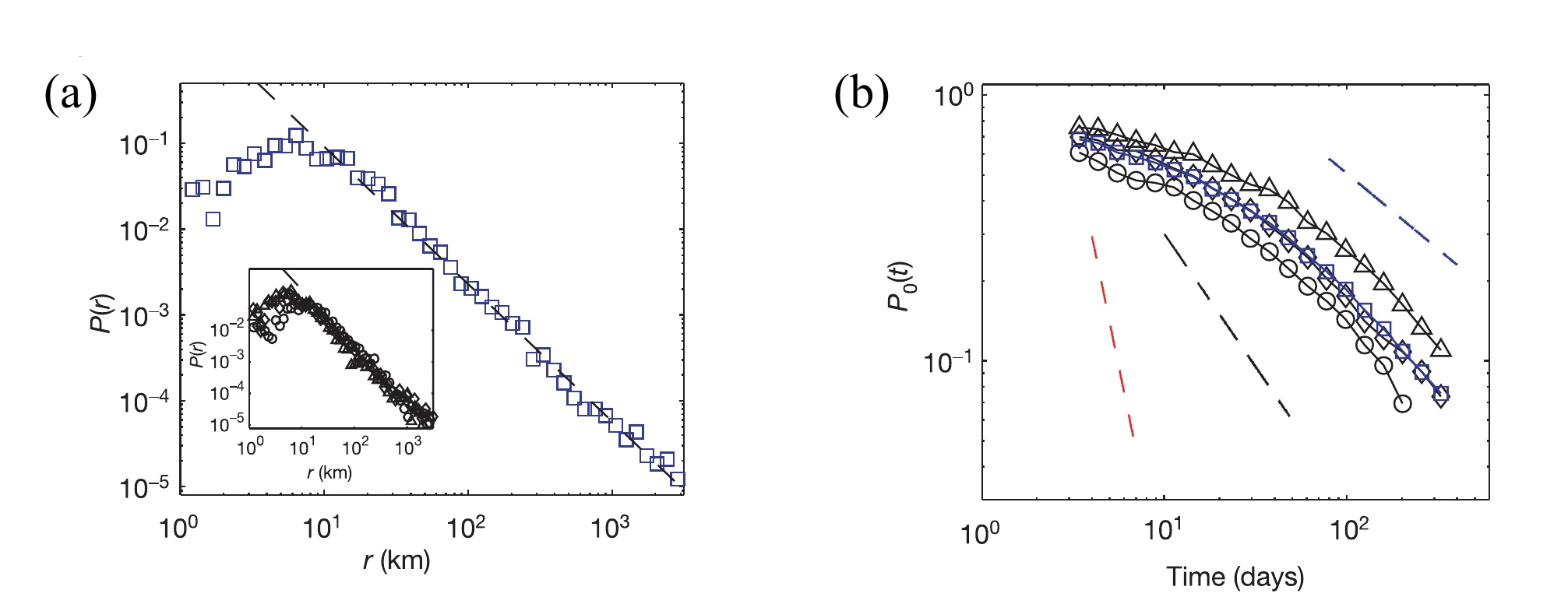}
    \caption{Human mobility patterns: An example of displacement distribution $P(r)$ (a) and inter-event time distribution $P(\tau)$ (b) from the mobility dataset inferred from the circulation of bank notes in the United States of America. Both distributions are described by power-law decaying behaviour. (\emph{Source:} Adapted from Ref.~\cite{Brockmann2006Scaling} by permission from Macmillan Publishers Ltd: Nature, 439:462--465, copyright (2006).)}
    \label{fig:empirical_mobility}
\end{figure}

\subsection{Human mobility}\label{subsect:mobility}

An important aspect of human dynamics addresses their mobility in geographical space as well as other abstract space. Their replacement is typically driven by everyday routines such as going to work, returning home, or go shopping, or on a larger spatial scale when sometimes they migrate to another city or country. Such commuting and travel patterns emerge as a multiscale spatiotemporal phenomenon, which may exhibit bursty patterns in space and/or time. Here events indicate individual movements, thus each event may be described by a time and a distance of the individual's displacement $r$. One common observation is that the distribution of displacement of individuals follows a power law as:
\begin{equation}
    P(r)\sim r^{-\mu},
\end{equation}
sometimes with exponential cutoff~\cite{Brockmann2006Scaling, Gonzalez2008Understanding, Song2010Modelling, Asgari2013Survey}. Some values of $\mu$ are summarised in Table~\ref{table:mobility}, while note that evidences for exponentially distributed replacement has also been found in other datasets~\cite{Liang2012Scaling, Kang2012Intraurban}. Spatial dynamics with power-law distributed displacements is commonly called L\'evy-flights and was found to characterise the mobility of humans~\cite{Gonzalez2008Understanding, Brockmann2006Scaling} and foraging of animals~\cite{Viswanathan1999Optimizing} in spatial space, and even in mental space as well~\cite{Baronchelli2013Levy}.

\begin{table}[!ht]
    \caption{Empirical findings of human mobility patterns. In case when the displacement distribution $P(r)$ is a power law, $\mu$ denotes the power-law exponent characterising the displacement distribution. Other notations are the same as in Table~\ref{table:individual}.}
    \label{table:mobility}
    \begin{tabular}{m{3.5cm} | m{6.1cm} | m{4.7cm}}
        \hline
        Reference & Dataset & Finding \\
        \hline
        Brockmann \emph{et al.}~\cite{Brockmann2006Scaling} & Circulation of bank notes in the United States of America & $\mu=1.59(2)$ and $\alpha=1.60(3)$ \\ \hline
        Gonzalez \emph{et al.}~\cite{Gonzalez2008Understanding} & Mobile phone call dataset & $\mu=1.75(15)$ \\ \hline
        Jiang \emph{et al.}~\cite{Jiang2009Characterizing} & GPS data of taxis' positions in four cities/towns in Sweden & $\mu=2.5$ for intracity, $4.6$ for intercity \\ \hline
        Song \emph{et al.}~\cite{Song2010Modelling} & Mobile phone call dataset and users of a location-based service & $\mu=1.55(5)$ and $\alpha=1.8(1)$ \\ \hline
        Kang \emph{et al.}~\cite{Kang2012Intraurban} & Mobile phone dataset from 8 Chinese cities & exponential $P(r)$ \\ \hline
        Liang \emph{et al.}~\cite{Liang2012Scaling} & GPS datasets of taxis in Beijing & exponential $P(r)$, $\alpha\in [0.5,2.5]$ for individual taxis \\ \hline
        Yan \emph{et al.}~\cite{Yan2013Diversity} & Travel diaries of hundreds of volunteers & $\mu=1.05$ \\ \hline
    \end{tabular}
\end{table}

Individual mobility patterns can also be characterised by the radius of gyration $r_g$, measuring how far an individual trajectory is from its center of mass. For this analysis, the individual trajectory can be described by a sequence of locations, $\{\vec r_1,\cdots, \vec r_n\}$, to calculate the radius of gyration as follows:
\begin{equation}
    r_g \equiv \sqrt{\frac{1}{n}\sum_{i=1}^n |\vec r_i - \vec r_{_{\textrm{CM}}}|^2},
\end{equation}
where the centre of mass of the trajectory is defined as
\begin{equation}
    \vec r_{_{\textrm{CM}}}\equiv \frac{1}{n}\sum_{i=1}^n \vec r_i.
\end{equation}
The distribution of $r_g$ is found to decay as a power law~\cite{Gonzalez2008Understanding}, providing another evidence for heterogeneous mobility patterns of humans.

Recently, the trajectory or the sequence of locations is measured with high time resolution, enabling one to analyse such event sequences using the methods introduced in the previous Chapter. However, there are only several empirical results for inter-event time distributions. As each event denotes a displacement, the inter-event time indicates the staying time in a location or the time interval between two consecutive displacements. In all these cases heavy-tailed distributions $P(\tau)$ were found, some of them with power-law tails. For such cases the estimated values of $\alpha$ exponents are presented in Table~\ref{table:mobility}. These results together with the observations on heterogeneous replacement imply that human mobility is bursty in terms of time and space as well.

\begin{table}[!h]
    \caption{Empirical findings of bursty patterns of animals. The notations are the same as in Table~\ref{table:mobility}.}
    \label{table:animal}
    \begin{tabular}{m{3.5cm} | m{6.1cm} | m{4.7cm}}
        \hline
        Reference & Dataset & Finding \\
        \hline
        Sorribes \emph{et al.}~\cite{Sorribes2011Origin} & Walking activities of flies, \emph{Drosophila melanogaster} & Weibull distribution of $P(\tau)$ \\ \hline
        Boyer \emph{et al.}~\cite{Boyer2012Nonrandom} & Movements of capuchin monkeys & $\mu=2.7$ and $\alpha=1.6$ \\ \hline
        Proekt \emph{et al.}~\cite{Proekt2012Scale} & Movements of adult mice & $\alpha=1.7$ \\ \hline
        Wearmouth \emph{et al.}~\cite{Wearmouth2014Scaling} & Waiting time of ambush predators & $\alpha_w=1.58(36)$ for wild fish, $1.59(38)$ for captive fish \\ \hline
    \end{tabular}
\end{table}

\subsection{Animal behaviours}\label{subsect:animal}

Here we briefly discuss some similarities found in studies of bursty behaviour of animals with that of humans. The bursty behaviour is also observed in temporal patterns of monkeys, mice, and fruit flies, as summarised in Table~\ref{table:animal}. Sorribes~\emph{et al.}~\cite{Sorribes2011Origin} found the walking activity of \emph{Drosophila melanogaster} to show Weibull distributions of inter-event times. They argued that the bursty dynamics of fruit flies are similar to that of humans in terms of positive burstiness parameter $B$ in Eq.~(\ref{eq:burstiness_param}) and near-zero memory coefficient $M$ in Eq.~(\ref{eq:memory_coeff}). Boyer~\emph{et al.}~\cite{Boyer2012Nonrandom} found by analysing the displacements of capuchin monkeys that the power-law exponents are $\mu=2.7$ and $\alpha=1.6$, respectively. On the other hand in case of human mobility, the power-law exponents were found to be $\mu\approx 1.6$ and $\alpha\approx 1.8$, as presented in Ref.~\cite{Song2010Modelling}. Furthermore, Proekt~\emph{et al.}~\cite{Proekt2012Scale} observed that displacements of adult mice show the power-law $P(\tau)$ with exponent $\alpha=1.7$. This value of $\alpha$ turns out to be similar to the values for monkeys and humans, which could imply the same underlying mechanisms or a kind of universality. 


\chapter{Models and mechanisms of bursty behaviour}
\label{chapter:model}


Bursty dynamical patterns characterise human behaviour not only at the individual level but also at the level of dyadic interactions and even when it comes to collective phenomena at the network level. To get insight and capture these phenomena at multiple scale, a number of models have been introduced, which sometimes lead to seemingly conflicting interpretations. In this Chapter our aim is to give a comprehensive summary of all these efforts and let the reader to judge which one of them seems to be the most suitable explanation of the same phenomena.

\section{Models of individual activity}
\label{sec:indivmodels}

The first observations of human bursty patterns were commonly addressing the activity of individuals, although many observations were made from the datasets based on social interactions at the dyadic level. All these studies were reporting heterogeneous non-Poissonian dynamical patterns characterised by broad inter-event time distributions, which was explained in various ways, namely (i) due to intrinsic correlations via decision making mechanisms or (ii) due to independent actions influenced by circadian patterns or (iii) due to some other underlying mechanisms like reinforcement or temporal correlations. In addition several combinations of these modelling directions were proposed together with phenomenological models, which did not, however, address the possible reasons behind the observed dynamics but only aimed at reproducing signals with similar temporal features. Below we address in details all of these modelling directions.

\subsection{Queuing models of bursty phenomena}
\label{sec:PriQueMod}

As we have observed in Chapter~\ref{chapter:meas}, the bursty temporal patterns in human dynamics can be characterised in terms of broad inter-event time or waiting time distributions, which in many cases have power-law tails with exponent $\alpha$ and $\alpha_w$, respectively. The values of $\alpha$ and $\alpha_w$ turn out to be very diverse, implying that there could be various underlying mechanisms behind the observed scaling behaviours. To understand the differences in the scaling behaviour, Barab\'asi proposed in his seminal paper the idea that consecutive rational actions of an individual are driven by the execution of prioritised tasks~\cite{Barabasi2005Origin}. Here the model considers an agent with a priority list of $l$ tasks, each of them assigned with a priority value $x_i$ drawn from a distribution, denoted by $\eta(x)$. The priority values allow the agent to rank the tasks and execute them in the rational order based on their priorities. Then the central quantity of this model is the waiting time $\tau_w$ for a task to be spent between its insertion to the queue and its execution.

\subsubsection{Cobham priority queuing model}

The priority queuing model was first introduced by Cobham~\cite{Cobham1954Priority}, where the priority list can contain an arbitrary number of tasks and the priority of each task is an integer drawn from some distribution. The tasks are set to arrive with the rate $\lambda$ following a Poisson dynamics with exponential arrival time distribution and they are executed with rate $\mu$ by always choosing the one with the highest priority. Since then, the waiting time distribution for some case has been obtained~\cite{Abate1997Asymptotics}, and also discussed in Ref.~\cite{Vazquez2006Modeling}, as follows:
\begin{equation}
    P(\tau_w) \sim \tau_w^{-3/2} \exp\left( -\frac{\tau_w}{\tau_c} \right) \hspace{.3in} \mbox{with} \hspace{.3in} \tau_c=\frac{1}{\mu(1-\sqrt{\rho})^2},
    \label{eq:Ptw}
\end{equation}
where $\rho=\lambda/\mu$ denotes the control parameter of the process. If $\rho<1$, the task list is typically short as tasks are executed right after their arrival. Then $P(\tau_w)$ is reduced to an exponential distribution as $\rho\rightarrow 0$. On the other hand, in the limit of $\rho=1$ the waiting time distribution appears as a power-law function with exponent $\alpha_w=3/2$ and an exponential cutoff. In this case most of the tasks are executed shortly after their insertion, but some low priority tasks may be stuck in the list introducing heterogeneity in the waiting time distribution. It has been shown that the queue length $l$ performs a one-dimensional random walk with a bound at $l=0$, implying a return time distribution $P(t_{r})\sim t_{r}^{-3/2}$ that gives the origin of the same exponent value for $P(\tau_w)$~\cite{Vazquez2006Modeling}. Finally, if $\rho>1$, the average queue length is increasing linearly as $\langle l(t) \rangle = (\lambda-\mu)t$, and thus a fraction of tasks $1-\rho^{-1}$ will remains in the list forever. Nevertheless, the waiting time distribution of the executed tasks still follows the form of Eq.~(\ref{eq:Ptw}). We note that the predicted exponent value $\alpha_w=3/2$ is found to fit well with the empirical observations of letter correspondence activities~\cite{Oliveira2005Human}, as were found typically with $\rho=1.1$.  Later, Grinstein and Linsker~\cite{Grinstein2006Biased} obtained the analytic solutions for the continuous distribution of priority: $\eta(x)=1$ for $x\in [0,1]$.

\subsubsection{Barab\'asi priority queuing model}
\label{sec:barabquemod}

Another version of the model, motivated by the finite capacity of immediate memory of humans~\cite{Miller1956Magical}, operates with a priority list of fixed size $l$. In addition it assumes that occasionally agents may decide to perform a task with a lower priority before doing all the high priority ones. This is realised by introducing a probability $p$ that in the current iteration the agent performs the highest priority task, otherwise, with probability $(1-p)$ it selects a task randomly from the task list. In the limit $p\rightarrow 1$ the agent tends to choose always the task with the highest priority. This model results in a process that has a power-law tailed waiting time distribution with exponent $\alpha_w=1$, and it matches with the empirical observations reported in Refs.~\cite{Barabasi2005Origin, Oliveira2005Human, Vazquez2006Modeling}. On the other hand, if $p\rightarrow 0$, the agent performs a fully random selection strategy, in which case the waiting times appear with an exponential distribution.

Note that in order to classify bursty systems based on these early modelling results V\'azquez~\emph{et al.}~\cite{Vazquez2006Modeling} suggested two universality classes with the above mentioned two different exponent values characterising power-law inter-event time distributions. Nevertheless, this picture turned out to be not completely consistent as further empirical evidences and modelling results of other bursty systems were found with various different exponent values as we have seen in Chapter \ref{chapter:emp} and will discuss below.

\begin{figure}[!t]
\centering
    \includegraphics[width=\columnwidth]{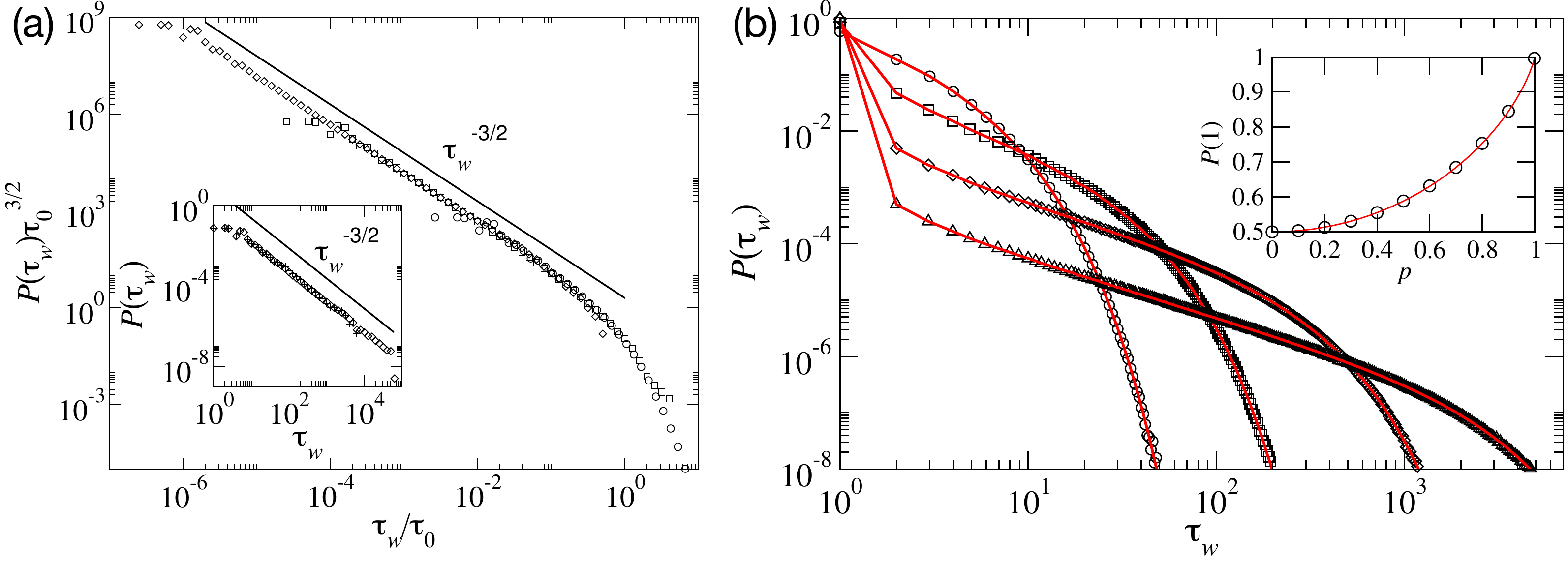}
\caption{Waiting time distributions of tasks in models with (a) arbitrary length and (b) fixed length ($l=2$) priority queues. In panel (a) the waiting time distribution with different values of $\rho$, i.e., $0.9$, $0.99$, and $0.999$, are scaled together and can be approximated with exponent $\alpha_w=3/2$ (solid line). In the inset $P(\tau_w)$ with $\rho=1.1$ is shown. In panel (b) the results of the model with fixed queue length of $l=2$ are shown with $p=0.9$ (squares), $0.99$ (diamonds), and $0.999$ (triangles). Solid lines depict the solution in Eq.~(\ref{eq:BPGMsol}). (\emph{Source:} Adapted from Ref.~\cite{Vazquez2006Modeling} under Copyright (2006) by the American Physical Society.)}
\label{fig:figurewtq}
\end{figure}

An exact stationary solution of this model for $l=2$ was provided by V\'azquez~\cite{Vazquez2005Exact} with the general form of the waiting time distribution derived as
\begin{equation} 
P(\tau_w) =
  \begin{cases}
    1-\frac{1-p^2}{4p} \mbox{ln} \frac{1+p}{1-p}, & \quad \tau_w=1\\
    \frac{1-p^2}{4p(\tau_w-1)}\left[ \left(\frac{1+p}{2}\right)^{\tau_w-1} -  \left(\frac{1-p}{2}\right)^{\tau_w-1}\right], & \quad \tau_w>1.
  \end{cases}
\end{equation}
which turned out to be independent of the priority distribution $\eta(x)$. In the limit of $p\rightarrow 0$, this solution reduces to the exponential form as
\begin{equation} 
    \lim_{p\rightarrow 0} P(\tau_w) = \left( \frac{1}{2}\right)^{\tau_w}, 
\end{equation} 
while in the limit of $p\rightarrow 1$ the solution reads
\begin{equation} 
\lim\limits_{p\rightarrow 1} P(\tau_w) =
  \begin{cases}
      1 + \mathcal{O}\left(\frac{1-p}{2} \ln (1-p)\right), & \quad \tau_w=1\\
      \mathcal{O}\left(\frac{1-p}{2} \right) \frac{1}{\tau_w-1}, & \quad 1<\tau_w\ll \tau_c,
\end{cases}
\label{eq:BPGMsol}
\end{equation} 
where $\tau_c=1/ \ln (2/(1+p))$ and the distribution is decaying with $\alpha_w=1$. Finally in the case when $0<p<1$, one finds a power-law distribution with exponential cutoff as follows:
\begin{equation}
P(\tau_w)\sim \frac{1-p^2}{4} \tau_w^{-1} \exp \left(-\frac{\tau_w}{\tau_c}\right)
\end{equation}
where the exponential cutoff is shifted towards larger $\tau_w$ values as $p\rightarrow 1$ (Fig.~\ref{fig:figurewtq}). Note that an exact non-stationary probabilistic description of this model was provided by Gabrielli and Caldarelli~\cite{Gabrielli2007Invasion} showing that for $0<p<1$ the system relaxes exponentially fast to the stationary solution. As $p\rightarrow 1$ the relaxation slows down and the system shows a non-stationary dynamics with a different exponent as described above.

The general stationary solution of the model with arbitrary but fixed length $l$ was provided by Anteneodo~\cite{Anteneodo2009Exact}. Here a master equation formalism was used to obtain the solution for the waiting time distribution as
\begin{equation}
    P(\tau_w)=\int_0^1 dR(x) r_{\tau_w}(x),
    \label{eq:Ptwexact}
\end{equation}
where $R(x)$ is the probability that a new inserted task has a priority smaller than $x$, and $r_{\tau_w}(x)$ assigns the probability that a new task, which was inserted at $t=t_0$ with priority $x$, will be executed at time $t=t_0+\tau_w$. This latter probability can be approximated as follows:
\begin{equation}
r_{\tau_w}(x) \simeq [1-r_1(x)][1-f(x)]^{\tau_w-2}f(x)
\end{equation}
where $f(x)$ is the average probability that a task is executed at time $t>t_0+1$. If $f(x)=(1-p)/l$ and $p=0$ the integral in Eq.~(\ref{eq:Ptwexact}) correctly results in the solution with exponential decay as $P(\tau_w)=(1-1/l)^{\tau_w-1}/l$, while if $p\rightarrow 1$ it leads to the asymptotic solution of a power-law with exponential cutoff $P(\tau_w)\sim \tau_w^{-1} \exp (-\tau_w/\tau_c)$ with $\tau_c=1/\ln [ l/(l-1+p)]\sim l/ (1-p)$. This solution also shows that the characteristic time $\tau_c$ of the exponential cutoff is shifted to larger values if $p\rightarrow 1$ or when $l$ is increased.

\paragraph{\textit{Waiting times vs. inter-event times.}} Empirical observations typically provide either the waiting time $\tau_w$ of a single task like a response to the received letter, or the inter-event times $\tau$ between similar tasks like mobile phone calls. The difference between these observables has been discussed in Section~\ref{sec:iet_rt_wt}. The models proposed above concern only the waiting times, while assuming that inter-event times are characterised by the same form of distributions. However, the relation between the two quantities and their distributions is not necessarily obvious. Following the arguments of Vazquez~\emph{et al.}~\cite{Vazquez2006Modeling}, while in the data we monitor the activities of an individual regarding a specific task only, in contrast the model simulates the execution of all tasks by an individual. Labelling tasks in the model process and reinserting them after execution would allow us to measure their inter-event time distribution, which would scale similarly with the waiting time distribution. A further argument says that inter-event times in communication depend on the activity patterns of a pair of interacting individuals. In case when they both prioritise their task lists, the effective inter-event time distribution would show the same scaling form as $P(\tau_w)$. Supporting these arguments Li~\emph{et al.} have reported an empirical study using a dataset of letter correspondence to confirm the matching exponents of the inter-event and waiting time distributions~\cite{Li2008Empirical}.

\paragraph{\textit{Criticism.}} After the seminal paper by Barab\'asi a few criticising comments were published~\cite{Stouffer2005Comment} that raised concerns about the data analysis, claiming that the observed inter-event time distribution in Ref.~\cite{Barabasi2005Origin} is better approximated by a log-normal distribution rather than a power-law  and that the model gives unrealistically high preference to execute newly arriving tasks while keeping low priority tasks extremely long in the queue. Further concerns were expressed in Ref.~\cite{Kentsis2006Mechanisms}, criticising that the proposed model completely disregards the semantic content of an individual correspondence and the social context in which this correspondence takes place. As a response~\cite{Oliveira2006Mechanisms}, it has been argued that it is impossible to detect semantic and social context of correspondence as the content of the messages are not available due to privacy reasons. However, arguably prioritising should yet play a role in human correspondence as not only the context but also the deadlines are driving individual decisions to perform a task. Moreover, one does not need to obtain any knowledge about the prioritisation mechanisms as the power-law waiting time distribution in the models is emergent regardless of the functional form of the priority distribution.

\paragraph{\textit{Extensions.}} After the initial observations, other studies~\cite{Li2008Empirical, Karsai2012Universal} have identified empirical systems with diverse exponent values for the inter-event time distributions. Motivated by these observations some other extended models have been proposed. Masuda \emph{et al.}~\cite{Masuda2009Priority} assumed that in a time iteration, task arrival does not  follow Poisson dynamics, similarly to Refs.~\cite{Cobham1954Priority,Barabasi2005Origin}, but $n$ tasks are added in each time step with a number sampled from a power-law distribution of $n^{-\gamma}$. This extension in turn led to the exponent values of the waiting time distribution depending on the exponent $\gamma$, the average number of $n$, and the execution probability $\mu$. Further extensions addressing interacting priority queues of links~\cite{Oliveira2009Impact} and coupled as a network~\cite{Min2009Waiting} were proposed providing alternatives to explain diverse exponent values. Their discussion will be the subject of Section \ref{sec:linkbrstmodel}. Finally, Gon\c{c}alves and Ramasco~\cite{Goncalves2008Human} suggested to extend the model such that multiple number of tasks are executed in each time steps. They showed that the execution of three tasks leads to an emerging exponent $\alpha=1.25$, which fits well with their observations for online browsing dynamics. Note that other extensions of the queuing model were proposed by Mryglod \textit{et al.}~\cite{Mryglod2012Editorial} and Jo~\emph{et al.}~\cite{Jo2012Timevarying}, where time-varying priority was considered to model heterogeneous dynamics of editorial review processes with or without peer-review processes. Finally, Cajueiro and Maldonado~\cite{Cajueiro2008Role} considered the cost of keeping a non-processed collection of tasks by introducing a discount factor, and they identified various protocols for executing tasks, depending on the discount factor, for minimising the cost function.

\subsubsection{Position based priority lists}

A somewhat different model was proposed by Vajna \textit{et al.}~\cite{Vajna2013Modelling} who defined a priority list model assuming that the priorities of tasks depend on their position in the list. More precisely they took a task list of size $l$ with ordered positions starting from $i=1$ to $l$. The list is filled with tasks of different types of activities. In each time step a task is chosen based on its position in the list with a probability $w_i$ that is decreasing as a function of $i$. Once a task is chosen, it jumps to the front of the list to trigger the corresponding activity, and it pushes the tasks that preceded it to the right. Once a task is in the front of the list it has the largest probability to be chosen again in the next iterations. In this way, the heterogeneous inter-event times between the consecutive executions of the same task are generated. Note that this model proposes the observations of the inter-event times rather than the waiting times.

The authors have shown that this model is capable of inducing power-law decaying inter-event time distributions with a tunable exponent between $\alpha\in [1,2]$ by using various $w_i$ distributions. Furthermore, they generalised their results by using a power-law decaying and an exponentially decaying priority distribution for $w_i$, as well as discussed the case of stretched exponential. In case of a finite list, they found that $P(\tau)$ decays as a power-law with an exponential cutoff, where the cutoff is the consequence of reaching the end of the list but it disappears in the $l \rightarrow \infty$ limit.
				
\subsection{Memory driven models of bursty phenomena}

Another modelling paradigm of bursty activity patterns concerns non-Markovian correlations between consecutive actions of an individual. It assumes that memory functions or reinforcement processes lay behind bursty signals in human behaviour. In the following we are going to walk through modelling examples contributing to this direction.

\subsubsection{Processes with simple memory functions}

One of the first models of this kind was proposed by Vazquez \textit{et al.}~\cite{Vazquez2007Impact} who aimed at modelling email and letter correspondence behaviour by assuming that the subsequent actions of an agent is influenced by its previous mean activity rate. Their model builds on the probability $\lambda(t)dt$ that an agent performs an action within the time window $[dt,t+dt]$ formulated as
\begin{equation}
\lambda(t)=\frac{a}{t}\int_0^t \lambda(t') dt'
\label{eq:VazqLamd}	
\end{equation}
where the parameter $a>1$ determines the degree and type of reaction to the past perception. If $a=1$, one obtains a stationary process with $\lambda(t)=\lambda(0)$, while if $a\neq 1$, the process is non-stationary either with acceleration ($a>1$) or with reduction ($a<1$). At the starting time $t=0$ one assumes that an agent does not consider what happened before and performs the actions for a period $T$. The general solution of Eq.~(\ref{eq:VazqLamd}) shows $\lambda(t)=\lambda_0 a (t/T)^{a-1}$, where $\lambda_0$ is the mean number of actions within $T$. $\lambda(t)$ is approximately a constant for short time intervals of $T$ and the dynamics follows a Poisson process with an exponential inter-event time distribution. However, if $a>1$, i.e., the system is in the accelerating regime, the inter-event time distribution exhibits a power-law form, $P(\tau)\sim (\tau /\tau_0)^{-\alpha}$, where $\tau_0=1/(a\lambda_0)$ and $\alpha=2+1/(a-1)$ if $\tau_0 \ll \tau < T$. On the other hand, in the reduction regime, if $1/2<a<1$ the $P(\tau)$ does not show a power-law scaling, while for $0<a<1/2$ it appears again to follow a power-law with the exponent $\alpha=1-a/(1-a)$ if $\tau \ll \tau_0$.

A somewhat similar model was introduced by Han \textit{et al.}~\cite{Han2008Modeling} to model actions like web-browsing or video game playing, which are arguably driven by the adaptive interest. In their paper they introduced two thresholds, $T_1$ and $T_2$ (where $T_1\ll T_2$), to model the increased and the depressed activity rates by focusing on the probability $r(t)$ that the given action will occur at time $t$. They measured the inter-event times between consecutive occurrences of actions. While employing discrete time steps if the $(i+1)$th event appeared at time $t$ the value of $r$ is updated as $r(t+1)=a(t)r(t)$, where $a(t)$ determines the actual activity rate. For convenience they choose the $\tau_i=t_{i+1}-t_i$ be the inter-event time between the $i$th and $(i+1)$th events. Then if $\tau_i \leq T_1$, $a(t)=a_0$ and the process evolves with depressed rate. On the other hand, if $\tau_i \geq T_2$, $a(t)=a_0^{-1}$ and the process evolved with an increased rate. Finally, if $T_1 < \tau_i < T_2$ or there was no event within time $t$ then $a(t)=a(t-1)$. This model was found to induce bursty activity dynamics, characterised by a power-law inter-event time distribution with exponent $\alpha=1$.

\subsubsection{Self-exciting point processes}
\label{sec:SelfExcPP}

Another family of memory driven models for bursty processes are based on self-exciting stochastic processes of Hawkes type~\cite{Masuda2013SelfExciting}. Such models are able to reproduce heterogeneously distributed inter-event times and short-term temporal correlations, commonly observed in case of human dynamics. The general definition of Hawkes processes concerns the activity rate $\lambda(t)$ defined as follows:
\begin{equation}
    \lambda(t)=\lambda_0+\sum_{i,t_i\leq t}\phi(t-t_i),
\end{equation}
where $\nu$ sets the ground activity level, while $\phi(t)$ is called the memory kernel, i.e., the additional rate incurred by the past events. For more comprehensive account of the Hawkes process, see the  review~\cite{Mehrdad2015Hawkes}. There are several definitions of the memory kernel function that have been considered to describe human bursty phenomena. For example Masuda \emph{et al.}~\cite{Masuda2013SelfExciting} assumed an exponential form $\phi(t)=\alpha e^{-\beta t}$ to model event clusters initiated by single events appearing with rate $\lambda_0$. Such event trains appeared with size $c=1/(1-(\alpha/\beta))$ on average and induced a stationary rate of events $\overline{\lambda}=c\lambda_0$ with the condition that $\alpha<\beta$. This way their model process was fully determined by the parameters $\lambda_0$, $\alpha$, and $\beta$, which could be estimated by the maximum likelihood methods from the empirical data. In their study they used two face-to-face conversation datasets recorded independently in two Japanese companies, and after estimating the parameters of the interaction sequences of active individuals they found surprisingly good match between the statistical characteristics of the empirical and modelled activity signals.

In another work Jo \textit{et al.}~\cite{Jo2015Correlated} applied a power-law memory kernel of the form, $\phi(t)=1/t$, to define a memory function for $t>t_w$ as follows:
\begin{equation}
m(t)=\sum_{i=1}^w \frac{1}{t-t_i},
\end{equation}
where the memory is kept only up to the $w$th latest events in order to take into account the finite capacity of memory. This is called sequential memory loss mechanism. In the model, the larger $m(t)$ is, it induces the higher probability of new events at time $t$. As a result, the heavy-tailed inter-event time distributions emerges, while long-range correlations between inter-event times are limited by the control parameter $w$. For more realistic consideration, instead of having $w$ fixed, $w$ can be a variable such that for each newly occurred event, the value of $w(t)$ is reset to $1$ with probability of $q[w(t)]=1-\left[w(t)/(w(t)+1)\right]^\nu$, otherwise $w(t)$ is set to be increased by $1$. Here the larger $w(t)$ implies the longer time for resetting the memory, hence more consecutive events. This is called preferential memory loss mechanism. Here for the intermediate range of $\nu$, they found more realistic features in terms of temporal heterogeneities and higher order correlations, as exemplified by a power-law bursty train size distribution $P(E)$.

Note that a power-law memory kernel was also used by Crane and Sornette~\cite{Crane2008Robust} to model a somewhat different phenomenon, namely the cascade of social influence diffusion in social networks. Such processes lead potentially to bursty cascades of information adoption events at the population level with long lasting relaxation times back to normal adoption rates.

\subsubsection{Reinforcement point processes}

\begin{figure}[!t]
\centering
  \includegraphics[width=.8\columnwidth]{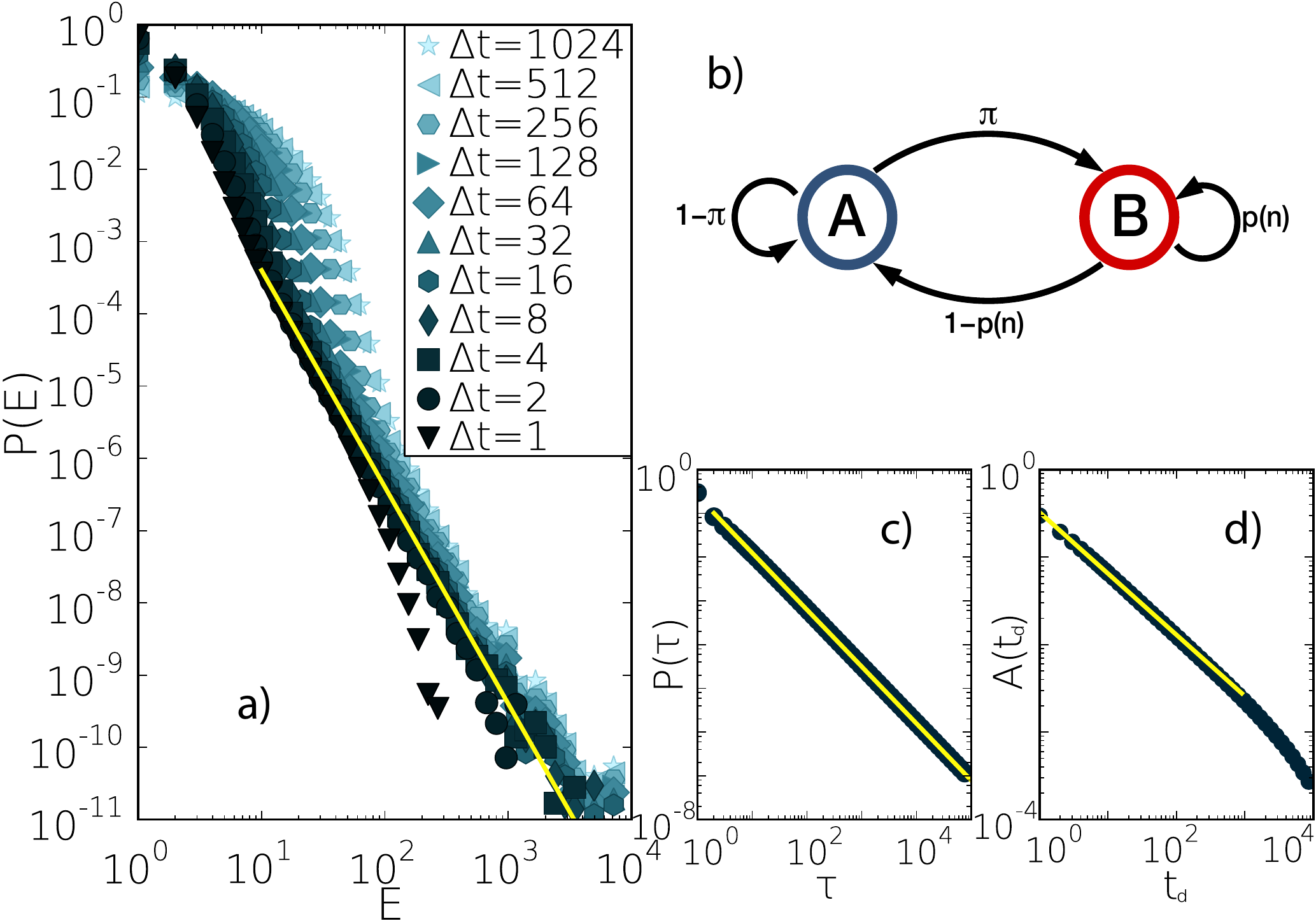}
\caption{The schematic definition and numerical results of the model in Ref.~\cite{Karsai2012Correlated}. (a) $P(E)$ distributions of the synthetic sequence after logarithmic binning with window sizes $\Delta t=1 \cdots 1024$ and fitted power-law exponent $\beta=3.0$. (b) Transition probabilities of the reinforcement model with memory. (c) Inter-event time distribution of the simulated process with a maximum inter-event time $\tau^{max}=10^6$ and emerging exponent value $\alpha=1.3$.  (d) The average autocorrelation function with the maximum lag of $t_d^{max}=10^4$ and emerging characteristic exponent $\gamma=0.7$. Results are averages over $1000$ independent realisations with parameters $\mu_A=0.3$, $\mu_B=5.0$, $\nu=2.0$, $\pi=0.1$, and $T=10^9$.}
\label{fig:Karsai2012Correlated5}
\end{figure}

Reinforcement mechanisms provide another way to consider memory effects in dynamical processes, and they propose a possible explanation for the emerging bursty patterns. Based on this idea Karsai \emph{et al.}~\cite{Karsai2012Correlated} introduced a model to capture not only the heterogeneous individual communication dynamics but also the correlated bursty event trains commonly observed in real systems. In their model they defined a two-state dynamics, by considering an agent who can be either in a normal state $A$, for which events are executed for longer time, or in an excited state $B$, where actions appear with a higher rate. The timings of the consecutive events were determined by a reinforcement process with the assumption that the longer the system waits for an event, the larger the probability that it will keep waiting. Note that similar assumption was taken in models of collective bursty dynamics~\cite{Stehle2010Dynamical, Zhao2011Social}, which will be discussed in Section \ref{sec:brstmodelgroups}. In the model, the inter-event times are induced by a reinforcement function of the form
\begin{equation}
    f_{A,B}(\tau)=\left(\frac{\tau}{\tau+1}\right)^{\mu_{A,B}}
    \label{eq:KarsaiMemory}
\end{equation}
that gives the probability to wait one time step longer in order to execute the next event after the system has waited already time $\tau$ since the last event. Here the exponents $\mu_A$ and $\mu_B$ control the reinforcement dynamics in states $A$ and $B$, respectively. If $\mu_A \ll \mu_B$ the characteristic inter-event times in states $A$ and $B$ become fairly different leading to the emergence of temporal inhomogeneities in the dynamics. In addition the actual state of the system is determined by transition probabilities as demonstrated in Fig.~\ref{fig:Karsai2012Correlated5}(b). To be more specific, the model is defined as follows: first the system performs an event in a randomly chosen initial state. If the last event was in the normal state $A$, it waits for a time induced by $f_A(\tau)$, after which it switches to an excited state $B$ with probability $\pi$ and performs an event there, or with probability $1-\pi$ it stays in the normal state $A$ and executes a new normal event. In the excited state the inter-event time for the actual event comes from $f_B(\tau)$ and the probability to perform the next event in the excited state is given by $p(n)=(n/(n+1))^{\nu}$ as determined by the number of excited events $n$ since the last event in state $A$ and by the reinforcement exponent $\nu$. Then the model induces power-law inter-event time distribution with exponent $\alpha=\mu_A+1$ as shown in Fig.~\ref{fig:Karsai2012Correlated5}(c) and also generates correlated bursty trains whose size distribution is written as $P(E)\sim E^{-\beta}$ with $\beta=\nu+1$, as can be seen in Fig.~\ref{fig:Karsai2012Correlated5}(a). Note that a similar model without memory was also introduced in Ref.~\cite{Kleinberg2002Bursty}, which will be discussed in Section \ref{sec:PoissInfAut}.

Somewhat similar model was proposed by Wang \textit{et al.} \cite{Wang2014Modeling} to model the blog-posting behaviour of individuals. Their objective was to introduce short term correlations to capture the heterogeneous distribution of inter-event times and the power-law decay of the memory coefficient as defined in Eq.~(\ref{eq:Mgen}). Their model assumes that in each time step an agent can select one from $n$ possible tasks in two possible ways, i.e., randomly with probability $r$ or with probability $1-r$ it selects a recently performed task again with probability $t_i/m$. Here $t_i$ assigns the number of times a given task $i$ was performed in the last $m$ time steps, which in turn defines the length of the memory. By varying $m$ they observe that for smaller values of $m$ the inter-event time distribution scales as a power-law with exponent $\alpha>2$, while for larger memory lengths the exponent increases and the distribution relaxes into an exponential form. In addition they argue that their model successfully reproduces the short term power-law decay of the memory function and deviates from the empirical observations only in the tail region.

\subsection{Poisson models of bursty phenomena}
	
\subsubsection{Infinite automatons}
\label{sec:PoissInfAut}

One of the early models of bursty phenomena in human dynamics was proposed by Kleinberg~\cite{Kleinberg2002Bursty}, whose aim was to understand heterogeneous and hierarchical patterns of topic appearances in document streams. His subsequent aim was to provide a better organisation principle for large document archives, such as emails and scientific publications. Based on the analogy of email correspondence he suggested a model using an infinite-state automaton, with states determining the actual rate of message arrival and with inter-state transitions determined by the upward difference between states.

More precisely, he takes an automaton $\mathcal{A}$, which can be in states $q_i$ and performs $n+1$ events over a period of $T$. First of all he assumes that events in the state $q_i$ occur with inter-arrival times sampled from a ``memoryless'' exponential distribution $f_i(\tau)=a_i e^{-a_i \tau}$ with $a_i>0$. In other words events in a given state behave as a Poisson process. Each state $q_i$ is characterised by the arrival rate of messages $a_i$ such that $a_i=(n/T)s^i$ for states $i=0,1,\cdots$, where $s>1$ is a scaling parameter. In addition, the automaton $\mathcal{A}$ can transfer from state $q_i$ to $q_j$ with cost $\kappa(i,j)$, where the cost is proportional to $(j-i)\ln n$ for $j>i$, or simply zero for $j<i$. Subsequently the ultimate aim here was to find a sequence of states $\bm{q}=(q_{i_1}, q_{i_2},...,q_{i_n})$  for a given sequence of inter-arrival times $\bm{\tau}=(\tau_1, \tau_2,...,\tau_n)$, such that the overall cost function defined as
\begin{equation}
    c(\bm{q}|\bm{\tau})=\left( \sum_{t=0}^{n-1} \kappa(i_t,i_{t+1}) \right) + \left( \sum_{t=1}^{n} -\ln f_{i_t}(\tau_t) \right)
\end{equation}
is minimal. A recursive solution of this model was provided in Ref.~\cite{Kleinberg2002Bursty} and was used to identify hierarchical structures in terms of state transitions (and thus in inter-arrival times) in document streams. Note that the overall framework developed in this paper can be viewed as drawing an analogy with models from queuing theory for bursty network traffic~\cite{Kelly1996Notes}, as well as the formalism of hidden Markov models~\cite{Rabiner1989Tutorial}. The principal aim 
of this model was not to reproduce heterogeneous inter-arrival sequences, but more to provide a possible reason behind their emergence and to give applicable solutions in order to organise better streaming of documents.

\subsubsection{Heterogeneous Poisson model}

In reflection to the model by Barab\'asi~\cite{Barabasi2005Origin} a simple explanation was proposed by Hidalgo~\cite{Hidalgo2006Conditions} to describe the emergence of a power-law inter-event time distribution by using Poissonian agents that change the rates at which they perform an event in a random or deterministic fashion. To be more precise, the event rate of an individual is denoted by $\lambda$ and its distribution at the population level by $f(\lambda)$. It has been shown that if $f(\lambda)$ is heterogeneous, the asymptotic behaviour of the emergent 
inter-event time distribution reads as
\begin{equation}
    P(\tau)=\frac{f(1/\tau)}{\tau^2}.
\end{equation}
Assuming a uniform distribution of the form $f(\lambda)\sim U\left[0,L\right]$, the inter-event time distribution appears as $P(\tau)\propto \tau^{-\alpha}$ with $\alpha=2$. If $f(\lambda)\sim \lambda^{\nu}$, then one gets $P(\tau)\propto \tau^{-(\nu+2)}$, which implies $\alpha=\nu+2$. Similar scaling would also hold if we assume similar distributions of activity rates at the individual level~\cite{Hidalgo2006Conditions, Chatterjee2003Login}, or in case of periodically varying activity rates of individuals. Although this model was meant to describe natural phenomena it provides a simple explanation for bursty processes in cases when human individuals are assumed to be Poissonian agents.

\subsubsection{Bursty model with Poissonian cascades}

\begin{figure}[!t]
\centering
  \includegraphics[width=\columnwidth]{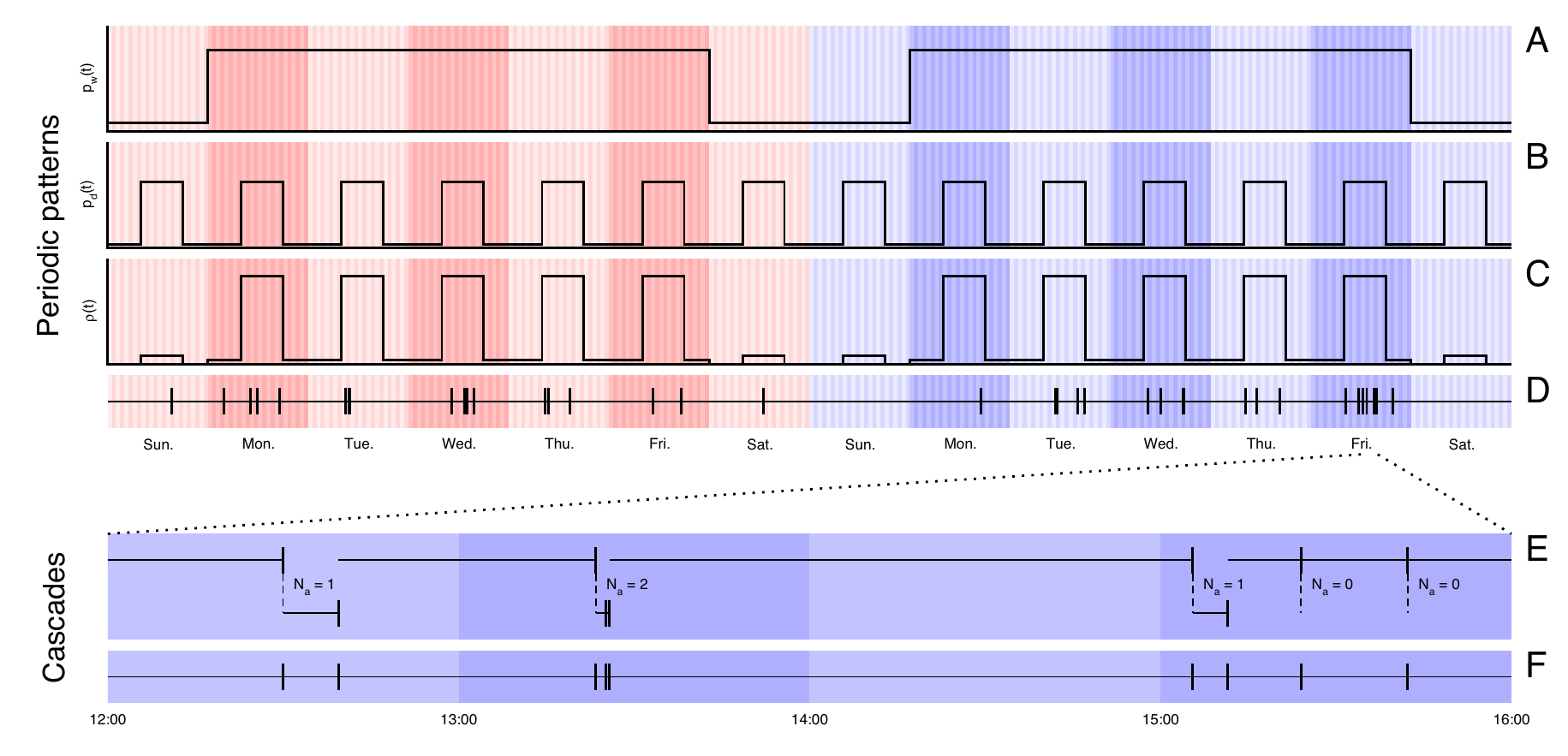}
  \caption{An example of a periodic and cascading stochastic process. (A) Expected probability for starting an active interval during a particular day of the week $p_w(t)$. (B) Expected probability for starting an active interval during a particular time of the day $p_d(t)$. (C) The resulting activity rate $\rho(t)$ for the non-homogeneous Poisson process. Here the form $\rho(t)=N_wp_w(t)p_d(t)$ is assumed, where the proportionality constant $N_w$ is the average number of active intervals per week. (D) A time series of events generated by a non-homogeneous Poisson process. Each event in this time series initiates a cascade of additional events, called an active interval. (E) Schematic illustration of cascading activity with $N_a$ additional emails sent according to a homogeneous Poisson process with rate $\rho_a$. (F) Observed time series. (\emph{Source:} Adapted from Ref.~\cite{Malmgren2008Poissonian} under Copyright (2008) National Academy of Sciences, U.S.A.)}
\label{fig:Malmgren2008A_1}
\end{figure}

An alternative and descriptive modelling framework of bursty phenomena in human interactions was proposed by Malmgren \emph{et al.}~\cite{Malmgren2008Poissonian,Malmgren2009Universality}. In this approach it is argued that ``human behavior is primarily driven by external factors such as circadian and weekly cycles, which introduces a set of distinct characteristic time scales, thereby giving rise to heavy tails''~\cite{Malmgren2008Poissonian}, instead of rational decision making and correlated activity patterns, 
proposed by Barab\'asi and others~\cite{Barabasi2005Origin, Oliveira2005Human, Vazquez2006Modeling}. They proposed a model, that captures individual email correspondence and builds on the intuition that our activities are strongly determined by circadian and weekly patterns, while they are grouped in cascades of actions in short active periods. For an illustration, see Fig.~\ref{fig:Malmgren2008A_1}. In this model the dynamics is defined as alternating non-homogeneous and homogeneous Poisson processes, which in turn gives rise to heterogeneous temporal behaviour with good correspondence with empirical observations.

More precisely, the model accounts for periodic activity patterns by using a non-homogeneous Poisson process with a time-dependent periodic rate of events $\rho(t)=\rho(t+W)$, with period $W$. This rate function captures the convolution of daily and weekly activity distributions of active interval initiation, $p_d(t)$ and $p_w(t)$ as follows
\begin{equation}
    \rho(t)=N_w p_d(t) p_w(t),
    \label{eq:NHPoissRate}
\end{equation}
where $N_w$ stands for the proportionality constant being the average number of active intervals within one period $W$ (here a week). Each event generated by $\rho(t)$ initiates a secondary process that is a cascade of activity or active period, modelled by a homogeneous Poisson process with the rate $\rho_a$. During an active period, $N_a$ additional events occur, after which the activity of an individual is again governed by the primary process defined in Eq.~(\ref{eq:NHPoissRate}). Here the number of events $N_a$ is drawn from some distribution $p(N_a)$. The inter-event time between events within an active period is determined by $\rho_a$, while the times between active periods are induced by $\rho(t)$. In this way the process is fully determined by the parameters $N_w$, $\rho_d(t)$, $\rho_w(t)$, $\rho_a$, and $p(N_a)$, which can be inferred from data by using simulated annealing. Fitting this model on individual activity sequences gives a very close match between the modelled and empirical inter-event time distributions~\cite{Malmgren2008Poissonian}, as shown in Fig.~\ref{fig:Malmgren2008A_2}. This suggests that Poissonian bursts provide an alternative description for individual bursty activity patterns. In addition, in a complementary work~\cite{Anteneodo2010Poissonian} it is argued that email correspondence patterns present no detectable correlations in terms of the ordering of events as compared to randomly reordered time series. This is in contrast to what has been suggested in Ref.~\cite{Barabasi2005Origin}. They also concluded that the proposed Poisson model is sufficient in describing the observed phenomena with spurious correlations. Finally the same authors studied the estimation of the functional form of the inter-event time and waiting time distributions in email activity logs~\cite{Stouffer2006Lognormal, Malmgren2009Universality}, with the conclusion that they can be better approximated by log-normal distributions or the superposition of two log-normal distributions rather than with a truncated power-law function as suggested in Ref.~\cite{Barabasi2005Origin}. They argue that the generative queuing model proposed by Barab\'asi may not describe the observed log-normal waiting-time distributions as it predicts power-law distributed waiting times.

\begin{figure}[!t]
\centering
  \includegraphics[width=\columnwidth]{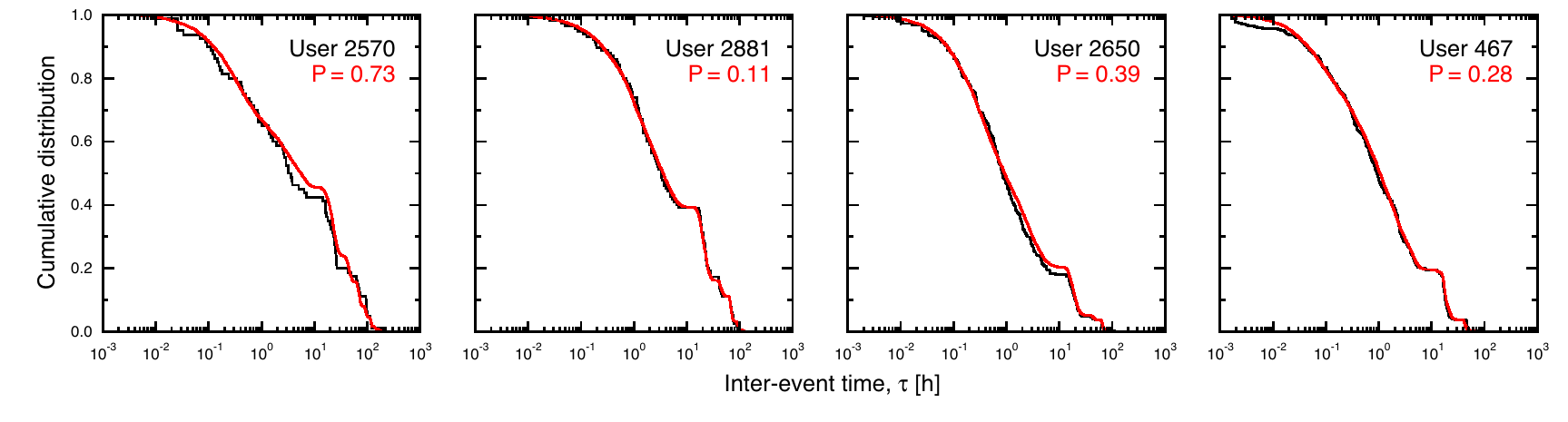}
\caption{Comparison of the predictions of the cascading non-homogeneous Poisson process (red line) with the empirical cumulative distribution of inter-event times $P(\tau)$ of email correspondence (black line) for selected users. (\emph{Source:} Adapted from Ref.~\cite{Malmgren2008Poissonian} under Copyright (2008) National Academy of Sciences, U.S.A.)}
\label{fig:Malmgren2008A_2}
\end{figure}

\paragraph{\emph{Criticism.}} Some criticism has been expressed about this modelling approach. Firstly, it has been argued that even though this model gives a very close approximation with the empirical data it is only descriptive. It does not provide any generative explanation about the emergence of heterogeneous human dynamics but (a) provides a quite precise approximation by fitting the model process using a large set of parameters; (b) assuming circadian fluctuations to induces heterogeneity in human activity patterns. However, it has been shown (see Section \ref{subsect:cyclic}) that even after removing effects of such periodic fluctuations the signal remains bursty, indicating that circadian patterns cannot be the generative reason behind this phenomena~\cite{Jo2012Circadian, Zhou2012Relative}; (c) assuming only two states might be an over-simplification of human behaviour, while assuming two or more types of active states give considerable better approximation of bursty behaviour at the individual level~\cite{Ross2015Understanding}; and (d) although the model assumes that the action dynamics of an individual consists of independent events, temporal correlations have been detected in such signals~\cite{Karsai2012Universal, Gandica2016Origin}.

\paragraph{\emph{Extensions.}} Recently two extensions of this model have been proposed. In one case, Jiang \emph{et al.}~\cite{Jiang2016Twostate} modelled the communication activity of an individual by a two-states Markov-chain Poisson process where an individual can be either in the normal state ($s_n$) or in the bursty state ($s_b$). The interaction dynamics of an individual is determined by two Poisson processes $\mathcal{P}_n$ and $\mathcal{P}_b$ with characteristic intensities $\lambda_n$ and $\lambda_b$ such that $\lambda_n < \lambda_b$. Assuming a normal initial state $s_n$, the dynamics of an individual is determined by $\mathcal{P}_n$ where an interaction is initiated with the probability $\lambda_n t_p$, where $t_p$ denotes a characteristic time. The state of the next call is determined by a conditional probability $p(j|i)=p(i,j)/p(i)$ to switch to the bursty state, (in case with $i=s_n$ and $j=s_b$), or remain in the normal state (in case with $i=s_n$ and $j=s_n$). In the bursty state once the next call occurs with probability $\lambda_b t_p$, the next state is determined by the conditional probability to switch to $s_n$ (where $i=s_b$ and $j=s_n$), or to remain in the bursty state (where $i=s_b$ and $j=s_b$). The parameters $t_p$, $\lambda_b$, and $\lambda_n$ can be estimated directly from empirical data just as in Ref.~\cite{Malmgren2008Poissonian}. Just like the model of Malmgren \textit{et al.}, this model can give a close approximation to the interaction dynamics and inter-event time distribution of an individual.

Another extension was proposed by Ross and Jones~\cite{Ross2015Understanding} who suggested a model based on observations on Twitter activity logs. In this model an individual can be in one inactive state $s_0$, or in two active states constituting a more bursty state $s_1$, corresponding to conversation type of communication, and a less bursty state $s_2$, corresponding to broadcasting type of communication. In the inactive state $s_0$, inter-event times are determined by an inhomogeneous Poisson process just like in Ref.~\cite{Malmgren2008Poissonian} with a time dependent intensity function $\lambda_0(t)$. After performing an event in $s_0$ the node may switch to one of the active states $s_1$ and $s_2$ with probabilities $p_1$ and $p_2$, respectively. The number of events in active states is sampled from a geometric distribution with parameters different for $s_1$ and $s_2$, and inter-event times are sampled from an arbitrary distribution $g(\tau|s_i)$. The authors considered $g(\tau|s_i)$ being an exponential, log-normal, or Weibull distribution, with parameters again depending on the actual active state. After fitting this multi-parameter model with the empirical data, they found closer match between the real and modelled inter-event time distributions as in the case of the two-state model~\cite{Malmgren2008Poissonian}.

\subsubsection{Non-homogeneous Poisson process with decreasing interest}

There is yet another model proposed by Guo \textit{et al.} that studies inter-event time distributions of dynamical systems, where the interest of people in doing something is dependent on time~\cite{Guo2011Weblog}. Their hypothesis is introduced by an event rate
\begin{equation}
    \lambda(t)=a+\frac{\alpha}{bt+1},
\end{equation}
which is decreasing until it reaches a stationary value $a$ of personal interest. In addition they show that in case of a non-homogeneous Poisson process with event rate $\lambda(t)$ and independent stationary increments, the cumulative distribution of the inter-event times appears as follows:
\begin{equation}
    1-F(\tau) \sim b^{-\frac{\alpha}{b}}\left( \tau+\frac{1}{b}\right)^{-\frac{\alpha}{b}}e^{-a \tau}  \hspace{.3in} \text{as} \hspace{.3in} \tau\rightarrow \infty
\end{equation}
for positive constants $a$, $b$, and $\alpha$. This is a mixed distribution with exponential and power-law features that approximates the Gamma distribution. To demonstrate their analytic findings they come up with good fits between their model and the inter-event time distribution of blog posts of four users of a popular Chinese blog space.

\subsection{Other type of models}
\label{sec:otherindivmods}

\paragraph{\emph{Models of rational bursty consumers.}}

In their study Maillart \emph{et al.}~\cite{Maillart2011Quantification} mapped the priority queuing process onto the economic theory of consumption. In economic theory, the consumer is assumed to maximise the total utility from consuming some units of wealth under the constraint of the limited wealth, namely under budget constraint. Similarly, in priority queuing processes the agent tries to maximise the total utility from consuming time $\tau_w$ for solving a task under the constraint of the limited total time, i.e., the time budget, which is expressed by 
\begin{equation}
    \sum_{i=1}^{N(T)}\tau_{w,i}\leq T,
\end{equation}
where $N(T)$ is the number of tasks in a given time period of $T$. Based on this mapping, the authors use several strategies of executing tasks to find realistic waiting time distribution with power-law tails.

In their paper Jo \emph{et al.}~\cite{Jo2012Optimized} introduced an alternative economics-inspired model, where an agent in an uncertain situation tries to reduce the uncertainty by communicating with information providers, while the agent has to wait for responses. Here the waiting time can be considered as cost. The authors showed that the optimal choice for the waiting time under uncertainty gives rise to the bursty dynamics, characterised by a power-law distribution of the optimal waiting time. More precisely, the risk-averse utility function is assumed to be $u(x_t)=-\exp(-x_t)+a$ with $a\geq 0$, where the uncertainty of the state $x_t$ is described by a normal distribution with the zero mean and variance of $\sigma^2/t^\eta$. Here the parameter $\eta$ controls the speed of decreasing uncertainty. That is, the uncertainty decreases with time as the agent waits for the information, while the cost of the spent waiting time is modelled as $c(t)\propto \sigma^{-2/\nu}t$, where the parameter $\nu$ controls the cost per unit time. Then the expected utility is obtained as $E[u(x_t)]-c(t)$, which is optimised to obtain the optimal waiting time as follows:
\begin{equation}
    \tau_w(\sigma)=C_1\sigma^{2/\eta}W(C_2\sigma^{2(\nu-\eta)/[\nu(\eta+1)]})^{-1/\eta}
\end{equation}
with coefficients $C_1$ and $C_2$. Here $W$ denotes the Lambert function. Then using the distribution of uncertainty $P(\sigma)=e^{-\sigma}$, one obtains the optimal waiting time distribution $P(\tau_w)$ with power-law exponent, e.g., $\alpha_w=1-\eta/2$ if $\nu=\eta$.

\paragraph{\emph{Independent models.}}

The simplest way to model bursty activity sequences is by sampling inter-event times from a given distribution. Note that although this method provides heterogeneous activity patterns, it does not provide neither any understanding about the roots of bursty phenomena nor it induces correlations between consecutive events, which in turn remain independent. This method has been used in Ref.~\cite{Horvath2014Spreading, Jo2014Analytically} to study the effects of node and link burstiness on the speed of information spreading in temporal networks, or in Ref.~\cite{Karsai2012Universal} to highlight the spurious behaviour of the autocorrelation function in case of heterogeneous but independent activity signals and to detect the presence of temporal correlations common in empirical cases.

\paragraph{\emph{Voter model.}}

An alternative definition of the voter model has been suggested by Fern\'andez-Gracia \emph{et al.}~\cite{FernandezGracia2013Timing} where temporal heterogeneities arise as a consequence of new update rules. In the model each node gets updated with a probability that depends on the time since the last event of the node took place. Here, an event can be an update attempt (exogenous update) or a change of state (endogenous update). In their paper they find that both update rules can give rise to power-law inter-event time distributions. If the update probability of a node is given as $p(\tau)=b/\tau$, where $\tau$ is the time since the last update, then the inter-event time distribution emerges in the form $P(\tau)\sim \tau^{-b}$. In addition it is shown that for the exogenous update rule and the standard update rules the voter model does not reach consensus in the infinite size limit, while for the endogenous update there exists a coarsening process driving the system toward consensus configurations.
	
\paragraph{\emph{Rank shift model.}}

This model, proposed in Ref.~\cite{Ratkiewicz2010Characterizing}, addresses popularity dynamics of online contents, which emerge with heterogeneous patterns in terms of the number of citation events. In this model each task is in a list and assigned with a popularity, implemented as a citation probability that decays as a power-law as the actual position of the given task in the list. In addition the model accounts for exogenous effects that potentially changes the popularity of a task suddenly and drastically. The simplest way to implement this mechanism is by introducing in the ranking model a re-ranking probability. In this case at each iteration every item is moved to a new position towards the front of the list, which is chosen randomly with equal probability between $1$ (the top position) and the task's current rank $j$. As a consequence of these two mechanisms the model induces power-law distributed number of citations of tasks with exponent close to empirically observed ones in Wikipedia citations and online crawling data. It should be noted that this model does not propose explanation for emergent bursty patterns in terms of time, and also that it is somewhat similar to the one in Ref.~\cite{Vajna2013Modelling} with an important difference that its definition is not based on a priority queue.

\paragraph{\emph{Random reference models.}}

A set of models has been recently proposed that are not generative but yet address temporal bursty behaviour. These models~\cite{Karsai2011Small,Miritello2011Dynamical,Holme2012Temporal} apply various random shuffling techniques on real event sequences to obtain statistical reference models where selected temporal or structural correlations are vanished from the system. This modelling techniques are commonly used to remove bursty activity patterns from empirical signals to study their effects on data-driven models of dynamical processes. These models, in a way, ``Poissonise'' a temporal event sequence either by shuffling times between events or by assigning a random time to each event selected uniformly from a given period $T$. The use and advantage of these models will be discussed in details in Chapter~\ref{chapter:processes}.

\paragraph{\emph{Self-organised critical systems.}}

Finally, there is a very interesting proposition by Tang \textit{et al.}~\cite{Tang2010Stretched} in order to model retrospectively the 
bursty dynamics of the emergence of wars in ancient China. They argue that the dynamics of wars is driven mostly by short term correlations between the last and next following events and they can be related to the Bak-Sneppen evolutionary model with self-organised criticality.

\section{Models of link activity}
\label{sec:linkbrstmodel}

\subsection{Interacting priority queues}

The bursty models we have discussed so far in Section \ref{sec:indivmodels} attempt to model the action dynamics of single individuals, while neglecting the fact that the tasks are commonly carried out in human-to-human interactions. Examples can be found in any type of communication or communication driven activities, where as a consequence bursty patterns appear between connected peers and thus they are associated more to links~\cite{Karsai2012Correlated} rather than to individual dynamics. This problem was addressed by Oliveira and Vazquez~\cite{Oliveira2009Impact} who introduced a model based on the definition by Barab\'asi~\cite{Barabasi2005Origin} but considering two priority queues $A$ and $B$ with fixed sizes $l_A$ and $l_B$, respectively. They assumed two types of tasks to be present in each queue, a single interacting task $I$ and $l_j-1$ non-interacting tasks $O$ with $j=A,B$. Each task is assigned with a random priority $x$ drawn from the uniform distribution in $[0,1]$ to obtain
\begin{equation}
    f_{ij}(x)=\left\{
        \begin{array}{ll} 1,& i=I\\
            (l_j-1)x^{l_j-2} ,& i=O,\\
        \end{array}
        \right.
\end{equation}
where $f_{Oj}(x)$ denotes the highest priority among $l_j-1$ non-interacting tasks. Then $l_j-1$ non-interacting tasks with priorities uniformly distributed in $[0,1]$ can be reduced to one non-interacting task with priority $f_{Oj}(x)$ as only the highest priority task is relevant. Initially, the priorities are assigned to the tasks as described earlier. In each time step, both agents select the task with highest priority in their lists. If both agents select the task $I$ then it is executed, otherwise each agent executes the task $O$. Each executed task is assigned with a new priority drawn from the distribution $f_{ij}(x)$. This process leads to bursty patterns of task execution, which in turn induces power-law distributed inter-event times with an exponent $\alpha$ depending on the length of the priority queues. The exponent follows qualitatively the relation $\alpha=1+1/\max\{l_j-1\}$. Its maximum is $\alpha=2$ if the queues consists of two tasks, then $\alpha=3/2$ for three tasks leading to $\alpha=1$ as $l$ increases. This suggests that there are not only two ``universal'' exponent values, e.g., as proposed in Ref.~\cite{Vazquez2006Modeling}, but $\alpha$ can take several other values depending on the length of the priority queues. Note that the authors provided a definition of coarse grained models to achieve large scale simulations with large inter-event times and reliable scaling exponent estimations. They also showed that the emerging cutoff of the inter-event time distribution is a consequence of the finite simulation window thus the power-law functional form of $P(\tau)$ is asymptotically true.

An extended definition of the above model was proposed by Min \textit{et al.} who considered two scalable interaction protocols and a network of individuals with priority queues~\cite{Min2009Waiting}. They identified the above model definition as the AND-type protocol where $I$ tasks are executed only if they obtain the largest priority at the same time. They show that this protocol commonly leads to frozen states when applied to queues connected in a network. They argue, however, that an OR-type protocol would be more reasonable for the tasks, which require simultaneous actions of two or more individuals though the action can be initiated primarily by one of them. Examples are phone calls or instant messages where the task of answering of an incoming interaction jumps usually to the top of one's priority queue immediately when one receives a call or a message. The iteration of the model starts by choosing a random node $i$. If its highest priority task is $I_{ij}$, the two tasks $I_{ij}$ and $I_{ji}$ are executed regardless of the priority value of $I_{ji}$; in the other case if $O_i$ is the highest priority task, only that is executed. Priorities of all the executed tasks are randomly reassigned. This model process does not drive to a globally frozen state yet the $P(\tau)$ exhibits power-law tails, however with an exponent that depends on the network size $N$ as well as the network topology in a diverse way.

\subsection{Models with combined mechanisms}

Wu \emph{et al.}~\cite{Wu2010Evidence} proposed a combined model of Poissonian and priority induced bursts to explain the bimodal shape of the inter-event time distribution, typical in SMS communications. They argue that this phenomenon is a consequence of the interplay between processes effective at different time scales and determined by three important ingredients, namely (a) a Poisson process responsible for the initiations of bursts, (b) execution of competing tasks of an individual, and (c) interactions. They identify two types of tasks, an interaction task (I) and other tasks (O), and they consider each interaction task whether it is an initiation or a response action. Based on data analysis of individual SMS interaction patterns, they found that the inter-event time distribution of an individual can be described best by a power-law distribution if $\tau<\tau_0$ and by an exponential if $\tau>\tau_0$, where $\tau_0 \simeq 20$ min.
	
Based on these observations they propose a model defined as two interacting priority queues to mimic the interaction dynamics of two individuals. They first consider the priority queues of tasks of individuals in which the tasks in the queue are executed one by one with the probability $\Pi=x^{\alpha}$ in a ranked order by their randomly chosen priority $x\in (0,1)$. In addition they introduce a processing time $t_i$ determining as the time scale by which tasks are executed and added to the list. Interacting tasks (I) are added to the list with a small rate $\lambda_i=\lambda t_i$ in a Poissonian fashion. Next they consider the interaction between individuals: This occurs when one of the agents A (or B) executes an I-task, which will add an I-task to the list of B (A) with a corresponding probability $P_B$ (or $P_A$). All the I-tasks are randomly initiated by an individual and responding to other tasks will be put to the waiting list with a random priority $x$, subsequently competing for the execution with the O-tasks. Thus, the model is controlled by three important parameters for each user, i.e., $\lambda_i$, $\alpha_i$, and $P_i$, each of which is related to the Poisson process, decision making, and interaction, respectively. Fitting these parameters with empirical sequences allows the model successfully reproducing the bimodal shape of inter-event time distributions between interactions of an individual. This suggests that all the three ingredients are necessary in explaining this phenomenon.

\section{Network models of bursty agents}

\subsection{Zero-crossing random walk model}

Beyond a single node or link dynamics, other models have been proposed to simultaneously capture the topological and temporal features of agents interacting in a larger network. The first among these models was proposed by G\"{o}tz \emph{et al.}~\cite{Goetz2009Modeling} whose aim was to simulate the posting dynamics of bloggers, which in turn induces a reference network between blogs and postings with particular topological features. In their model they associate a blogger with a random walker in one dimensional space, who posts each time when it returns to its original position, hence it is called zero-crossing model. In the beginning of the process a blogger $A$ starts a walk from position $0$ and in each time step, with probability $1/2$, adds or subtracts a unit from its position. Whenever, the position of $A$ becomes $0$ it creates a post $P$ which, with probability $1-p_L$, is a new conversation, or otherwise a comment on another post. In the latter case, with probability $1-p_E$, the blogger comments on one of the posts of a blog it has already commented on (exploitation mode), or otherwise chooses a new blog to comment on (exploration mode). Subsequently in the selected blog $B$ to refer a post $Q$ is chosen with a probability weighted by the number of times this post has been earlier referred. Finally for each post $R$ reachable from post $P$, for each path $p$ from $P$ to $R$ create a link from post $P$ to post $R$ with probability $p_{LE}^{|p|}$. Here $p_{LE}$ denotes the probability for expanding a link and $|p|$ is the path length. The authors show that the structure of the simulated blog post network emerges with a power-law in-degree distribution and cascade size, while the inter-event times between consecutive posts of a blogger are distributed as $P(\tau)\sim \tau^{-3/2}$. In addition the emerging activity signals are self-similar with fractal dimension 0.5, comparable to the empirical observations presented by the authors.

\subsection{Reinforcement models of group formation}
\label{sec:brstmodelgroups}

A model based on reinforcement mechanisms was proposed by Stehl\'e \emph{et al.}~\cite{Stehle2010Dynamical} and Zhao \emph{et al.}~\cite{Zhao2011Social} to simulate social group formation in bursty contact networks. Their model simulates interacting agents forming disconnected groups, which evolve by successive mergings and splittings. These actions are driven by underlying reinforcement processes summarised by the authors as ``the longer an agent interacts with a group, the less that agent is likely to leave the group and the more an agent is isolated the less likely the agent is to interact with a group.'' More precisely their model considers $N$ agents, which either can be isolated or belong to a group defining an instantaneous contact network. Each agent is characterised by two variables: the number $p_i$ of actually contacted agents (its group size minus one) and the time $t_i$ when $p_i$ changed for the last time. At each time step $t$ an agent $i$ is randomly chosen. If the agent is isolated, it changes state with probability $b_0f(t,t_i)$ and chooses another isolated agent $j$ with probability $\Pi(t,t_j)$, such that they form a pair and update their state variables $p_i$, $t_i$, $p_j$, and $t_j$. On the other hand if $i$ is part of a group it changes its state with probability $b_1f(t,t_i)$. When the state changes, the agent can become isolated with probability $\lambda$, or otherwise it introduces an isolated node $j$ selected with probability $\Pi(t,t_j)$. If a node leaves or a new node is introduced to a group, all participating nodes update their state variables accordingly. The parameters $b_0$ and $b_1$ determine the tendency of the agents to change their state between being isolated and in a group, while $\lambda$ controls the tendency either to leave groups or in contrary to make them grow. In addition, the model dynamics strongly depends on the functions $f$ and $\Pi$. For simplicity, choosing them to be identical and to decay as a power-law like
\begin{equation}
    f(t-t_0)=\Pi(t-t_0)=(1+\tau)^{-1}, \hspace{.2in} \tau=(t-t_0)/N
\end{equation}
leads to system dynamics governed by reinforcement processes. This way the modelled system can reproduce several realistic features observed in real interaction data, such as power-law distributed interaction and inter-event times and duration of triadic interactions, and that the stability of groups decreases with their size.

\subsection{Evolving networks with interacting priority queues}

Jo \textit{et al.}~\cite{Jo2011Emergence} introduced an evolving network model, which integrates different interaction strategies, inspired by the Kumpula model for social network evolution~\cite{Kumpula2007Emergence}, with interacting priority queues defined above. In their model $N$ agents are given, each with a priority queue of two tasks $I$ and $O$ with priorities randomly assigned from a uniform distribution. At each time step $t$ every node selects its highest priority task. If it is an $I$-task, the node $i$ selects a target node for interaction either (a) from the whole population with probability $p_{_{GA}}$, (b) from its next nearest neighbours with probability $p_{_{LA}}$, or (c) from its neighbours with probability $1-p_{_{GA}}-p_{_{LA}}$ weighted by their link weight $w_{ij}$. The next nearest neighbour $j$ of the node $i$ is defined as a node satisfying $\{ t_{ik},t_{jk} \} =\{ t-2, t-1 \}$ with an intermediate node $k$, implying that $i$ and $k$ interacted at time $t-2$, and $k$ and $j$ interacted at time $t-1$. In both cases of (a) and (b), links between nodes are created with unit weight either randomly, representing the {\em focal closure} mechanism, or by closing a triangle, representing the {\em triadic closure} mechanism. The case of (c) represents the reinforcement mechanism as existing links are selected and their weights are reinforced. After a target node $j$ is selected an interaction between $i$ and $j$ takes place if the target $j$ has not been involved in any event at this time step $t$. After an interaction the priority of $I$-task of node $i$ is updated. In addition, at each time step each node can forget all of its existing connections with probability $p_{_{ML}}$, i.e., memory loss, to become isolated. By measuring the inter-event times between two consecutive $I$-tasks of a given node the system exhibits a broad inter-event time distribution with an exponential cutoff. In addition the emerging network structure shows several realistic features such as Granovetterian community structure~\cite{Granovetter1973Strength}, high clustering, assortative degree correlations, and broad link-weight distributions. It is worth to note that similar interaction dynamics has been observed in a variant of this model without priority queues, possibly suggesting that the source of heterogeneous dynamical behaviour could be also a consequence of the link-weight reinforcement process.

\subsection{Dynamic networks with memory}

A conceptually different type of evolving network model was proposed by Colman and Greetham~\cite{Colman2015Memory} who used a different memory kernel to induce bursty interaction sequences of agents in an evolving network. To generate event sequences with power-law distributed inter-event times they defined a discrete-time stochastic process, which generates an infinite sequence of binary random variables $X_t$ taking values $1$ ($0$) if an event takes place at time $t$ (or not). To determine $X_t$ an agent has a memory capacity of size $M$, represented by $m_n(t)$ for $n=1,2,\cdots,M$. Each $m_n(t)$ can have a value of $1$ or $0$. Using the definition of $k_t=\sum_{n=1}^M m_n(t)$, the kernel $f(k_t)$ determines the probability to execute an event in time step $t$. The new event occurs, i.e., $X_t=1$, with probability $f(k_t)$, otherwise no event occurs, i.e., $X_t=0$. The authors introduced two possible memory updating mechanisms: One is for a random $n'$ to set $m_{n'}(t+1)$ by the value of $X_t$, while keeping all others, i.e., $m_n(t+1)=m_n(t)$ for $n\neq n'$. The other is basically shifting $m_n(t)$ by one position, i.e., $m_n(t+1)=m_{n+1}(t)$ for $n=1,\cdots,M-1$, and setting $m_M(t+1)=X_t$. In this way, the memory is kept up to the $M$th latest realisations, similarly to the sequential memory loss mechanism proposed in Ref.~\cite{Jo2015Correlated}. They propose to use a linear probability kernel
\begin{equation}
    f(k_t)=\frac{k_t+x}{M+x+\epsilon},
    \label{eq:fkt}
\end{equation}
where $x$ and $\epsilon$ are positive real numbers. If $x$ is large the system approaches a Bernoulli process, while if $x$ is small relative to $M$ the inter-event time asymptotically follows a power-law as $P(\tau)\sim \tau^{-(2+x)}$. Using this dynamics for each agent they introduce an evolving network model of $N$ nodes and $E$ edges, where each node is assigned with a fitness value $x$ sampled from a probability distribution $\rho(x)$. The network is originally a random structure and at each step a node $i$ is randomly selected with a probability given by its attachment kernel $\Pi(i,x_i)$, a second node $j$ is selected in the same way and an edge is created between them. At the same time the oldest edge is removed from the network keeping the average degree constant. Considering the attachment kernel as
\begin{equation}
    \Pi(i,x_i)=\frac{k_i+x_i}{\sum_j(k_j+x_j)}
    \label{eq:Piix}
\end{equation}
with $k_i$ denoting the degree of $i$, they show that if $\rho(x)$ follows a power law the model process induces a scale-free structure and since always the oldest link is removed, $M=E$. In case of setting $x_i+\epsilon=N\langle x \rangle/2$, Eqs.~(\ref{eq:fkt}) and~(\ref{eq:Piix}) become equivalent. Thus if the fitness distribution is chosen such that $\langle x \rangle \ll \langle k \rangle$, the interacting nodes will exhibit bursty interaction patterns.

\subsection{Activity driven network models with bursty nodes}

Activity driven models of time-varying networks is a family of generative temporal network models, which can be used to simulate synthetic interaction sequences of model agents with arbitrary level of complexity. In its simplest definition~\cite{Perra2012Activity} the model assumes that there are $N$ independent agents, all assigned with an activity potential $a_i$ drawn from an arbitrary distribution. The activity potential describes the probability that an agent initiates an interaction with a randomly selected other agent at each time step. Initiating the model with disconnected agents and simulating their interactions over a transient period one can cumulate the emerging interaction structure and obtain a generative network structure. It has been shown that this cumulated network structure emerges with a degree distribution, which scales as the originally assumed distribution of activities. Further extension of the model leads to emerging weight heterogeneities~\cite{Karsai2014Time}, communities, weight-topology correlations~\cite{Laurent2015From}, etc. just to mention a few examples to demonstrate the potential of this modelling framework.

This model has been extended in two ways to consider agents with bursty activity patterns, in order to understand the effects of non-Poissonian dynamics on the emerging network structure. Although burstiness is not an emergent property in any of these models, yet we briefly discuss them as they may be useful to study in the future the effects of burstiness on emerging structures or ongoing dynamical processes.

In one definition of Moinet \textit{et al.}~\cite{Moinet2015Burstiness, Moinet2016Aging} the timings of node activities are determined by a renewal process. Each agent $i$ is assigned with a time-dependent activity $a_i(\tau)$, which depends on the time $\tau$ passed since its last activation. The activation of each node follows a renewal process governed by a waiting time distribution $P(\tau, c_i)$, where $c$ is a parameter determining the heterogeneity of the activation rate of the agents, and assumed to be randomly sampled from a distribution $\eta(c)$. Assuming a power-law waiting time distribution
\begin{equation}
P(\tau,c)=\alpha c(c\tau+1)^{-(\alpha+1)}, \hspace{.2in} 0<\alpha <1
\end{equation}
and a power-law heterogeneity distribution
\begin{equation}
\eta(c)=\frac{\beta}{c_0}(c/c_0)^{-(\beta+1)}
\end{equation}
with $\beta>\alpha$, the degree distribution of the emerging network emerges as
\begin{equation}
P_t(k)\sim (c_0 t)^{\beta}(k-\langle r \rangle_t)^{-1-(\beta/\alpha)}
\label{eq:ptk}
\end{equation}
where $\langle r \rangle_t$ is the average number of times a node becomes active up to time $t$. Equation~(\ref{eq:ptk}) leads to a relation between the power-law exponents of the degree, waiting time, and heterogeneity distributions as $\gamma=1+\beta/\alpha$, indicating dependencies between the topological properties of the network and the distribution of renewal events. Based on this the authors show that the model is affected by ageing effects when the waiting time distributions have the power-law tail with $\alpha<1$, which they demonstrate using numerical simulations and empirical results measured in scientific co-publication networks.

In another work Ubaldi \textit{et al.}~\cite{Ubaldi2016Burstiness} also build on the activity driven network but extended it with two mechanisms. First of all they introduced bursty dynamics directly by assuming that inter-event times $\tau_i$ for node $i$ were drawn from a power-law distribution of the form
\begin{equation}
P(\tau_i)=\frac{\alpha}{\xi_i^{-\alpha}}\tau_i^{-(1+\alpha)}, \hspace{.2in} \tau_i \in [\xi_i,+\infty ).
\end{equation}
Here $\xi_i$ is a lower time cutoff for the minimum inter-event time of node $i$, which in turn determines its characteristic time-scale as its activity $\xi_i\sim 1/a_i$. If $\xi_i$ is distributed as a power-law $P(\xi_i)\sim \xi_i^{\nu-1}$, for small $\xi_i$ values the induced node activities will also be  power-law distributed with a corresponding exponent of $-(\nu+1)$. In this way the burstiness directly governs the evolution of the network. Another mechanism considered by the authors is a memory driven tie allocation process that enhances repeated interactions of already existing ties. They show analytically and by means of numerical simulations that the simultaneous control of the relative strength of burstiness and the tie reinforcement leads to a non-trivial phase diagram determined by the interplay of the two processes. They found two different dynamical regimes, one in which the burstiness governs the evolution of the network, and another in which the dynamics is completely determined by the process of tie allocation. Interestingly, if the reinforcement of previously activated connections is sufficiently strong, the burstiness governs the network evolution even in the presence of large inter-event time fluctuations. 


\chapter{Dynamical processes on bursty systems}
\label{chapter:processes}


The bursty temporal patterns in human interactions are important not only for understanding the dynamics of the egocentric and global social networks but also because they have indisputable effects on the evolution of dynamical processes taking place on them like random walks, information diffusion, epidemic and social contagion, or various types of evolutionary games, just to mention a few. In the earlier studies of these processes it was commonly assumed that they evolve over static structures. In these cases links representing interactions between nodes, were always present in the network, while the question was about the effects of structural heterogeneities and correlations on the final outcome of the process in question~\cite{Barrat2008Dynamical}. However, the recent availability of large digital datasets recording temporally detailed interactions of individuals led to the advent of the new field of temporal networks~\cite{Holme2012Temporal}. In this representation interactions are not taken to be static, but assumed to vary in time and allow information to pass between connected nodes only at the time of their interactions. Parallel to the foundation of the methodologies, models, and theories of temporal networks, several studies addressed the effect of time-varying interactions on the evolution of dynamical processes.

Importantly, it has been found that the bursty nature of human interactions has dramatic effects on the unfolding of several modelled processes. First reported observations were the results of data-driven simulations, where synthetic dynamical processes were simulated on real and inherently bursty interaction sequences~\cite{Karsai2011Small, Miritello2011Dynamical, Rocha2011Simulated}. These observations together with early theoretical results~\cite{Vazquez2007Impact, Kivela2012Multiscale, Iribarren2009Impact} disclosed a main puzzle, as burstiness was found to slow down the emergence of several types of global phenomena, while in some other cases it appeared to show opposite effects, leading to faster scenarios as compared to the Poissonian case. In order to address these seemingly contradicting observations two general modelling directions have been considered. On one hand, for data-driven simulations a new modelling concept using random reference models (RRMs) has been proposed (for a brief discussion see Section~\ref{sec:otherindivmods}). These models define several ways of shuffling interaction sequences to remove temporal and structural correlations in a controlled way for identifying their effects on the simulated dynamical processes. On the other hand, more formal approaches consider the effects of bursty characteristics, like the heterogeneous inter-event time distribution, residual times, and local temporal correlations to explain the observed behaviour.

All these results tend to draw a very heterogeneous picture, with some comprehensive understanding about the effects of bursty interactions on dynamical processes, but leaving some other problems to be opened in several ways for further research. We have learned that the observed effects strongly depend on the actual datasets in use and the model we chose to investigate. Thus, instead of providing a closed theory about the effects of bursty patterns on dynamical processes, we first discuss the different characteristics of bursty behaviour, which were found to be relevant in various studies, and then we present the main findings on different types of dynamical processes, which were investigated on bursty temporal networks of human interactions.

\section{Bursty characteristics controlling dynamical processes}

\subsection{Inter-event time and residual time distributions}

In human interactions the bursty dynamics has been characterised by a broad inter-event time distribution $P(\tau)$, which commonly appears as a power-law, potentially with an exponential cutoff or in a log-normal form, as discussed in Chapter~\ref{chapter:emp}. It indicates that individual dynamics are typically non-Poissonian with events being separated by heterogeneous inter-event times, unlike in case of Poisson dynamics with exponentially distributed inter-event times. In this Section we are mostly interested in effects induced by non-Poissonian dynamics, while the corresponding Poissonian system will be used as a reference. Any dynamical process that unfolds in bursty temporal networks can be effected by the broad $P(\tau)$ such that short inter-event times tend to help the rapid update of interacting nodes while long inter-event times act in an opposite way, keeping information locally stuck for long period of times.

\begin{figure}[!t]
\centering
    \includegraphics[width=.8\columnwidth]{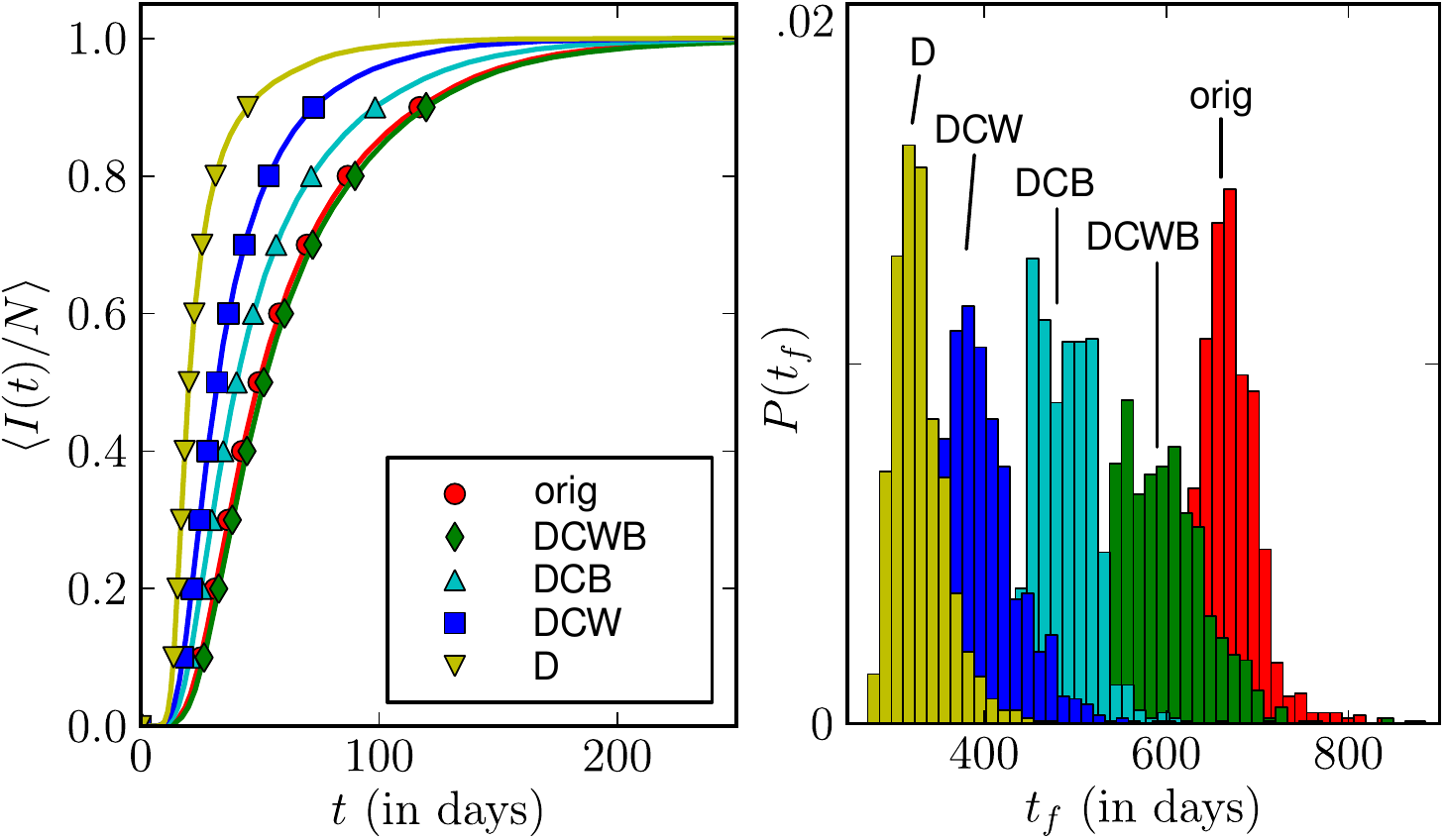}
    \caption{Demonstration of RRMs in case of Susceptible-Infected or SI spreading on mobile communication networks. (a) Average fraction of infected nodes $\left<I(t)/N\right>$ at each point in time for the original event sequence ($\circ$) and null models:  equal-weight link-sequence shuffled DCWB ($\lozenge$), link-sequence shuffled DCB ($\vartriangle$), time-shuffled DCW ($\square$) and configuration model D ($\triangledown$). (b) Distribution of full prevalence times $P(t_f)$ due to randomness in the initial conditions. (\emph{Source:} Adapted from Ref.~\cite{Karsai2011Small} under Copyright (2011) by the American Physical Society.)}
\label{fig:Malmgren2008A_2}
    \label{fig:SbSW}
\end{figure}

These effects can be easily verified by using appropriate random reference models in data-driven simulations on bursty temporal networks. One of the most frequently used RRMs takes the event sequence of interacting individuals and shuffle the interaction times between events~\cite{Karsai2011Small, Kivela2012Multiscale, Miritello2011Dynamical}. Shuffling in this way destroys any temporal correlation in the original event sequence, including the bursty temporal patterns, as it assigns a random time to each event over the observation time window $T$. Note that an equivalent method would be to pick a random time for each event from the window $T$. Using both of these RRMs one can obtain a sequence of interactions, which shows Poissonian dynamics and exponentially distributed inter-event times, while keeping the network structurally unchanged. This removal of bursty patterns in most observations fastens the emergence of a global phenomena, as demonstrated in Fig.~\ref{fig:SbSW}(a) (time-shuffled curve with blue squares), which in turn suggests that burstiness actually slows down the dynamical process~\cite{Karsai2011Small, Kivela2012Multiscale, Miritello2011Dynamical, Gauvin2013Activity, Delvenne2015Diffusion, Karimi2013Threshold, Backlund2014Effects}. On the other hand, some exceptions have also been reported~\cite{Rocha2013Bursts, Rocha2011Simulated, Takaguchi2013Bursty}, where the same procedure indicates that burstiness accelerates some diffusion processes. As we will discuss below, heterogeneous inter-event times have different effects on the early and late stage dynamics, which tends to give to some extended explanation of these seemingly contradicting observations.

\subsubsection{The waiting-time paradox}
\label{sec:wtp}

One simple mathematical argument, known as the \emph{waiting-time paradox} (a.k.a the bus paradox, or hitch-hiker's paradox) provides a simple explanation about the effect of temporal heterogeneity on the speed of any dynamical processes. It concerns a single point process capturing the interaction dynamics of an individual $i$ (or a social tie, or the arrival of buses to a stop, etc.), where events are assumed to be independent, following each other with inter-event times sampled from the distribution $P(\tau)$. The waiting-time paradox states that if information (random walker, virus in epidemics, rumour, etc.) arrives to node $i$ from another node, it needs to wait on average longer than the half of the average inter-event time before it can leave node $i$ and pass to another node $j$. This is true for point processes with any level of temporal heterogeneity, including Poisson and bursty systems, even if the arrival time is uniformly distributed during two consecutive events of node $i$.

In order to understand better this paradox we need to recall the dependence we already discussed in Section \ref{sec:iet_rt_wt} between the residual time $\tau_r$ and the inter-event time distribution $P(\tau)$~\cite{Kivela2012Multiscale}. Let us assume that we have two connected nodes $i$ and $j$ and $i$ receives information at a uniformly random point in time $t_0$. In this setting the residual time $\tau_r$ is defined as the random variable that represents the time between this random time $t_0$ of information receiving and the time of the next event occurring between $i$ and $j$. As we have seen in Eq.~(\ref{eq:rst_iet}), the residual time $\tau_r$ of the link is obviously determined by the inter-event time distribution of events of the actual link. Note that the inter-event times and residual times have the same distributions when the process is Poissonian, while for bursty processes they both appear with power-law tails with exponents related as $P(\tau)\sim\tau^{-\alpha}$ and $P(\tau_r)\sim\tau_r^{-(\alpha-1)}$ \cite{Lambiotte2013Burstiness}.

We have also seen in Eq.~(\ref{eq:taurderiv}) that the residual time can be derived as 
\begin{equation}
\langle \tau_r \rangle=\frac{\langle \tau^2 \rangle}{2\langle \tau \rangle}.
\end{equation}
Consequently, the value of $\tau_r$ depends heavily on the first and the second moment of the inter-event time distribution, or equivalently, on the average and fluctuation of event rates taking place on a link. In case the point process is maximally regular, i.e. $P(\tau)$ is a delta function, $\langle \tau^2 \rangle = \langle \tau \rangle^2$ and we obtain the intuitive result
\begin{equation}
\langle \tau_r \rangle = \frac{\langle \tau \rangle}{2}.
\end{equation}
However, in case of a Poisson process with $P(\tau)=\langle \tau \rangle^{-1} e^{-\tau / \langle \tau \rangle}$ we obtain $\langle \tau^2 \rangle = 2 \langle \tau \rangle^2$, which leads to a twice larger mean residual time when compared to the regular case. If the inter-event time distribution is broader than exponential, e.g., a power-law distribution, the deviation from the regular case is even bigger. We consider $P(\tau)=(\alpha-1)\tau_0^{\alpha-1}\tau^{-\alpha}$ with the lower bound of inter-event times, $\tau_0$. If the power-law exponent $\alpha$ is larger than $3$, both $\langle \tau \rangle$ and $\langle \tau^2 \rangle$ are finite, and the relation
\begin{equation}
\langle \tau_r \rangle = \frac{\alpha-2}{2(\alpha-3)}\tau_0
\end{equation}
is obtained. On the other hand, if $\alpha\leq 3$, the diverging $\langle \tau^2 \rangle$ leads to the diverging mean residual time $\langle \tau_r \rangle$~\cite{Kivela2012Multiscale}.

As we are primarily interested in the effects of the shape of the inter-event time distribution on $\langle \tau_r \rangle$, it is natural to use the Poisson model as a reference. Thus we may consider a normalised mean residual time as it has been introduced in Eq.~(\ref{r_definition}). This quantity measures the ratio of the second moment to the square of the first moment of the inter-event time distribution. Generally, the broader the distribution is, the larger the second moment is as compared to the square of the first moment. Hence, Eq.~(\ref{r_definition}) indicates that the more bursty an event sequence is, the longer the residual times are. In case of a power-law $P(\tau)$ distribution, this ratio becomes infinite when the power-law exponent $\alpha=3$ while it decreases with increasing $\alpha>3$, reaching $1$ when $\alpha=2+\sqrt{2}$. Thus, for power-law inter-event time distributions in this regime, the mean residual times are longer than those for the Poissonian reference case.

\subsubsection{Ordering of events}
\label{sec:OE}
In social networks the ties may show very different activity levels, which in turn can lead to different residual time distributions for each link. This has a consequence for several dynamical processes where the ordering and the timing of interactions determine the path of diffusion.

\emph{Random walk processes} are considered as generic models for diffusion and are commonly studied on static or temporal networks. One of the variants of these models is defined on temporal networks and is called \emph{greedy random walk}, where a single random walker is diffusing in the network hopping from one node to another, only at the time of their temporal interactions. The walker is greedy because after arriving to a node $i$ it leaves immediately via the next event towards some node $j$. In this way the probability that the walker at node $i$ will end up to a specific neighbour $j$ depends on one hand on $P_{ij}(\tau_r)$, but also on the residual-time distribution $P_{ik}(\tau_r)$ of any other $k$ neighbour of $i$. If an event towards $k$ appears earlier than towards $j$, the random walker will necessarily hop to node $k$ instead of node $j$. The probability that the random walker will end up on $j$, can be written as:
\begin{equation}
P^{RW}_{ij}(\tau_r)=P_{ij}(\tau_r) \prod_{k\neq j}\int_{\tau_r}^{\infty}P_{ik}(\tau'_r)d\tau'_r,
\label{eq:rwtransp}
\end{equation}
where the product denotes the probability that no event appeared earlier than the one with $j$.

\emph{Spreading processes} are also largely influenced by the ordering and timing of the interactions~\cite{Scholtes2014Causality}, which determine time-respecting paths in a temporal structure, along which information, disease, or rumor can travel. Spreading processes are commonly modelled by assuming that the nodes of a network can be, e.g., in three mutually exclusive states: Susceptible (S), infected (I), or recovered (R). The susceptible node (S) can become infected (I) with the infection rate $\tilde \beta$ due to the interaction with an infected neighbour. The infected node can spontaneously recover with the recovery rate $\tilde \mu$, corresponding to the transition from I to R. In other model definitions, the infected node can return back to the susceptible pool. Thus, what matters for spreading is that an interaction event of an infected node $i$ with a susceptible node $j$ occurs earlier than an infected node recovers. This happens with the probability
\begin{equation}
P^{SIR}_{ij}(\tau_r)=P_{ij}(\tau_r) \int_{\tau_r}^{\infty}r_i(t)dt,
\end{equation}
where $r_i(t)$ is the probability that the infected node $i$ recovers after time $t$.

Consequently, in case of a random walk the relative behaviour of the residual time distributions on neighbouring links is important, while in case of spreading the relative behaviour of residual time distribution and recovery time distributions. It indicates that not only the heterogeneous temporal behaviour but also the ordering of events are crucial~\cite{Lambiotte2013Burstiness}. If ties with low activity and bursty interaction dynamics occupy important positions in the network (like bridges between communities), they may have a large impact on the final outcome of spreading as they are able to keep spreading local inside well connected communities with active links.

\subsubsection{Early and late time effects of burstiness}

Heterogeneous inter-event times may have different effects when considering the early and late time behaviour of a dynamical process. Recently much effort has been devoted to clarify how the burstiness of events influences the spreading speed, partly by using empirical data analysis~\cite{Vazquez2007Impact, Karsai2011Small, Iribarren2011Branching, Miritello2011Dynamical, Rocha2011Simulated, Gauvin2013Activity} and partly by model calculations~\cite{Holme2012Temporal, Vazquez2007Impact, Iribarren2009Impact, Rocha2013Bursts, VanMieghem2013NonMarkovian, Jo2014Analytically}. In those studies the bursty character of an event sequence was found to slow down the late time dynamics of spreading. However, for the early time dynamics, conflicting results have been reported~\cite{Masuda2013Predicting}. In studies by Vazquez~\emph{et al.}~\cite{Vazquez2007Impact} and Karsai~\emph{et al.}~\cite{Karsai2011Small} the burstiness is found to slow down spreading, while other works point towards the opposite direction~\cite{Iribarren2011Branching, Rocha2011Simulated, Rocha2013Bursts}. In the following we address separately the early and late time effects of heterogeneous temporal behaviour by means of modelling the deterministic Susceptible-Infected (SI) processes at these two extremes. The SI processes are a specific case of SIR models where recovery is not possible ($\tilde \mu=0$), thus once a node is infected it keeps its state until the end of the process. More specifically, we consider a deterministic SI process where infection pass between connected nodes with probability $1$, which corresponds to the fastest possible spreading scenario, determined exclusively by the ordering and timing of temporal interactions.

\paragraph{\emph{Early time effects:}}

The early stage dynamics of a spreading process is mainly driven by small inter-event times, which generally leads to the faster spreading for non-Poissonian dynamics as compared to Poisson-like cases. Since at the early time of spreading most of the nodes are still susceptible one can safely assume in the modelling that finite size effects do not play a role and an infected node can always find a susceptible neighbour. To understand this limit we follow the argumentation of Jo \emph{et al.} presented in details in Ref.~\cite{Jo2014Analytically}.

\begin{figure}[!t]
\center
    \includegraphics[width=.8\columnwidth]{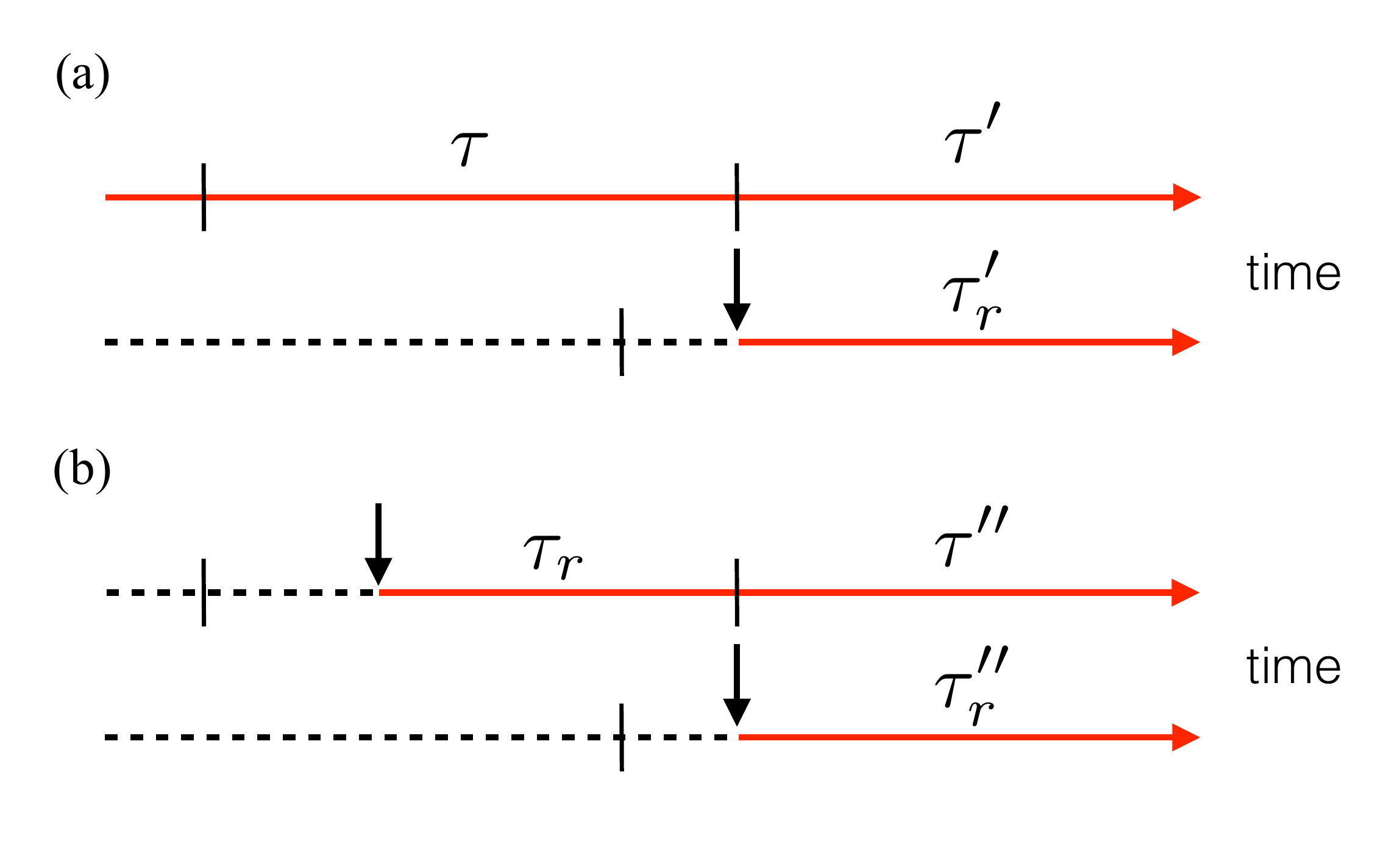}
    \caption{Schematic diagram of the infections by an already infected node (a) and by a newly infected node (b). Vertical lines and vertical arrows denote activation timings of nodes and infections from infected nodes (solid horizontal line) to susceptible nodes (dotted horizontal line). Inter-event times $\tau$, $\tau'$, and $\tau''$ are independent of each other, and so are residual times $\tau_r$, $\tau_r'$, and $\tau_r''$.}
    \label{fig:branching}
\end{figure}

Let us consider a system with $N$ nodes, which perform instantaneous interactions with dynamics modelled by a renewal process~\cite{Feller1971Introduction} with an arbitrary inter-event time distribution $P(\tau)$, which is the same for every node in the whole population. Note that $P(\tau)$ determines only the activation times of nodes, irrespective of whether the nodes are susceptible or infected. Whenever an infected node becomes active, it chooses randomly another node from the remaining $N-1$ nodes and if the chosen node is susceptible, then it becomes infected. Here the probability of choosing a susceptible node is $1$ in the infinite size system as the dynamics starts from a single infected node. The newly infected node remains inactive as long as its residual time $\tau_r$ before it becomes active and selects randomly a node to infect. The early stage of the spreading dynamics is sensitive to the variation of the initial distribution of active or inactive nodes. Note that this model is related to a class of Bellman-Harris branching processes~\cite{Harris2002Theory, Iribarren2009Impact, Iribarren2011Branching}, which have been used to address similar phenomena.

We investigate the spreading dynamics starting from one infected and active node at time $t=0$. Hence the number of infected nodes is initially $I_0(t)=1$ and remains unchanged for a time interval $\tau$ until the next event of the initially infected node takes place. At $t=\tau$, $I_0(t)$ can be written as the sum of two numbers: One is for the infecting node and its subsequent infected nodes, which can be denoted by an independent and identical copy of $I_0$ but starting at $t=\tau$, i.e., $I'_0(t-\tau)$. The other is for the newly infected node and its subsequent infected nodes, similarly denoted by $I'_1(t-\tau)$, where $I'_1$ is an independent and identical copy of $I_1$, to be defined below. Thus we get
\begin{eqnarray}
  I_0(t)&=&\left\{\begin{tabular}{ll}
    $1$ & \textrm{if}\ $t<\tau$,\\
    $I_0'(t-\tau)+I_1'(t-\tau)$ & \textrm{if}\ $t\geq \tau$.
  \end{tabular}\right.
\end{eqnarray}
Since the newly infected node must wait a residual time $\tau_r$ as in Fig.~\ref{fig:branching}(b), the number of infected nodes starting from one infected and inactive node can be written as
\begin{eqnarray}
  I_1(t)&=&\left\{\begin{tabular}{ll}
    $1$ & \textrm{if}\ $t<\tau_r$,\\
    $I_0''(t-\tau_r)+I_1''(t-\tau_r)$ & \textrm{if}\ $t\geq \tau_r$,
  \end{tabular}\right.
\end{eqnarray}
where $I''$s are independent and identical copies of $I$. The generating function for $I_0(t)$ is defined as $F_0(z,t)=\sum_{k\geq 0}\Pr[I_0(t)=k]z^k$, and we get
\begin{equation}
  F_0(z,t)= \left\{\begin{tabular}{ll}
    $z$ & \textrm{if}\ $t<\tau$,\\
    $F_0(z,t-\tau)F_1(z,t-\tau)$ & \textrm{if}\ $t\geq \tau$,
  \end{tabular}\right.
\end{equation}
where $F_1(z,t)$ is the generating function defined for $I_1(t)$. By taking the expectation value over $\tau$ with $P(\tau)$, we obtain
\begin{equation}
  F_0(z,t)=z\int_t^\infty P(\tau)d\tau+\int_0^t F_0(z,t-\tau)F_1(z,t-\tau)P(\tau)d\tau.
\end{equation}
We can use the generating function to calculate the average number of $n_0(t) \equiv  \langle I_0(t)\rangle$ as
\begin{equation}
  n_0(t)=\left.\frac{\partial F_0(z,t)}{\partial z}\right|_{z=1}=\int_t^\infty P(\tau)d\tau+\int_0^t [n_0(t-\tau)+n_1(t-\tau)]P(\tau)d\tau
\end{equation}
where $n_1(t)\equiv \langle I_1(t)\rangle$. Taking the Laplace transform gives
\begin{eqnarray}
  \tilde n_0(s)&=& \frac{1-\tilde P(s)}{s}+[\tilde n_0(s)+\tilde n_1(s)]\tilde P(s),\\
  \tilde n_1(s)&=& \frac{1-\tilde P_r(s)}{s}+[\tilde n_0(s)+\tilde n_1(s)]\tilde P_r(s),
\end{eqnarray}
where $\tilde P(s)$ and $\tilde P_r(s)$ denote the Laplace transforms of $P(\tau)$ and $P(\tau_r)$, respectively. This straightforwardly leads to
\begin{equation}
  \tilde n_0(s)=\frac{1}{s}+\frac{\tilde P(s)}{(s-\langle \tau \rangle^{-1})[1-\tilde P(s)]},
\end{equation}
where we have used the relation $\tilde P_r(s)=\frac{1}{\langle \tau \rangle s}[1-\tilde P(s)]$. Then, $n_0(t)$ can be calculated by taking the inverse Laplace transform of $\tilde n_0(s)$. Note that this solution has been obtained for arbitrary inter-event time distributions, which enables us to evaluate the effect of burstiness on spreading for both Poissonian and non-Poissonian cases.

In order to investigate the effect of the lower bound of inter-event times, we consider the shifted power-law distribution with exponential cutoff defined as
\begin{equation}
  \label{eq:shiftedNonPoisson_P0}
  P(\tau)= \frac{\tau_c^{\alpha-1}}{\Gamma(1-\alpha,y)} \tau^{-\alpha}e^{-\tau/\tau_c}\theta(\tau-\tau_0),
\end{equation}
where $\Gamma$ is the upper incomplete Gamma function, $\theta$ is the Heaviside step function, and $y\equiv \tau_0/\tau_c$ with $\tau_0$ and $\tau_c$ being the lower bound and the exponential cutoff of $ P(\tau)$, respectively. The Laplace transform of $P(\tau)$ in Eq.~(\ref{eq:shiftedNonPoisson_P0}) is as follows:
\begin{equation}
  \label{eq:shiftedNonPoisson_P0s}
  \tilde P(s)=(s\tau_c+1)^{\alpha-1}\frac{\Gamma(1-\alpha,y(s\tau_c+1))}{\Gamma(1-\alpha,y)}.
\end{equation}
To investigate the early time dynamics of $n_0(t)$, we consider the case when $s\gg 1$. By expanding the incomplete Gamma function, we obtain
\begin{equation}
  n_0(t) \approx 1+A\left(e^{\frac{t-\tau_0}{\langle \tau \rangle}}-e^{-\frac{t-\tau_0}{\tau_c}}\right)\theta(t-\tau_0),
\end{equation}
where $A=\frac{1}{x+y}\frac{y^{1-\alpha}e^{-y}}{\Gamma(1-\alpha,y)}$ with $x\equiv \tau_0/\langle \tau \rangle$. The spreading rate $C_0$ at $t=\tau_0^+$ is obtained as
\begin{eqnarray}
  \label{eq:shiftedNonPoisson_C0}
  C_0(x,y,\alpha) \equiv \langle \tau \rangle \left.\frac{dn_0}{dt}\right|_{t=\tau_0^+} =\frac{1}{x}\frac{y^{1-\alpha}e^{-y}}{\Gamma(1-\alpha,y)}.
\label{eq:C0}
\end{eqnarray}
By considering the inter-event time distribution in Eq.~(\ref{eq:shiftedNonPoisson_P0}), a Poissonian dynamics corresponds to the case when $\alpha=0$, while we obtain non-Poissonian interaction dynamics of $\alpha>0$. Using this parameterisation, Eq.~(\ref{eq:C0}) leads to
\begin{equation}
    C_0(x,y,\alpha)\geq C_0(x,y,0),
\end{equation}
suggesting that non-Poissonian bursty activity always accelerates the early time spreading dynamics as compared to the shifted Poissonian case with the same mean $\langle \tau \rangle$ and lower bound $\tau_0$ of the inter-event time distribution~\cite{Jo2014Analytically}.

Note that the accelerating effect of bursts on the short-term dynamics of SI processes has been observed by means of numerical simulations in independent studies by Rocha \emph{et al.}~\cite{Rocha2013Bursts} and Horv\'ath \emph{et al.}~\cite{Horvath2014Spreading}.

\paragraph{\emph{Late time effects:}}

As we mentioned earlier, a spreading process may behave differently in the late time limit as long inter-event times may slow down the process to reach the full prevalence. In order to better understand this limit we present here the argumentation of Min, Vazquez, and others~\cite{Vazquez2007Impact, Min2011Spreading, Min2013Burstiness}, which utilises branching processes in a somewhat similar way as we discussed for the early time limit.

In this case the deterministic SI process is diffusing on a temporal network with an underlying static tree-like structure. Its dynamics if determined by the generation time $\Delta$, which is defined as the time interval between the infection of a node and the transmission of the infection to one of its neighbours. In this model if an infection starting from a single node at time $t=0$, the average number of new infected nodes at time $t$ can be expressed as
\begin{equation}
n(t)=\sum_{d=1}^{D}z_d g^{*d}(t),
\label{eq:ntltb}
\end{equation}
where $z_d$ is the average number of nodes in $d$ contacts away from the seed node, and $D$ is the maximum of $d$. $g^{*d}(t)$ is the $d$th order convolution of $g(\Delta)$ corresponding to the probability density function of the sum of $d$ residual times, i.e., $g^{*1}(t)=g(t)$ for the immediate neighbour of the seed, and $g^{*d}(t)=\int_0^tg(t')g^{*d-1}(t-t')dt'$ in general.

The long time behaviour of $n(t)$ can be obtained by using Eq.~(\ref{eq:ntltb}), e.g., for the case where $g(\Delta)\sim\Delta^{-\nu}$ with $1<\nu<2$, corresponding to the stable L\'evy regime~\cite{Min2011Spreading, Min2013Burstiness}, we obtain $g^{*d}(t)\sim t^{-\nu}$ in the limits of $t\rightarrow \infty$ and $d\gg 1$ independently of the network structure. This corresponds to the asymptotic scaling of the prevalence as
\begin{equation}
n(t)\sim t^{-\nu},
\end{equation}
which means that in case of heterogeneous activity patterns the system slows down in the long-time limit with prevalence, which decays with the same exponent as the generation time distribution.

Next we assume that the interaction dynamics of nodes is dictated by a renewal process, generating independent events with an arbitrary inter-event time distribution $P(\tau)$. In this case if a node $i$ is infected at time $t_i$ and having a susceptible neighbour $j$, the generating time that the node $j$ receives the infection from node $i$ corresponds to the residual time $\tau_r$ between $t_i$ and their next interaction at $t_j$. As we shown in Eq.~(\ref{eq:rst_iet}), the residual time distribution can be easily derived from the inter-event time distribution. Therefore, for the activity patterns of uncorrelated events with $P(\tau)\sim \tau^{-\alpha}$ ($2<\alpha <3$) we obtain
\begin{equation}
n(t)\sim t^{-(\alpha-1)}.
\end{equation}
This can be compared to the case where the renewal process follows a Poisson dynamics with an exponentially decaying prevalence, which can be obtained along the same logic.

Note that the same conclusion can be drawn following the same argumentation used for the early time behaviour~\cite{Jo2014Analytically}, provided that the network size is assumed to be finite. In addition several numerical studies have confirmed this result~\cite{Vazquez2007Impact, Min2011Spreading, Min2013Burstiness, Jo2014Analytically} or more generally, showed that SI spreading slows down in the long time regime due to burstiness~\cite{Karsai2011Small, Miritello2011Dynamical, Vazquez2007Impact, Min2011Spreading}.

The non-stationarity of the interaction dynamics provides another way to address the early versus late time behaviour of dynamical processes. As it has been shown by Rocha \emph{et al.}~\cite{Rocha2013Bursts} a non-stationary contact dynamics may induce a rapid SI spreading with more infected nodes for early times as compared to a system with Poisson dynamics. The same was concluded by Horv\'ath \emph{et al.}~\cite{Horvath2014Spreading} who showed that power-law governed, non-stationary processes of young age can cause very rapid spreading, even for power-law exponents that would result in slow spreading in the stationary state. Consequently, the age of the processes has a strong influence on the outcome of spreading if the inter-event time distribution is heavy-tailed.

\subsection{Triggered event correlations}
\label{sec:triggev}

Another important character of bursty temporal networks, influencing dynamical processes, is the existence of causal correlations between the events sharing at least a node in common. Such triggered event pairs are the responsible for the emergence of mesoscopic temporal motifs~\cite{Kovanen2011Temporal} in which a larger number of correlated events are performed in a bursty fashion between two or more individuals. This kind of behaviour has been observed in human communication systems~\cite{Karsai2011Small, Kivela2012Multiscale, Miritello2011Dynamical,Scholtes2014Causality} and were shown to enhance the diffusion of information locally, and to accelerate globally the spreading dynamics at the early time limit.

Random reference models provide a straightforward way to study the effects of triggered event correlations on the global spreading dynamics. Taking an empirical temporal network one can obtain the sequence of interactions on each link. In order to remove only triggered event correlations between neighbouring links we can shuffle the network by re-assigning the entire interaction sequence of each link to randomly selected other links with the same number of events~\cite{Karsai2011Small}. In this way the synchronisation of events, i.e., triggered causal correlations, between neighbouring links are destroyed, while the system remains otherwise unchanged. Note that another equivalent method would be to add a random offset time to each event time on a link while applying temporal periodic boundary conditions~\cite{Backlund2014Effects}. As it is shown in Fig.~\ref{fig:SbSW}(a) and (b) (DCWB model assigned with green rhombus), triggered event correlations turn out to accelerate the process in the early stage while slowing down in the long run.

This effect has been identified in several works using numerical modelling and analytical calculations. Kivel\"a \emph{et al.}~\cite{Kivela2012Multiscale} have studied the case of an SI spreading process between three nodes connected by two links. In this toy system an event on one link may induce a triggered event on the other link with a given probability $p$, or the events are performed independently otherwise. Interesting quantity here is the average triggered relay time $\langle \tau_t \rangle$, which indicates the average time that information needs to wait to pass over the second link if it arrived at an earlier time on the first link. They show that
\begin{equation}
\langle \tau_t \rangle=\left( 1-\frac{p}{2}\frac{n-1}{n} \right) \langle \tau_r \rangle,
\label{eq:taut}
\end{equation}
where $n$ is the number of events on the second link. Equation~(\ref{eq:taut}) indicates that if $p=0$, i.e., all events are independent on the two links, the mean triggered relay time is equal to the mean residual time. However, for $p>0$, the greater the number of triggered events is, the shorter the triggered relay times are on average, which indicates that information spreading takes place faster.

Miritello \emph{et al.} have addressed the same problem~\cite{Miritello2011Dynamical} using data-driven simulations on real interaction sequences of mobile phone communication events. They considered a Susceptible-Infected-Recovered (SIR) spreading process with the infection rate $\tilde \beta$ and a constant recovery time $\Delta t$. They addressed the effects of causally correlated event pairs taking place within a short time on two neighbouring links $(*,i)$ and $(i,j)$. They have considered the case when the spreading reaches the node $i$ from an arbitrary neighbour $*$ (other than $j$) at time $t_0$, after which it could infect the node $j$ if any event occurred between $i$ and $j$ before it is recovered at time $t_0+\Delta t$. They mapped this problem to a static link percolation problem~\cite{Newman2002Spread} and showed that the average transmissibility, i.e., the probability that the infection passes over a triggered event on link $(i,j)$ is
\begin{equation}
\mathcal{T}_{ij} (\tilde \beta,\Delta t)=
\begin{cases}
	\tilde \beta \langle n_{ij} \rangle & \hspace{.1in} \text{for}\  \hspace{.1in} \tilde \beta \ll 1 \\
	1-P_{ij} & \hspace{.1in} \text{for}\  \hspace{.1in} \tilde \beta \simeq 1
    \end{cases}
\label{eq:transmiss}
\end{equation}
where $\langle n_{ij}\rangle$ denotes the average number of events between $i$ and $j$ after $i$ becomes infected and before it is recovered, and $P_{ij}$ denotes the probability that there is no such event during the period of $\Delta t$. 

Using the data analysis and random reference models they have shown that due to the correlation between events on neighbouring links, the number of events in a tie following an incoming call is always larger for the real-time data than for the time-shuffled case, corresponding to a Poisson process. Consequently, for small $\tilde \beta$, the average transmissibility and the size of the epidemic cascades are always larger in the real case than the time-shuffled case. In contrast, the bursty nature of the communication makes the tail for the real inter-event time distribution heavier than an exponential distribution found in time-shuffled case. Thus, if the recovery time is large enough, $P_{ij}$ is larger in the real-time data than in the shuffled-time data, leading to smaller spreading cascades.

In another study, Starnini \emph{et al.} modelled random walk processes on empirical temporal networks of face-to-face interactions~\cite{Starnini2012Random}. They found that the random walker explores more slowly the network with longer mean first-passage time on the empirical sequences than that for the mean-field solution assuming Poissonian dynamics. They have argued that the temporal correlations between consecutive conversations constitute a unique reason for this slowing down over the heterogeneously distributed conversation lengths. 

\subsection{Effects of link burstiness}

In the previous Section we have discussed the importance of triggering effects, and causal correlations between events on neighbouring links. However, causal correlations not only appear between events on different links of the same individual, but more commonly they evolve between events on the same social tie~\cite{Kovanen2011Temporal}. Such correlations are responsible for the emergence of long bursty trains of interactions (see Section~\ref{sec:PE} and~\ref{sec:egoburst}), which were shown to be induced by dyadic conversations rather than between a larger group of people. Consequently, causally correlated event trains reflect the characteristics of links rather than those of the nodes~\cite{Karsai2012Correlated}. Here we summarise the studies, which address the effects of bursty links rather than nodes on dynamical processes.

In their modelling study of SIR and SIS processes on a real temporal network, Holme and Liljeros~\cite{Holme2014Birth} considered whether the bursty link dynamics or merely the life span of links matters. Assuming a finite observation time window $T$ they considered two interpretation scenarios of link dynamics: The ongoing link picture assumes that observed link has been created earlier and survived longer than the observation time window. In this case the important temporal structure is the time between events over the link. On the other hand, the link turnover picture suggests that a large fraction of links are created and broken during the observation. This picture is motivated by the observations that the time between the beginning of $T$ and the first event on a given link, and equivalently between the last event and the end of $T$ is longer than one would expect from the observed inter-event time distribution $P(\tau)$. Then the question is whether the life span of links or the precise timing of bursty interactions matters more for the final outcome of the simulated spreading process. They argued by defining null models and performing large-scale numerical simulations on $12$ empirical temporal networks. They found that by assuming that the events on a link occur regularly with the same inter-event time over $T$, i.e., by destroying the inter-event time distribution, the epidemic outbreak size does not change considerably. On the other hand, what matters more are the beginning and ending of the life span of a given link. If these are destroyed by letting all links begin or end simultaneously, the epidemic outbreak size changes radically. This alone does not disqualify the ongoing link picture with burstiness being important in disease spreading, but suggests that the creation and dissolution of ties should also be considered in studying epidemic model as they may considerably affect the final outcome of the epidemics.

In another work by Saram\"aki and Holme~\cite{Saramaki2015Exploring}, they simulated a greedy random walks on $8$ empirical temporal networks. This process is particularly sensitive to temporal-topological patterns involving repeated contacts between sets of nodes. This is evident by the small coverage a random walker takes when compared to a temporal reference model. This shows that in empirical temporal networks greedy walks often get stuck within a small set of nodes. This is because of non-Markovian contact patterns on single links, such as bursty trains of so-called ping-pong callings between two individuals. 

\subsection{Other bursty characters}

As we have discussed earlier, data analysis and modelling studies suggested that \emph{periodic circadian fluctuations} could potentially explain the fat-tailed inter-event time distributions of human interactions. We have also discussed some pro and con arguments and concluded that such periodic patterns may not play deterministic roles. In order to evaluate the exclusive impact of daily patterns on spreading dynamics, some work has been done by using mobile phone communication sequences and random reference models~\cite{Karsai2011Small}. From the call data sequences a weighted aggregated network structure can be obtained by taking individuals (as nodes) and link them if they called each other during the observation period, with link weights defined as the number of their dyadic interactions. In order to study the effects of circadian fluctuations one can use this static structure and generate an interaction sequence on each link by two Poisson processes that conserve the original link weights: One is a homogeneous Poisson process with a constant rate $\lambda$, and the other is a non-homogeneous Poisson process whose instantaneous rate $\lambda(t)$ follows the daily pattern as calculated from the call statistics on the hourly basis. Simulating SI dynamics for both cases reveals that the difference between the spreading curves on networks with homogeneous and non-homogeneous Poisson link dynamics is negligible, demonstrating that the daily pattern has only a minor impact on the spreading speed.

We would like to point out that it is not only the microscopic bursty features that can influence the dynamical processes but also the heterogeneous temporal characters, which appear in the interaction dynamics at the system level. As described in Chapter~\ref{chapter:meas}, the \emph{temporal sparsity} of a network can capture its overall burstiness by measuring for a given time window the effective number of links divided by that for a reference system, where the timings of events on each link are randomised. Perotti~\emph{et al.}~\cite{Perotti2014Temporal} have shown that spreading velocity of an SI process is strongly correlated with the sparsity of the underpinning temporal network. They found that the smaller the temporal sparsity of the network is, the more heterogeneous the bursty temporal patterns appear to be. This has a direct implication on the dynamical processes on a temporal network. They simulated an SI spreading dynamics on various kinds of empirical temporal networks and measured the slow down coefficient defined as the actual spreading speed divided by that for the reference systems. The smaller value of slow down coefficient implies the slower spreading. They observed that the slow down coefficient is almost linearly dependent on the temporal sparsity, which indicates that the burstiness at the system level slows down the spreading dynamics.

Medvedev and Kert\'esz~\cite{Medvedev2017Empirical} studied how \emph{bridging interactions} between nodes in a population speed up the SI spreading on temporal networks of mobile phone communication. They categorised people into three groups: White nodes (customers of the provider with ZIP code), grey nodes (customers of the provider without ZIP code), and black nodes (customers of other providers). For the spreading dynamics they considered only grey and black nodes who have at least two connections to white nodes as they can be identified as bridges for spreading processes between white nodes. They found that such bridges speed up the spreading even if their interactions are bursty, independently of the city population.

\subsection{Dominant characters}

Having discussed various characteristics of bursty behaviour, which were shown to influence the dynamical processes, we yet need to consider which of them are the most dominant. This is not an easy problem and sometimes it turns out to drive to seemingly contradicting observations on different datasets. The real interaction sequences are not only bursty, but also correlated in time and with the interaction structure that consists of communities and ties with heterogeneous activities. All these correlations play some roles simultaneously during the unfolding of dynamical processes. Hence to say something exclusively about the effects of burstiness is challenging.

Once again a straightforward approach to distinguish between the effects of different structural and temporal correlations is provided by random reference models. Comparing simulation results on random reference networks after removing some correlations from the temporal network would tell us which bursty characteristics affect the outcomes of dynamical processes and how. This type of analysis~\cite{Karsai2011Small, Kivela2012Multiscale, Miritello2011Dynamical, Holme2015Information, Holme2016Temporal, Rocha2011Simulated} has shown that the most dominant character for controlling the speed of epidemic spreading is the temporal heterogeneity (burstiness) of interactions. As we have explained via the waiting-time paradox, any level of temporal heterogeneity has an overall slowing down effect. However, when compared to Poissonian systems, bursty interactions accelerate spreading for early times while slowing down for the later time dynamics~\cite{Karsai2011Small, Kivela2012Multiscale, Miritello2011Dynamical, Rocha2011Simulated}. At the same time triggered event correlations were found to be somewhat less dominant in enhancing the spreading behaviour at the early time limit~\cite{Karsai2011Small, Kivela2012Multiscale, Miritello2011Dynamical}, while they were found to slow down the diffusion of a random walker~\cite{Saramaki2015Exploring, Starnini2012Random}. In terms of the structure, the weight-topology correlations were found to be important~\cite{Karsai2011Small, Kivela2012Multiscale, Rocha2011Simulated} as high activity links located inside communities may enhance spreading, while low activity links, which are responsible for bridging communities and connecting the network together, may have the opposite effects by keeping information local due to their infrequent interactions~\cite{Granovetter1973Strength}.

Recently Delvenne~\emph{et al.}~\cite{Delvenne2015Diffusion} have addressed a similar question regarding whether temporal inhomogeneities or structural properties influence more diffusion on a temporal network. To answer this question they provided a mathematical framework to describe diffusion in linear multi-agent systems with $N$ interacting nodes as follows:
\begin{equation}
    D\vec x = L\vec x,
\end{equation}
where the vector $\vec x$ consists of variables $x_i$ denoting the state of node $i$, $L$ is an $N\times N$ matrix describing the interaction structure between nodes, while $D$ captures the time evolution of variables. Assuming a random walker diffusing on the network, the temporal inhomogeneity can be incorporated into a waiting time distribution $\rho(\Delta t)$, implying that the random walker hops from one node $i$ to its neighbouring node $j$ after waiting the time $\Delta t$ on the node $i$. Then the above equation reads in terms of the argument $s$ of the Laplace transform:
\begin{equation}
    \left(\frac{1}{\rho(s)}-1\right)\vec x(s)= \left(\frac{1}{\rho(s)}-1\right)\frac{1}{s}\vec x(t=0) + L\vec x(s),
\end{equation}
with Laplacian $L$. Here the mixing time $\tau_{\textrm{mix}}$, i.e., the relaxation time to stationarity, can be approximated as
\begin{equation}
    \tau_{\textrm{mix}}\approx \max\{\mu\epsilon^{-1},\ \frac{\sigma^2-\mu^2}{2\mu},\ \tau_c\},
\end{equation}
where $\mu$, $\sigma^2$, and $\tau_c$ are respectively the mean, the variance, and the exponential cutoff of the waiting time distribution, while $\epsilon$ is the spectral gap of $L$, representing the structural property of the system. Therefore, the mixing time of  diffusion on such temporal networks can be dominated either by the temporal inhomogeneity or by the structural properties. They analysed several empirical datasets and concluded that in the absence of some temporal correlations, the characteristic times of the dynamics are dominated either by temporal or by structural heterogeneities, as those observed in real-life systems. In systems where correlated temporal patterns are the dominating factor, the aggregation of communities are not necessarily relevant in general, but the temporal characteristics impose the natural description levels of the dynamics.

\section{Dynamical processes on bursty temporal networks}

In the second part of this Chapter we will discuss the representative dynamical processes, which were investigated for bursty systems. Earlier we have discussed the effects of different bursty characteristics on the dynamical processes. Here our focus is more on identifying the dependencies of dynamical processes in the bursty temporal patterns. We will summarise how the process-specific characteristics depend on the bursty behaviour of the underpinning temporal network. We will address five different classes of dynamical processes without going in details about their definitions and critical behaviour. However, we refer the interested reader to books~\cite{Barrat2008Dynamical, Porter2016Dynamical} and a review paper~\cite{PastorSatorras2015Epidemic}, where these processes and their dynamics on static networks are addressed in detail.

\subsection{Epidemic spreading}

As we have already discussed, several epidemic models like SI~\cite{Karsai2011Small, Kivela2012Multiscale, Vazquez2007Impact, Min2011Spreading, Min2013Burstiness, Horvath2014Spreading, Rocha2013Bursts, Gueuning2015Imperfect, Starnini2013Immunization, Jo2014Analytically}, SIR~\cite{Iribarren2009Impact, Miritello2011Dynamical, Rocha2013Bursts, Zhu2014Effect, Holme2015The, Lambiotte2013Burstiness} and SIS~\cite{Lambiotte2013Burstiness, Holme2014Birth} have been studied on bursty temporal networks. These processes are commonly characterised by the infection rate $\tilde \beta$ and the recovery rate $\tilde \mu$. Their long-term dynamics has been described by a ratio $R_0\equiv \tilde \beta/\tilde \mu$, called the basic reproduction number. This ratio gives the average number of infections that a single infected node generates in a population. This number can also be used to characterise whether the process is in a subcritical phase ($R_0<1$), where the epidemic process vanishes spontaneously, or in a supercritical/endemic phase ($R_0>1$), where a considerable fraction of the population is infected to evolve into a stationary state. We have seen earlier that heterogeneous temporal interaction patterns may influence the dynamics of a spreading process, thus it is straightforward to ask how they behave as the function of $R_0$ in a bursty system.

Iribarren~\emph{et al.}~\cite{Iribarren2009Impact} addressed this question by modelling information propagation based on observations from an online email recommendation experiment. They interpreted the spreading process in terms of a Bellman-Harris branching process model. Precisely, in their model the average fraction of infected nodes at time $t$ is written as
\begin{equation}
i(t)=1-G(t)+R_0\int_0^t i(t-\tau_r)P(\tau_r )d\tau_r,
\label{eq:itBH}
\end{equation}
where $G(t)=\int_0^tP(\tau_r)d\tau_r$ is the cumulative distribution of the residual time distribution. They showed that for processes with $P(\tau_r)$ decaying slower than exponential, including bursty processes with log-normal and power-law tails, if $R_0<1$ then Eq.~(\ref{eq:itBH}) is reduced to $i(t)\sim \frac{1-G(t)}{1-R_0}$. This indicates that the spreading depends mostly on those individuals whose residual time is the longest. Thus temporal heterogeneity has a profound impact on the dynamics of information spreading. It does not depend on the mean value of $\tau_r$ but on the tail of its distribution $G(t)$, which drastically slows down the propagation of information. Interestingly, large temporal heterogeneity has the opposite effect above the epidemic threshold ($R_0 > 1$). In this case the Bellman-Harris model predicts an initial exponential growth of the epidemic spreading where information shows faster spreading than expected.

In another work, Miritello~\emph{et al.}~\cite{Miritello2011Dynamical} studied the effects of heterogeneous residual times and triggered events on SIR processes. As we discussed earlier in Section~\ref{sec:triggev}, they found that in random networks the basic reproduction number\footnote{In their work Miritello \emph{et al.}~\cite{Miritello2011Dynamical} called the basic reproduction number as the secondary reproduction rate, $R_1$, and defined as the average number of secondary infections produced by an infectious individual, which is the definition of $R_0$. Moreover, in their definition they referred to other works~\cite{Barthelemy2004Velocity,Newman2002Spread}, which concerns $R_0$, thus we decided to adopt the notation $R_0$ in Eq.~(\ref{eq:MirBRN}), rather than $R_1$ as in the original paper.} is dependent on the transmissibility, defined in Eq.~(\ref{eq:transmiss}), as
\begin{equation}
R_{0}(\tilde \beta,\Delta t)=\frac{\langle (\sum_j \mathcal{T}_{ij})^2\rangle_i - \langle \sum_j \mathcal{T}^2_{ij} \rangle_i}{\langle \sum_j \mathcal{T}_{ij} \rangle_i}.
\label{eq:MirBRN}
\end{equation}
In case of homogeneous dynamics ($\mathcal{T}_{ij}=\mathcal{T}$) Eq.~(\ref{eq:MirBRN}) recovers the common result $R_0=\mathcal{T} (\langle k^2_i \rangle / \langle k_i \rangle -1)$ found in random networks. This is an important result as $R_0$ can be used to determine the critical point of the SIR spreading even in bursty systems, while its value scales proportionally with the speed of the epidemics.

Rocha~\emph{et al.}~\cite{Rocha2013Bursts} addressed various characteristics of spreading processes as a function of the system's stationarity and its temporal heterogeneity. In their systematic study they defined a temporal network model where nodes are activated by an independent renewal process with exponential (homogeneous case) or power-law (heterogeneous case) inter-event time distributions and contact each other randomly. In addition they introduced node turnover processes by replacing nodes with disconnected new ones with a given rate in order to ensure that the system reaches a stationary state. As for the fabric of this temporal network they simulated SI and SIR models and measured the peak and volume of prevalence, and then estimated $R_0$ and the distribution of the epidemic outbreak sizes. They showed that the prevalence curve at the early stages is characterised by a faster and steeper growth of infected nodes in the case of heterogeneous contact patterns, while at the later stages its characteristics depend more on the epidemics model, turnover rate, and other parameter values. In the absence of replacement of nodes, the prevalence of the infection is generally higher for homogeneous contact patterns, however, for later times the heterogeneous contact patterns slow down the spread of the infection. For some configurations of the SIR dynamics, heterogeneous patterns provide a way to decrease the global impact of the epidemic. In terms of $R_0$ they found that it depends both on the heterogeneity and the node turnover rate of the network. In general, heterogeneous temporal patterns tend to result in higher values of $R_0$, with the exception in the case of stochastic SIR dynamics with the infection probability around $1$. Note that a similar picture has been presented by Zhu~\emph{et al.}~\cite{Zhu2014Effect} from SIR simulation results on temporal scale-free networks.

Gueuning~\emph{et al.}~\cite{Gueuning2015Imperfect} studied the SI spreading process on temporal networks but with the probability $p$ for the successful infection. This $p$ is essentially related to the infection rate $\tilde\beta$ and it also determines the average residual time as
\begin{equation}
\langle \tau_r \rangle =\frac{\langle \tau^2 \rangle}{2\langle \tau \rangle} + \frac{1-p}{p} \langle \tau \rangle.
\end{equation}

In case when $p=1$, the deterministic SI process is recovered, where the spreading dynamics is determined by the waiting-time paradox as mentioned with Eq.~(\ref{eq:taurderiv}) in Section~\ref{sec:iet_rt_wt}. On the other hand, if $p<1$, the slowing-down effect of heterogeneous interaction dynamics becomes weaker. As $p\to 0$, what determines the spreading is the average residual time rather than the tail part of residual time distribution. In addition, the transmissibility of interactions decreases for the increasing $p$ in bursty cases, indicating their important effects in hindering the epidemic spreading.

Another important characteristics of an SIR spreading is the quantity $\Omega$ that describes the fraction of infected nodes in the population after the outbreak has passed and the process reached its disease-free absorbing state. In static networks a unique deterministic relation exists between $R_0$ and $\Omega$, while Holme~\emph{et al.}~\cite{Holme2015The} have found that the relation is violated in case of temporal networks. They showed that the different pairs of $\tilde\beta$ and $\tilde\mu$, leading to the same value of $R_0$, may lead to different outbreak sizes. Hence the question is which structural and temporal features of a temporal network determine the most the correlation between $\Omega$ and $R_0$. It has been found that as a temporal quantity, the burstiness parameter $B$ (defined in Chapter~\ref{chapter:meas}) determines dominantly the correlation between these two quantities. Results showed that the more heterogeneous the inter-event time distribution is, the less predictable the value of $\Omega$ is from the corresponding $R_0$.

Finally, it should be noted that there have been some studies considering bursty behaviour in order to design efficient immunisation strategies. While only system-level effects of burstiness have been addressed by using random reference models in Ref.~\cite{Starnini2013Immunization}, an immunization strategy has been proposed in Ref.~\cite{Lee2012Exploiting}, which exploits heterogeneous temporal behaviour by immunizing the last interacting neighbour of a randomly selected node at a random time. This strategy has been shown to be effective in data recording face-to-face interactions, where the turnover of relationships is large.

\subsection{Random walks}

Random walks serve as a model dynamics that has extensively been used to study bursty temporal networks. As we have discussed in Section~\ref{sec:OE}, a special model variant called \emph{greedy random walks} has lately attracted much attention as it is defined on temporal networks and its dynamics is sensitive to temporal heterogeneity. The temporal network is commonly defined as a static structure with interaction dynamics on links defined as renewal processes with an arbitrary inter-event time distribution but with parameters characteristic to each link. A single greedy random walker is diffusing on such a network by moving between nodes via temporal interactions whenever it is possible, i.e., after arriving to the node $i$ it always takes the first emerging link to move to another node. Two variants of greedy random walks have been proposed by Speidel~\emph{et al.}~\cite{Speidel2015Steady}: 
\begin{enumerate}[(a)]
    \item In case of the \emph{active random walk}, after the walker arrives at a node $i$, it re-initialises the inter-event times of all the links. Then, the residual time, i.e., the time a walker waits on a node before the link appears, is equivalent to the inter-event time.
    \item In case of the \emph{passive random walk}, the re-initialisation of each link is not assumed. Instead, a new inter-event time is chosen only for the link through which the walker arrived at the node. Then the transition rates of the passive random walk depend on the trajectory that the walker has taken, implying that one has to account for the entire trajectory of the random walker to accurately evaluate its behaviour~\cite{Speidel2015Steady}. 
\end{enumerate}
Note that if the dynamics of links are driven by Poisson processes with exponentially distributed inter-event times, the active and passive random walks are identical and reduce to the usual continuous-time random walk on the static network.

\subsubsection{Active random walks: generalised mean-field equations}

In order to characterise these two models, we are interested in their steady state behaviour and the mean recurrence time $\langle T_{i|i}\rangle$, i.e., the average duration it takes for the random walker having initiated from the node $i$ to return to $i$ for the first time. The steady state behaviour of the active random walk problem was studied by Hoffman~\emph{et al.}~\cite{Hoffmann2012Generalized, Hoffmann2013Random}. They introduced the probability that a random walker makes a step from node $j$ to $i$ accounting for all other processes on $j$ in a similar way as in Eq.~(\ref{eq:rwtransp}) as
\begin{equation}
T_{ij}(\tau)=P_{ij}(\tau)\times \prod_{k\neq i}\left( 1- \int_{0}^{\tau}P_{kj}(\tau')d\tau'\right),
\end{equation}
where $P_{ij}(\tau)$ denotes the distribution of waiting times on a link between nodes $i$ and $j$. Using this probability they introduce a generalised Montroll-Weiss master equation~\cite{Montroll1965Random} describing the evolution of the probability mass function $n_i(t)$ for a walker to occupy node $i$ in time $t$. In general this can be written as
\begin{equation}
n_i(t)=\int_0^t \phi_i(t-t')q_i(t')dt',
\end{equation}
where $q_i(t')$ is the probability that the walker arrived at the node $i$ in time $t'\leq t$ weighted by the probability $\phi_i(t-t')$ of not leaving the node since then. The Laplace transform reduces $n_i(t)$ to a product in the Laplace space as
\begin{equation}
\hat{n}_i(s)=\hat{\phi}_i(s)\hat{q}_i(s).
\label{eq:nlt}
\end{equation}
Here an expression for $\hat{\phi}_i(s)$ can be obtained by taking the probability distribution $T_i(t)=\sum_{j=1}^{N}T_{ij}(t)$ to make a step from node $i$ to any other node, which leads to the probability density function of remaining at $i$ for a time $t$:
\begin{equation}
\phi_i(t)=1-\int_0^t T_i(t')dt'
\end{equation}
with a Laplace transform as
\begin{equation}
\hat{\phi}_i(s)=s^{-1}(1-\hat{T}_i(s)).
\label{eq:philt}
\end{equation}

One can obtain an expression for $\hat{q}_i(s)$ by considering that $q_i(t)=\sum_{k=0}^{\infty}q_i^{(k)}(t)$ where $q_i^{(k)}(t)$ is the probability to arrive at the node $i$ in time $t$ in exactly $k$ steps. Taking its Laplace transform and summing it over all $k$ the authors yield
\begin{equation}
\hat{q}(s)=\left( I - \hat{T}(s)\right)^{-1}n(0),
\label{eq:qlt}
\end{equation}
where $I$ is the identity matrix and $q$ and $n$ are vectors. After substituting Eq.~(\ref{eq:philt}) and Eq.~(\ref{eq:qlt}) into Eq.~(\ref{eq:nlt}) they obtained a generalised Montroll-Weiss master equation~\cite{Montroll1965Random} that applies to arbitrary network structures:
\begin{equation}
\hat{n}(s)=s^{-1}\left( I-\hat{D}_T(s)\right)\left( I-\hat{T}(s)\right)^{-1} n(0),
\end{equation}
where the components of the diagonal matrix are given as $\left( \hat{D}_T \right)_{ij}(s)\equiv \hat{T}_i(s)\delta_{ij}$. Taking the inverse Laplace transform leads to
\begin{equation}
\frac{dn}{dt}=\left( T(t) \ast \mathcal{L}^{-1}\left\{ \hat{D}^{-1}_T(s)\right\}-\delta (t)\right)\ast K(t)\ast n(t)
\end{equation}
where $\mathcal{L}^{-1}$ denotes the inverse Laplace transform and $\ast$ is the convolution respect to time. Here the memory kernel $K$ characterises the amount of memory in the system. Because of the convolution they conclude that the temporal evolution of $n_i(t)$ depends on the states of the system at all times since the initial setting. For further details on the derivation see Refs.~\cite{Hoffmann2012Generalized, Hoffmann2013Random}.

The authors further obtained an effective transmission matrix $\mathbb{T}_{ij}$ for the whole network and found that if the dynamics of links are dictated by a Poisson process, a random walk on the temporal network is equivalent to a Poisson continuous-time random walk on a static network with links weighted by the number of interactions. They also concluded that in the Poissonian case the stationary solution of the random walk is a uniform vector. In contrast, if the dynamics of links is non-Poissonian, e.g., bursty, the stationary solution appears only in the limit $\tau_r \to \infty$ and it is not uniform. In terms of mean recurrence time, Speidel~\emph{et al.}~\cite{Speidel2015Steady} found that if inter-event times on different links are identically distributed then $\langle T_{i|i}\rangle \propto 1/k_i$, i.e., it is inversely proportional to the degree of nodes, thus determined by the structure and not by the dynamics of the network.

\subsubsection{Passive random walks}

In case of the passive random walk problem, the inter-event and residual time distributions are not identical but related to each other as shown in Eq.~(\ref{eq:rst_iet}). If $P(\tau_r)$ has a heavy tail, the inter-event time picked for the last active link, which transferred the walker from node $j$ to node $i$, will most likely to be shorter than the residual times on other links of the node $i$, leading that the walker will most likely return back to node $j$. This behaviour of getting stuck in conversations between two nodes has somehow already been observed empirically by Saram\"aki and Holme~\cite{Saramaki2015Exploring}. This mechanism makes the system non-Markovian as the destination of the walker at any node $i$ depends on its origin and not only its actual state. Furthermore, it was shown that unlike for the active random walk, the approximated steady state of the passive random walk is the uniform distribution for any network and distribution of inter-event times. Neither in this case the mean recurrence time depends on the distribution of inter-event times as it appears as $\langle T_{i|i}\rangle \propto N\langle \tau \rangle/k_i$. It has also been shown that the active random walk produces smaller mean recurrence times for each node than the passive walk does when the inter-event time follows the power-law distribution. In contrast, the mean recurrence times are larger for the active random walk than for the passive random walk when inter-event time follows a less heterogeneous Weibull distribution.

\subsection{Threshold models}

Threshold-driven contagion models define a family of dynamical processes, where the infection of an individual is conditional to some individual threshold of pathogen concentration or social influence, etc. In these systems individual thresholds together with the temporal and topological structure of the network determine the spreading dynamics. This is fundamentally different from the case of epidemic spreading, where the process is stochastic and controlled by a single rate of infection, characteristic to the modelled disease and not to the individual. Threshold models are important not only due to their epidemiological relevance, but also because they capture mechanisms that are recognised to drive social contagion phenomena, such as the spreading of memes, adoption of innovations, and decisions to join collective actions~\cite{Centola2007Complex}.

A widely known threshold model for static networks was proposed by Watts~\cite{Watts2002Simple}, which was recently extended for temporal networks~\cite{Backlund2014Effects, Karimi2013Threshold, Karimi2013Temporal, Karimi2015Tightly}. The model of temporal networks assumes that nodes can be in two mutually exclusive states, susceptible or infected (also called adopted). Initially each node is susceptible except a randomly selected seed node, which is set to be in infected state. During simulations we follow the set of contacts in timely order and let each contact be an opportunity for the nodes to learn about the state of their neighbours and to potentially change state. A node $i$ changes from susceptible to infected state if the $\phi_i$ number (or fraction) of its observed infected neighbours overcomes a given threshold $\Phi$. However, nodes remember the state of their observed neighbours only for a finite time window $\theta$. Thus a node gets infected at time $t$ only if it has $\phi_i>\Phi$ within a time frame $[t-\theta,t]$.

Karimi~\emph{et al.}~\cite{Karimi2013Threshold} have studied two versions of this model simulated on six different empirical temporal networks and on the corresponding random reference models where they shuffled the interaction times to eliminate burstiness. In one case, they defined $\phi_i$ as the fraction of infected neighbours among all neighbours observed in $\theta$ and found that the size of the infection cascade decreases by $\theta$. As they explained, longer memory time window means larger number of observed neighbours who are mostly susceptible in the beginning of the process, thus they decrease the probability of infection of the central node. In addition, they also showed that burstiness slows down the emergence of infection cascades. On the other hand, in the model variant where they define $\phi_i$ as the absolute number of infected neighbours observed in $\theta$, the cascade size increases with $\theta$ and cascades evolve faster due to burstiness in most of the empirical networks.

Backlund~\emph{et al.}~\cite{Backlund2014Effects} have studied yet two other model variants, where they assumed that $\phi_i$ is defined as the fraction of a number of infected neighbours of node $i$ with static degree $k_i$ observed in $\theta$. They simulated the process on large empirical temporal networks of mobile calls, SMS, emails, and face-to-face interactions. Similarly to Karimi~\emph{et al.} they used the time-shuffled random reference model to address the effects of burstiness and in addition a random offset model (see Section~\ref{sec:triggev}) to eliminate triggered event correlations. In one model variant, which they called \emph{stochastic threshold model}, they assumed a linear correspondence between $\phi_i$ of node $i$ and the probability of its getting infected. Although the threshold rule does not directly count the number of contacts from the same adopted neighbour, these interactions still contribute indirectly because the stochastic rule is activated whenever a contact occurs. The authors observed that this indirect effect of burstiness hinders the infection rate because of increased waiting times on links and redundant repeated events. Here multiple adopted neighbours drive the adoption, which are unlikely if the bursty periods evolve only on links or between limited number of people. In this case time shuffling destroys burstiness, and spreads events more evenly across time, thus giving rise to an increased number of temporal paths ending at nodes within short time windows.

History-dependent contagion is a slightly different type of threshold model, which was studied by Takaguchi~\emph{et al.} on empirical bursty temporal networks~\cite{Takaguchi2013Bursty}. In this model each node $i$ of a network is assigned with an internal variable $\nu_i$, which represents, e.g., the concentration of pathogen in the individual, or her actual interest in adopting something. An initially susceptible node becomes infected once the concentration $\nu_i$ reaches a threshold $\nu_{\textrm{th}}$, and keeps this state until the end of the process. The initially zero concentration of a node, i.e., $\nu_i=0$, is increased by unity each time the node interacts with an infected neighbour, and it is decreasing exponentially with the rate $\tau_d$, otherwise as the function of the time between consecutive interactions (for more precise definition, see Ref.~\cite{Takaguchi2013Bursty}). In order to study the effect of burstiness on this process, simulations have been carried out on real face-to-face interactions and email networks and on corresponding random reference systems where all temporal heterogeneity were removed by shuffling interaction times. By measuring the final infection size as a function of $\nu_{\textrm{th}}$ and $\tau_d$ in both real and randomised networks, it has been shown that in the original bursty temporal network the spreading evolves faster, it reaches more nodes in its final state, and the parameter space of global contagion is expanded, all compared to the reference systems. However, the reachability ratio~\cite{Holme2012Temporal} of nodes were found to be smaller. As the authors explained, this inconsistency may be caused by the competition between two opposite effects by randomisation, which increases the reachability ratio of each node to enhance spreading but eliminates the burstiness to suppress the epidemic. 

\subsection{Evolutionary games}

There is yet another set of dynamical processes that have been studied on bursty temporal networks, namely different evolutionary games. All of the related studies employed empirical temporal networks and random reference models to understand how bursty temporal patterns affect the emergence of cooperation. Cardillo~\emph{et al.}~\cite{Castellano2009Statistical} studied the Hawk-Dove game and the Prisoner's dilemma on face-to-face interaction sequences. They used a snapshot representation of the temporal networks~\cite{Holme2012Temporal}, such that each snapshot represents the set of interactions appearing within a unit time period between any individuals in the dataset. A random reference model was defined by shuffling the snapshots, which provided a null model where the number of interactions per link and circadian fluctuations were kept, but temporal heterogeneity and event correlations were destroyed. Simulating games on the original and shuffled temporal networks showed that the temporal dynamics of social ties has a dramatic impact on the evolution of cooperation. In fact they showed that the dynamics of pairwise interactions favours selfish behaviour, and the cooperation is seriously hindered when the agent strategy is updated too frequently with respect to the typical time scale of agent interaction, and when realistic link temporal correlations are present.

Similar conclusions were drawn by Li~\emph{et al.}~\cite{Li2016Evolution} who studied the Prisoner's dilemma on similar real temporal networks. First they showed that the temporal network enhances the emergence of cooperation when compared to corresponding static structures. They also concluded that removing burstiness by shuffling interaction times in the dataset leads to improved cooperations. Thus they found that burstiness actually slows down the emergence of cooperations just like in case of many other dynamical processes.

\subsection{Dynamical process induced bursty behaviour}

Finally we would like to mention some studies, which propose potentially reversed situations, where instead of burstiness influencing dynamical processes, it is induced by them. More precisely, it has been shown that certain processes, such as the adoption of products or information, may induce bursty patterns in the interaction behaviour of individuals. We have already discussed one model study of Ref.~\cite{FernandezGracia2013Timing} in Section~\ref{sec:otherindivmods}, where in a voter model due to exogenous and endogenous update rules bursty patterns of update frequencies occurred at the individual level. However, one can find real world examples for similar phenomena. Kikas~\emph{et al.}~\cite{Kikas2013Bursty} showed that in the online social network of Skype, the link creation dynamics of individuals evolve through long bursty trains, which are commonly triggered by the adoption of different services. This in turn evolves as a complex contagion process on the fabric of the emerging network~\cite{Karsai2016Local}.

In another study, Myers~\emph{et al.}~\cite{Myers2014Bursty} arrived at a similar conclusion by analysing link creation and removal bursts in Twitter. They identified bursts by comparing link addition rates to the average daily activity curves. They argue that link creation bursts are commonly induced by external processes like retweet-bursts, content download, or protests and helped people to evolve a more homogeneous egocentric network in terms of interest. Furthermore, they showed that most of the new links created in bursty periods closed triangles in the network, thus were responsible to shape the structure locally to help the formation of communities.

Finally another direction of modelling was recently proposed by \'Odor~\cite{Odor2014Slow}, showing that interaction dynamics of systems in the critical Griffiths phase \footnote{In a critical system disorder can smear the phase transitions, making a discontinuous transition continuous or generating Griffiths phase, in which critical-like power-law dynamics appears over an extended region around the critical point.} exhibits slow bursty dynamics with power-law interconnection times. Various static and dynamic network topologies including one-dimensional rings, generalised small-world, and ageing scale-free structures were considered. On the top of these structures dynamical processes were simulated, such as the contact process or susceptible-infected-susceptible (SIS) dynamics, which are all known to exhibit a Griffiths phase when the topological disorder exists. It was shown that the inter-communication time between neighbouring agents appears with a power-law tail with various exponent value depending on the system considered. These observations suggest that in the case of non-stationary bursty systems, the observed non-Poissonian dynamics can emerge as a consequence of an underlying hidden Poissonian network process that is either critical or exhibits strong rare-region effects.


\chapter{Discussion}
\label{chapter:concl}

In this monograph we have presented an up-to-date overview of dynamical systems of human behaviour that show bursty phenomena. These systems evolve through inhomogeneous temporal event sequences with periods of high event frequency alternating with low frequency periods. It is this dynamical feature that makes such systems very interesting yet very challenging to understand and explain. Systems that show bursty behaviour can not be characterised just as a Poisson process with a single temporal scale and exponential inter-event time distributions. Instead, bursty systems show strong temporal heterogeneities such that their dynamics is deemed to be non-Poissonian with broad inter-event time distributions. 

Indeed, the quest to understand the bursty behaviour is interesting because it occurs in a variety of systems of Nature but also in man-made systems. One of the best-known examples of bursty behaviour is the dynamics of earthquakes, where the shocks at a given location appear burstily with the frequency of aftershocks decreasing as a power law and leading to a broad inter-event time distribution of shocks. From the theoretical point of view the stochastic processes underlying this and some other apparently very different phenomena like solar flares show universal features having the distributions of sizes, inter-event times, and temporal clustering, explainable in general by the theory of self-organised criticality (SOC). An example of a bursty system at different scale is a single neuron or group of neurons firing spike trains with high frequency separated by intervals of low frequency of activities, which is proposed to be the result of integrate-and-fire mechanism, commonly assumed in case of SOC systems. Further examples of bursty patterns can be found in switching between contrasting activities as in case of sleep-wake patterns of animals and humans or stop-start motion of fruit flies.

These few examples of systems showing bursty temporal patterns were presented to define theoretical concepts and develop models towards understanding their behaviour. However, in case of human bursty behaviour these concepts and models may be different especially when it comes to the behaviour of a collection of mutually linked individuals forming a social connectome~\cite{SocialConnectome} or a social network. This issue of connectivity constituted yet another dimension and a challenge to study bursty temporal patterns of human sociality in terms of quantitative analysis and of phenomenological and quantitative theory. These systems have recently become attainable to quantitative studies due to digital communication technologies through which most human socio-economic transactions now occur and are recorded in large datasets. 
 
At the behavioural level the timings of individual actions present heterogeneous temporal patterns, while similar dynamics was observed in dyadic social interactions between individuals, or in collective social phenomena of groups, communities and societies. One of the first observations of this kind was made in a study of email correspondence that reported a broad inter-event time distribution with a power-law tail~\cite{Eckmann2004Entropy} and was explained using a priority queuing model~\cite{Barabasi2005Origin}. 
 
 At the group or societal level, one has observed bursty dynamics e.g., in the emergence of causal temporal behaviour motifs, the evolution of mass demonstrations, revolutions, global information cascades, and even wars of various kinds. In all these cases of human bursty phenomena there is a challenge to characterise and model them in a unified way. A step forward to this direction has been the proposition of human bursty behaviour belonging to one of the two universality classes with two different exponent values characterising the power-law inter-event time distributions and queuing models. However, this picture turned out not to be complete as further empirical evidences from some bursty systems were found to give rise to various different exponent values.

In describing the human bursty behaviour the perspective of the priority queuing model is that the bursty patterns are consequences of people prioritising their tasks in the order of perceived importance, inducing intrinsic correlations between different tasks and resulting in bursty patterns of completed activities. Alternatively human behaviour is considered to be driven by external factors like circadian and weekly cycles without any intrinsic correlations, introducing a set of distinct characteristic time scales and giving rise to heavy tails due to alternating homogeneous and non-homogeneous Poisson processes. As further alternative approaches in describing bursty patterns one has assumed strong correlations between consecutive events and employed memory functions, or self-exciting point processes, or reinforcement mechanisms. Yet there are other models that have been proposed to describe human bursty behaviour based on self-organised criticality, or local structural correlations, or random walk, or contact process, or voter model process to introduce heterogeneous temporal patterns at the individual or system level. 

The richness of emergent features in human bursty dynamics has generated the development of methodologies and models to ask even more complex scientific questions about the effects of non-Poissonian patterns of individuals on collective dynamical processes. A typical example is diffusion of information in a temporal social network, where individuals interact burstily but are connected together in a network where information can diffuse globally. Beyond the conventional modelling and simulation techniques of such processes, data-driven models and random reference systems were recently shown to be successful in addressing such questions.

As is evident from the above not at all comprehensive set of examples, there is still a lot of open directions to take towards the better understanding of the underlying mechanisms and processes that lead to burstiness appearing in the systems of human dynamics. In this monograph we have embarked on to the road of addressing these questions of the underlying mechanisms of bursty behaviour in terms of data analysis as well as using various theoretical and modelling approaches. In order to make our research endeavour of human bursty behaviour as a logically proceeding narrative we organised our review in six Chapters, including this Chapter. Starting with a general introduction in Chapter \ref{chapter:intro} we provided a broader overview on bursty phenomena observed in Nature and in human dynamics together with the general motivations and organisation principles for this monograph.

In Chapter \ref{chapter:meas} we presented the theoretical description and characterisation of bursty human dynamics. Starting from the description of discrete time series we went through all characteristic measures, like inter-event time distribution, burstiness parameter, memory coefficient, bursty train size distribution, autocorrelation function, and so on, which were borrowed or introduced to describe human bursty systems from the individual to the population level. With these quantities, we showed how to detect the temporal inhomogeneities and long-range memory effects in the event sequences of human dynamics. At the same time we also introduced methods of system-level characterisation, mainly in the frame of temporal networks, which have been intensively studied in recent years to describe temporal human social behaviour. Finally, as human dynamics intrinsically shows the cyclic patterns like the daily and weekly ones, the methods for deciphering the effects of such cycles were also described. 

In Chapter \ref{chapter:emp}, we made a comprehensive collection of a large number of empirical observations of human bursty systems recorded in various situations and stored in a number of datasets. We divided these observations into two main categories, i.e., individual activities and interaction-driven collective activities. In addition, we briefly discussed examples from human mobility, financial systems, and animal behaviour. Precisely, as for the interaction-driven case, we sorted out the empirical findings from different social interaction modalities like face-to-face interactions, mobile-phone based interactions, communication by posted letters and emails to web-based social interactions, as they may reflect the different degree of sociality between a pair of individuals. To make the overview for the reader easier to follow such a large set of empirical studies, we presented a systematic summary of all these observations in tables including a short description of each dataset, the observed values of some bursty characters, and the references to the original works.

Next in Chapter \ref{chapter:model} we summarised the main modelling directions, which have been studied for the understanding of the emergence of bursty human behaviour. We addressed three main modelling paradigms concerning priority queuing models, reinforcement and memory driven processes, and Poisson models of bursty phenomena. In addition, we summarised less recognised modelling directions together with random reference models, which have been used lately to highlight the effects of burstiness and temporal correlations in empirical event sequences. Furthermore, we discussed several generative models at the individual level, where activity dynamics of a single person were in focus, but we also summarised models of bursty dyadic interactions, and network models with emergent bursty behaviour.

In Chapter \ref{chapter:processes} we summarised studies addressing any type of dynamical processes of bursty human interaction networks. Bursty human interactions have indisputable consequences on dynamical processes, as their heterogeneous timings largely control the possible transmission of any kind of information between interacting peers, or the timely connectedness of the temporal structure. To give a comprehensive review we first discussed all possible bursty characters like the inter-event and residual time distributions, ordering of events, triggered event correlations, node and link burstiness, etc., which were shown to play important roles in the early and late time behaviour of collective dynamical phenomena. In the second part of Chapter \ref{chapter:processes} we went through all the main families of dynamical processes studied so far in bursty interaction networks to understand how process specific behaviour is dependent on the heterogeneous dynamics.

\section{Future directions and methodological approaches}

As stated before this monograph is meant to serve as an up-to-date overview of what has been learned so far about human bursty phenomena. However, we need to ask what is next, what can one learn more, should one try to combine different perspectives, and what should one focus on? Here our perspective to study human burstiness has largely been that of statistical physics, at least when it comes to methodology. This includes the analyses of various kinds of small or large datasets to learn about the dynamical characters and other basic properties of human burstiness at the level of individuals, pairs of individuals, and networks of individuals. This is followed by the theory, building plausible models, and doing the actual computational modelling to understand and explain how these properties at different levels could have emerged. This is also important in describing on one hand the processes leading to burstiness and on the other the dynamics of it. This perspective is very much data-driven but also data-limited, since we are dependent on the availability of data.

One of the recent ICT-related developments in studies of social systems is continuous and automated app-based data collection using smart mobile phones and wearable devices~\cite{Aledavood2015Daily,Stopczynski2014Measuring,Eagle2006Reality,Karikoski2010Measuring}. As today's smart phones include a number of sensors, it is possible to continuously collect lots of different types of data from single individuals, such as their activity times by monitoring phone screen on/off sequences, location, accelerometer data, calls, messages, data from apps and services usage, social network data (e.g., Facebook and Twitter), and data from wearables. Together with the  smart phones' built-in facility to make online surveys and questionnaires, it is possible to collect qualitative and quantitative truly multidimensional ``social diary" data from a single individual and from interacting individuals in groups or in larger social networks. The smart phone based research approach can join seemingly different viewpoints to eventually one research perspective, that could be called ``computational social science" and once again aiming at getting even deeper understanding of the processes involved in human behaviour in general and human bursty behaviour in more detail.

As for data-driven observations of human dynamics, especially when the system behaves burstily, we raise yet another open question, namely the stationarity of the process. The reason for this is the fact that human activities are predominantly driven by circadian rhythms with a characteristic time scale of one day, meaning that every day a new process is started. This is particularly important if the inter-event time distribution, the dynamic system produces, is fat tailed, because then it takes infinite time for the process to become stationary. This in turn induces a non-stationary bursty dynamics at the individual level, but also at the network level where it is entangled with the evolving network structure with created and broken ties and with nodes arriving and leaving the system. Hence this question is indeed crucial at any level of human dynamical systems and ongoing dynamical processes. However, despite its importance, the non-stationarity has so far neither been well-demonstrated from data nor been systematically considered in the framework of modelling, except in few cases~\cite{Vazquez2007Impact,Guo2011Weblog,Rocha2013Bursts,Horvath2014Spreading,Delvenne2015Diffusion,Krings2012Effects}. Thus the non-stationary nature of the dynamics of the system together with the somewhat better characterised feature of higher-order temporal correlations between events and inter-event times still remain as questions to be answered for more comprehensive and deeper understanding of bursty human dynamics.

Our quest has so far been mostly concentrated on gaining understanding and explaining human bursty behaviour based on direct observations and analysis of related data. In order to learn more there is need to reach out and consider whether bursty behaviour in other complex systems show behavioural similarities, patterns, and universalities rather than differences, variation, and specifics. The former viewpoint is considered Platonic and it is pondered to be akin to physics while the latter is considered Aristotelian and pondered to be akin to biology~\cite{Ball2017Complexity}. In studies of complex systems both viewpoints are of course needed as they complement each other. According to the Platonic view one tends at least implicitly to assume---on the basis of observed regularities---that there should be some kind of, yet uncovered, governing laws that would lead to the behaviour of the complex system or at least give us some insight what kind of plausible processes or mechanisms could be involved. As these regularities appear in various complex natural, social, and man-made systems and at different scales, one is led to believe that there are observable similarities that can be characterised with the same type of mathematical relationships, scaling laws, and behavioural models. With this type of over-arching perspective one could take the next step to uncover the similarities and even possible universalities as well as some kind of governing principles of human bursty behaviour at the multiple levels ranging from individuals to social networks. This could be achieved on one hand with data-analytics approach and on the other hand with computational modelling, which could be seen to constitute a physics approach to decipher human bursty behaviour in terms of structure, function, and response.

However, at this point one should ask whether the above described Platonic viewpoint of research is too one-sided. Is it too simplistic and overly self-assured as well as falling short of addressing some key properties of human burstiness while ignoring some of the possibly important details of the complex system of interest? This is a specifically relevant question in case of humans who can be observed as individuals or as members of larger social networks of various kinds as well as in a society with all sorts of cultural and socioeconomic ramifications. These issues are traditionally the realms of Cognitive or even Neurocognitive Science, Psychology, Social Psychology, Social Sciences as well as even Political Science, carrying their own research perspective(s) and methodological approaches. In them the details, especially behavioural differences, variation, and specifics matter thus making the perspective more akin to Aristotelian viewpoint. These systems are studied using various kinds of experimental methodologies to observe individual behaviour and alternatively surveys targeted to groups of individuals.

This Aristotelian approach may provide also advances in the future. By focusing on individuals and pairs of individuals using various brain research methodologies, one may be bound to get deeper insight into more of the ``micro'' level properties of human burstiness. In addition in case of small groups using observational cognitive research methodologies one may understand better human burstiness as part of social gathering. Although these experiments, due to being specifically set up for certain purpose, carry a kind of ``in vitro'' flavour, they help us to build more realistic models of human dynamics. The same can be said about the survey or questionnaire studies of hundreds or thousands of individuals, which can be carried out with new digital platforms. Such studies may have features of individual subjectivity, which can be tuned to investigate better certain social situations in a controlled way. So rather than saying that one of these two viewpoints, either Platonic or Aristotelian, is more important than the other, we emphasise their equal importance and their mutual methodological complementarity in building deeper insight into human behaviour and its dynamics in general. This kind of complementary and joint Platonic and Aristotelian perspective can be expected to shed light to the governing laws (if any) or functional rules of human bursty behaviour and to possible behavioural similarities and universalities. 


\pagebreak
\section*{Acknowledgement}
\small{Although this monograph has three authors it was written with the direct and indirect help of many others. First of all we owe many thanks to our collaborators who motivated and followed us to explore aspects of human bursty dynamics. They are Jari Saram\"aki, J\'anos Kert\'esz, Albert-L\'aszl\'o Barab\'asi, Mikko Kivel\"a, Lauri Kovanen, Raj Kumar Pan, Juan I. Perotti, Dashun Wang, Chaoming Song, Ginestra Bianconi, Nicola Perra, Enrico Ubaldi, Raffaella Burioni, Alessandro Vezzani, Riivo Kikas, Marlon Dumas, Eunyoung Moon, and Eun-Kyeong Kim. H.-H.J. personally thanks Woo-Sung Jung and Seunghwan Kim for their support.

We are especially thankful for Jari Saram\"aki and J\'anos Kert\'esz for the insightful discussions and for reading the manuscript and helping us to improve the clarity of the content and text of our monograph.

We are also very grateful for our editors Elisabeth A.L. Mol, Annelies Kersbergen, and Stephen Soehnlen, who accommodated our monograph at Springer from the beginning and provided support and useful advices throughout the writing and publication process. We would also like to thank to our anonymous reviewers who provided us several constructive comments at different stages of the process and helped us to open the perspective of our monograph.

M.K is thankful for the DANTE Inria team from the Laboratoire de l'Informatique du Parall\'elisme at the Ecole Normale Sup\'erieure de Lyon and the IXXI Rh\^one Alpes Complex System Institute to assure the stimulating environment. H.-H.J. and M.K are both thankful for the multiple visiting research grants from the Aalto Science Institute, which helped largely their joint work despite large distances. We are also thankful for the Complex Networks research team at the Department of Computer Science at Aalto University to host H.-H.J. and M.K on several occasions.

Finally we all owe the greatest thanks to our families for their continuous support, patience and encouragement during the writing process. They provided us time and the background without which this monograph would not have been written.}

\addcontentsline{toc}{chapter}{Bibliography}
\bibliographystyle{plain}



\printindex


\end{document}